\documentclass[12pt]{article}

\usepackage{scicite}
\usepackage{url}
\usepackage{bm}
\usepackage{times}
\usepackage{multirow}
\usepackage{graphicx}
\usepackage{subfigure}
\usepackage{enumerate}
\usepackage[table,xcdraw]{xcolor}
\renewcommand{\figurename}{Fig.}
\usepackage[]{caption2} 
\renewcommand{\captionlabeldelim}{.}
\usepackage[colorlinks, linkcolor=blue, anchorcolor=blue, citecolor=green]{hyperref}

\newenvironment{sciabstract}{%
\begin{quote} \bf}
{\end{quote}}

\title{Anger makes fake news viral online }

\author
{Yuwei Chuai$^{1}$, Jichang Zhao$^{1,~2,~\ast}$\\
\\
\normalsize{$^{1}$School of Economics and Management, Beihang University, China}\\
\normalsize{$^{2}$Beijing Advanced Innovation Center for Big Data and Brain Computing, China}\\
\normalsize{$^\ast$Correspondence to: jichang@buaa.edu.cn}
}

\date{}

\begin{document} 


\baselineskip24pt


\maketitle 

\begin{sciabstract}
Fake news that manipulates political elections, strikes financial systems, and even incites riots is more viral than real news online, resulting in unstable societies and buffeted democracy. The easier contagion of fake news online can be causally explained by the greater anger it carries. The same results in Twitter and Weibo indicate that this mechanism is independent of the platform. Moreover, mutations in emotions like increasing anger will progressively speed up the information spread. Specifically, increasing the occupation of anger by 0.1 and reducing that of joy by 0.1 will produce nearly 6 more retweets in the Weibo dataset. Offline questionnaires reveal that anger leads to more incentivized audiences in terms of anxiety management and information sharing and accordingly makes fake news more contagious than real news online. Cures such as tagging anger in social media could be implemented to slow or prevent the contagion of fake news at the source.
\end{sciabstract}

\section*{Introduction}
Fake news refers to information that is fabricated, misleading, and verifiably false \cite{Lazer2018,Vosoughi2018}. Most people broadly accept information instead of critically questioning its authenticity \cite{Lazer2018}. In particular, with the boom of social media, on which individuals can be simultaneously producers and consumers of information, ordinary people can easily participate in circulation and gain influence through posting (e.g., tweeting) and reposting (e.g., retweeting). Consequently, the impact of fake news on social media could be disproportionate \cite{Allen2020} and profound \cite{Aral2019}, especially in the political and economic fields \cite{Vosoughi2018,Aral2019,Bovet2019,Allcott2017,Grinberg2019}. In the first few months of the 2016 U.S. presidential election, on average, each adult was exposed to more than one fake news item that was not only widely spread but also deliberately biased \cite{Allcott2017}. Furthermore, fake news is more likely to appear in the highly uncertain conditions of emergencies, such as disease epidemics and outbreaks \cite{Spinney2019,Carey2020}, accidents and conflicts, which makes the spread of fake news a byproduct of the natural response that people have to disastrous events, and social media can be fertile ground for this spread \cite{2020} online.

Fake news is more viral than real (true) news online \cite{Vosoughi2018}. The mechanism underlying its fast spread, though critical, remains unresolved. Unique structural features in the circulation of fake news, such as long diameters of penetration, have been revealed and have been found to be platform independent \cite{DelVicario2016,Zhao2020,Johnson2019,Wang2019}. However, fake news is generally verified to be false after explosive circulation \cite{Iyengar2019}; thus, in the early spread, it is essentially not thought to be fake, so the structural uniqueness is the manifestation of its fast spread, rather than a cause that can fundamentally explain its viral proliferation. Individuals, either human or bots \cite{Shao2018}, posting and reposting fake news on social media are an alternative cause, in particular, the human that occupies the dominant partition \cite{2018}. The spread of news is associated with the friends and followers of the author. Nevertheless, user characteristics fail to sufficiently explain the easy contagion of fake news due to their greater effects on the dissemination of real news \cite{Vosoughi2018}. The content of fake news, which was also found to be entangled with spread \cite{Vosoughi2018,Scheufele2019}, could offer promising directions in probing the mechanism of its fast spread. More importantly, instead of examining spreading structures \cite{Vosoughi2018,DelVicario2016} and reposter demographics \cite{Guess2019} after the circulation was ignited, revealing the mechanism at the source that independent to user demographics would be powerful in inspiring new cures with the minimum invasion of privacy. Hence, we would rather differentiate fake news from real news at the very beginning of their spread through scrutiny of content to figure out new treatments against fake news that can be implemented without delay.

Online news content not only delivers factual information but also carries sophisticated emotional signals. The digital contagion of emotions is embedded in information spread, and involves individuals experiencing the same feelings on social media that they feel in face-to-face emotional exchanges that occur offline \cite{Kramer2014,Goldenberg2020}. Emotions further impact the spread of information, e.g., promoting the sharing of information \cite{Stieglitz2014} or shaping the path of the information \cite{Brady2017}. When the relevance between content quality and popularity is not strong \cite{Acerbi2019}, the emotions involved and their influence on psychological arousal may be key \cite{Vosoughi2018,Weeks2015}. Moreover, the spread of different emotions can inherently be distinguished \cite{Weeks2015}, implying that emotions conveyed by both fake and real news could offer comparative proxy measurements by which to examine the mechanisms underlying their circulation. In fact, fake news is preferentially injected with emotions such as anger for political attacks \cite{Higgins2017}. However, differentiating fake news from real news is rarely based on emotions delivered in the content and incentives beyond reposting in extant efforts. The discrepancy in users' perceptions between fake news and real news are unraveled in the emotions of the replies \cite{Vosoughi2018}, while the emotions that inherently carried by the news itself are not considered in explaining circulation. In fact, the negative emotions in content have been shown to cause positive responses (e.g., sympathy) \cite{Wang2017}, meaning emotions, particular in the negative parts, should be directly examined when studying the spread of fake news. At the same time, although social media content can be short, simplifying the emotions it carries into a single emotion might cause the emotional richness of the content to be missed \cite{Vosoughi2018,Du2014,Penz2011} and lead to a failure of emotional recognition and inconsistent results \cite{Goldenberg2020,Weeks2015,Berger2012}.

In this study, by successfully combining digital traces on social media and offline questionnaires, we aim to unravel the mechanism underlying the fast spread of fake news by answering three key questions: What are the differences in the emotional distributions of real and fake news? Can these differences explain why fake news is more infectious than real news? How do they affect the incentives behind news reposting?

\section*{Results}
Considering the diversity of news diffusion platforms, besides collecting a large dataset of both fake news and real news from Weibo, the most popular Twitter-like service in China, we also gathered datasets from Twitter and mainstream news media in the west (see Materials and Methods and SM S1). On the basis of the number of followers on behalf of the broadcasting potential of authors and the number of retweets on behalf of the spreading capability of news \cite{Wang2019}, we built a division model and assembled both categories of news into treatment and control groups. For example, taking fake news with low numbers of followers and high volumes of retweets (LHF news) as the treatment group, the controlled counterparts consist of either fake news with high volumes of followers and low numbers of retweets (HLF news) or true news with high volumes of followers and low numbers of retweets (HLT news) (see SM S2 for details). Accordingly, by intentionally selecting news that is weakly retweeted but posted by highly followed authors, the possible effects from users can be controlled to amplify the spread promotion resulting from the particular emotion content it carries. Moreover, although fake news is statistically more contagious (longer path, faster speed, lasts longer, and gets more retweets) than real news (see SM S2.3 and SM S3), not every fake news item is necessarily more viral than any real news item. For instance, the diffusion capability of highly retweeted true news is definitely more powerful than that of lowly circulated fake news. Therefore, we would compare LHF news with HLF news and HLT news first and then extend the comparison to the full spectrum of discrepancies between true (T) news and fake (F) news in terms of emotions.

Emotional signals carried in either fake or real news can be sophisticated, i.e., a combination of elementary compounds rather than a single one \cite{Penz2011}. The distribution of five emotions that represent basic human feelings \cite{Vosoughi2018,Sauter2010,Ekman1992}, namely, anger, disgust, joy, sadness and fear, is inferred for each news item in our data through a lexicon that is manually labeled to cover 87.1\% of news items with the remaining considered neutral (see Materials and Methods and SM S4-5). Emotions with a strong presence in the distribution are the feelings that the sender of the news wishes the receivers to experience \cite{Bollen}. The proportion of anger (Fig. 1A) in LHF news is expected to be significantly higher than that in both HLF and HLT news, while joy is expected to be lower (Fig. 1E). The comparison is then extended to a full spectrum between all fake news and real news, and consistent results, though with shrinking gaps for anger and joy, as expected, are obtained (Figs. 1B and F). Furthermore, the dominance of anger in fake news (especially highly retweeted news) and joy in real news (even lowly retweeted news) is further confirmed with better resolution in the distribution of emotions of keywords that precisely separate the treatment groups from control groups (see SM S6 and Fig. S10). These observations persistently suggest that fake news carries more anger yet less joy than real news and imply the possibility that anger might promote the fast spread of fake news online. The divergence in anger and joy between fake news and real news is robust and independent of emotion inference models and emotion distribution measures (see SM S7). Even in specific events like COVID-19, the dominance of anger to joy in highly retweeted fake news conformably suggests the promotion of anger in the fast spread of fake news (see SM S7.3). By contrast the near overlap in disgust between different types of news (Fig. 1C and D), the less occupation of sadness more than 0.5 in fake news than that in real news (Fig. 1H), and the more dominant position of fear in HLF news (Fig. 1I) indicate their less positive roles in the virality of fake news \cite{Stieglitz2014,Berger2012}. Therefore, significant gaps across news groups could also be independent of circulation, and well-controlled causal inference is accordingly necessary for anger and joy.

To causally infer and qualify the promotion of anger and the prevention of joy in the spread of fake news, internal factors related to content \cite{Suh}, user \cite{Vosoughi2018} and external shocks such as disaster events \cite{Spinney2019} should be comprehensively controlled. Specifically, internal factors, including mention, hashtag, location, date, URL, length, topic, other emotions, follower (number of followers), friend (number of reciprocal followers) and external shocks including emergency (a disaster event) constitute control variables (see SM S8) in the logit and linear inference models (see SM S9). The results of the logit model (see SM S9) for lowly retweeted true (LT) news (control group) and highly retweeted fake (HF) news (treatment group) show that the coefficient of anger is significantly positive and the coefficient of joy is negative (Table 1 (1)), implying that anger causally promotes the fast spread of fake news online. Other emotions are omitted (Table 1(1)) due to multicollinearity and their trivial impact on circulation. Moreover, for the logit model used to estimate all true and fake news, anger is positively associated with fake news, though with a smaller coefficient and narrower deviation, as anticipated (Table 1(2)). Recalling the gaps observed in emotion distributions across groups of news, all the results consistently suggest the positive promotion effect of anger in the circulation of fake news, particularly for news that is highly retweeted. The causally negative relationship between joy and fake news contrarily indicates its prevention in dissemination. To further qualify the influence of both anger and joy in the spread of fake news, a linear regression model with the number of retweets as the dependent variable is established (see SM S9). It is congruously found for fake news and all news that the coefficients of anger are significantly positive while the coefficients of joy are negative (Table 1 (3) and (4)), suggesting that anger can promote circulation and joy can prevent the spread. Specifically, supposing that other factors are fixed, increasing the occupation of anger by 0.1 and reducing that of joy by 0.1 in fake news leads to 5.8 more retweets, and 2.2 more retweets occur if anger is increased by 0.1 in place of other negative emotions but joy is fixed. The above causal relationships between emotions and circulation are robust to alternative emotion detection approaches such as competent machine learning models (see Table S17). For other significant factors, although mentions can promote the spread of news (Table 1(3) and (4)), the coefficient is not significant for LHF news (Table 1(1)) and even prevents the spread of fake news (Table 1(2)); emergency is significantly positive in the logit models (Table 1(1) and (2)) but inconsistently negative in the linear models (Table 1(3) and (4)) (see SM S8 for more details). Therefore, carrying more anger and less joy is the mechanism behind the fast spread of fake news that makes it more viral than real news online. More importantly, additional evidence from extensive datasets of English news on both Twitter and mainstream media further confirms that this mechanism is independent of the platform (see SM S10).

Negative stimuli such as anger elicit stronger and quicker emotional reactions and even behavioral responses than positive stimuli such as joy \cite{Baumeister2001}. The odds of being forwarded through e-mails are also causally impacted by the physiological arousal caused by emotional articles, and those evoking high-arousal positive or negative emotions could be more viral \cite{Berger2012}. In the spread of fake news, the incentives behind the action of reposting that reignite circulation are therefore hypothetically associated with the anger and joy the news carries. Taking LHF news as the treatment group and HLF news and HLT news as the control groups, the possible associations between reposting incentives and emotions are examined through offline questionnaires. By selecting 15 typical news items with keywords from these groups (see Materials and Methods and SM S11), questionnaires are implemented to investigate four motivations for news reposting on social media \cite{Sudhir2014}, including anxiety management, information sharing, relationship management, and self-enhancement. The subjects of the surveys are Weibo users, and the overlapping between offline subjects and online users is ensured (see SM S12). Preliminary results indicate that the motivation of anxiety management in LHF news is significantly higher than that in the control groups (Fig. 2A). Moreover, compared to HLT news, subjects are more intensively incentivized to share information when reposting HLF news and LHF news (Fig. 2B). Thus, fake news can stimulate strong motivation for information sharing; in particular, news that is widely disseminated can also strengthen the motivation for anxiety management. There is no significant variation in the motivation for relationship management across news groups (Fig. 2C), and the motivation for self-enhancement in HLT news is stronger than that in fake news (Fig. 2D). What is more interesting is that in questionnaires with keywords highlighted with marks, the unique stimuli of widely circulated fake news for anxiety management is strengthened (see Fig. S23A). The incentive of information sharing is similarly enhanced for fake news (see Fig. S23B). All these results imply that the responses to the anger carried by fake news are sharing information and even managing anxiety. To validate this finding, the news in questionnaires is further split into anger-dominated news and joy-dominated news (see SM S13.2) to directly probe the impact of emotions. Compared to the retweeting motivations of joy-dominated news, anger-dominated news stimulates stronger incentives for anxiety management (Fig. 2E) and information sharing (Fig. 2F). Joy-dominated news ultimately excites stronger self-enhancement (Fig. 2H) than anger-dominated news. Meanwhile, no significant difference is observed between anger and joy in terms of relationship management motivation (Fig. 2G). Shuffling emotions randomly further testifies to the significance of these observations (see SM S13.2). Therefore, the greater anger delivered in fake news leads to more incentivized audiences with respect to anxiety management and information sharing, resulting in a greater likelihood of retweets and, thus, more viral contagion.

\section*{Discussion}
Our findings emphasize the necessity of considering emotions, particularly anger, in understanding the spread of information online. On social media, the associations between information diffusion and embedded emotions have been noted for a long time; however, the profiles of the roles of both positive and negative emotions are inconsistent and even contradictory across diverse contexts \cite{Goldenberg2020}. Considering the heterogeneous influence on spreading from negative emotions such as anger and sadness \cite{Stieglitz2014,Berger2012}, the causal impact on information diffusion should be examined with respect to well-resolved negative emotions. Instead of simplifying emotions binarily into positive and negative emotions, more elementary emotions are considered in this study, and the distribution of five emotions is inferred to reflect the complete emotional spectrum of news online. This more detailed spectrum of emotions identifies anger’s unique role in provoking strong incentives of anxiety management and information sharing, which results in the virality of fake news online. From this perspective, emotions could be genes of fake news circulation, and similar to small mutations, they could make the virus go viral. Mutations that increase anger or reduce joy in fake news enhance its likelihood of being retweeted. Additionally, fake news is more focused on societal (including politics) and financial topics (see SM S8 and S10), which further implies that anger can promote the spread of fake news about these topics more efficiently. Distinguishing structures in the circulation of fake news could also be deciphered based on the anger such news predominantly carries since anger prefers weak ties in social networks \cite{fan2020} and may inherently forge the diffusion structure of fake news. Meanwhile, the role of joy in preventing spread, especially in fake news, underlines the fundamentality of considering negative emotions of fine granularity to control and deepen future explorations. Therefore, it is anticipated that insights from emotions will improve the extant understanding of online information spread.

The vigorous promotion in circulation from anger implies new weapons against fake news. Although structural signals can be sensed at an early stage to target fake news \cite{Zhao2020}, fake news spreads rapidly and reaches the peak of new retweets in less than one hour (see Fig. S7), so the negative impact has been exposed to a large population of audiences before identification. Moreover, it can take more than three days for a post to be rated as false by outside fact-checkers on Facebook. What is worse, like a cat-and-mouse game between manipulation and detection, features derived from content or users that were found to be helpful in machine learning on targeting fake news \cite{Shu2017} can be easily converted to inspire future countermeasures for fabricating more sophisticated false news. In particular, fake news related to emergencies is widely disseminated because of its clever combination with anger, which may explain why efforts to counter misperceptions about diseases during epidemics and outbreaks are not always effective \cite{Carey2020}. Inefficient or ineffective efforts to detect fake news and debunk misinformation by correcting both calls for new treatments and preventing the spread of anger could be a profound and promising direction. The early deviation in dissemination paths between fake news and real news suggests the rapid effect of anger in shaping retweeting \cite{Brady2017}. For example, platforms such as Facebook, Twitter, and Weibo should warn and discourage users as they try to retweet news that delivers too much anger and persuade them to assess the credibility of the information more critically. The trade-off between free speech and fake news prevention is the prime principle; however, a better balance would be achieved by tagging angry news (e.g., with an occupation of anger of more than 20\%, see SM S14 for more details) at the very beginning to make audiences and potential spreaders less emotional and more rational \cite{Bullock2015}.

\section*{Materials and Methods}
\subsection*{Data collection}
We collected eight datasets in total. The main dataset collected on Weibo from 2011 to 2016 includes 10,000 true news items (with 409,865 users) posted by credibly verified users and 22,479 fake news items (with 1,189,186 users) endorsed by an official Weibo committee after wide dissemination (see SM S1 for more details). Given the scale and representativeness of the dataset, we used it to conduct the whole research process. Meanwhile, with the proliferation of fake news during the COVID-19 epidemic and the persistence of political fake news, another dataset related to COVID-19 was collected from Weibo to validate the results in the background of emergency incidents (see SM S7.3 for more details). Besides, six more English datasets from Twitter and mainstream news media in the west were also collected for supplementary evidence. Specifically, Dataset S1-4, composed in total of 129,690 news items centered around two topics: COVID-19 and the 2016 United States election, were combined to examine the effects of emotions in information spread; Dataset S5-6, composed by 23,959 fake news items and 21,417 real news items, was employed to reveal the mechanism beyond fake news virality. More details can be found in SM S10. We further conducted offline questionnaires to profile retweeting incentives towards fake news. The design of the questionnaires ensures that other influencing factors on the interface are consistent, and only representative real or fake news information is used to stimulate users. A total of 1,291 valid responses from 1,316 questionnaires were collected from active Weibo users.

\subsection*{Methods}
According to the number of followers and retweets, a division model was built to divide the news into treatment and control groups. Author groups with more than $10^3$ followers have more influence than groups with less than $10^3$ followers (Fig. S3). For all true and fake news, approximately 97\% of the structural virality \cite{Goel2016} is lower than 2 when the number of retweets is less than 10 (Fig. S4). Meanwhile, fake news is more viral (longer average path) than true news (K-S test $\sim$ 0.159, P $\sim$ 0) in terms of structural virality, which is consistent with previous results on Twitter \cite{Vosoughi2018}, implying the universality of our dataset from Weibo. Six typical diffusion networks of both fake and real news are also shown in Fig. S5 to further illustrate this (see SM S2 for more details). Starting from the treatment and control groups, we analyzed the differences of emotion distributions between true and fake news. There are three ways to calculate the distribution of emotions, namely, emotion lexicon, machine learning models, and deep neural networks. For emotion lexicon, we segmented all the text into terms and composed a candidate set (see SM S4 for more details). Nine well-instructed coders screened 6,155 emotional terms through a WeChat applet, named Word Emotion. In the meantime, we conducted a statistical test of emotional differences using the K-S test. Considering control variables related to content, user profiles, and external shocks (see SM S8 for more details), we built logit models to verify the emotional differences between true and fake news and built linear models to analyze the influences of different emotions on the number of retweets (see SM S9 for more details). Aiming at identifying the effects of emotions carried in the tweets on users, we selected 15 typical news items for questionnaires with the help of K-Means clustering (see SM S11 for more details). For the responses to the questionnaires, we analyzed the differences in motivations among different groups after eliminating subjective bias (see SM S13 for more details).

\clearpage

\clearpage
\section*{Acknowledgments}
\textbf{Funding:} This work was supported by the National Key Research and Development Program of China (Grant No. 2016QY01W0205) and NSFC (Grant No. 71871006). \textbf{Author contributions:} YC conducted the analysis and wrote the manuscript. JZ conceived of the study, conducted the analysis, wrote the manuscript and oversaw the work. \textbf{Competing interests:} The authors declare no competing interests. \textbf{Data and materials availability:} All data and code used in this study are publicly available through the permanent link \url{https://doi.org/10.6084/m9.figshare.12163569.v2}.

\clearpage     
\section*{Supplementary materials}
Materials and Methods\\
Figs. S1 to S25\\
Tables S1 to S27\\

\clearpage
\begin{figure}[ht]
	\centering
	\includegraphics[scale=0.4]{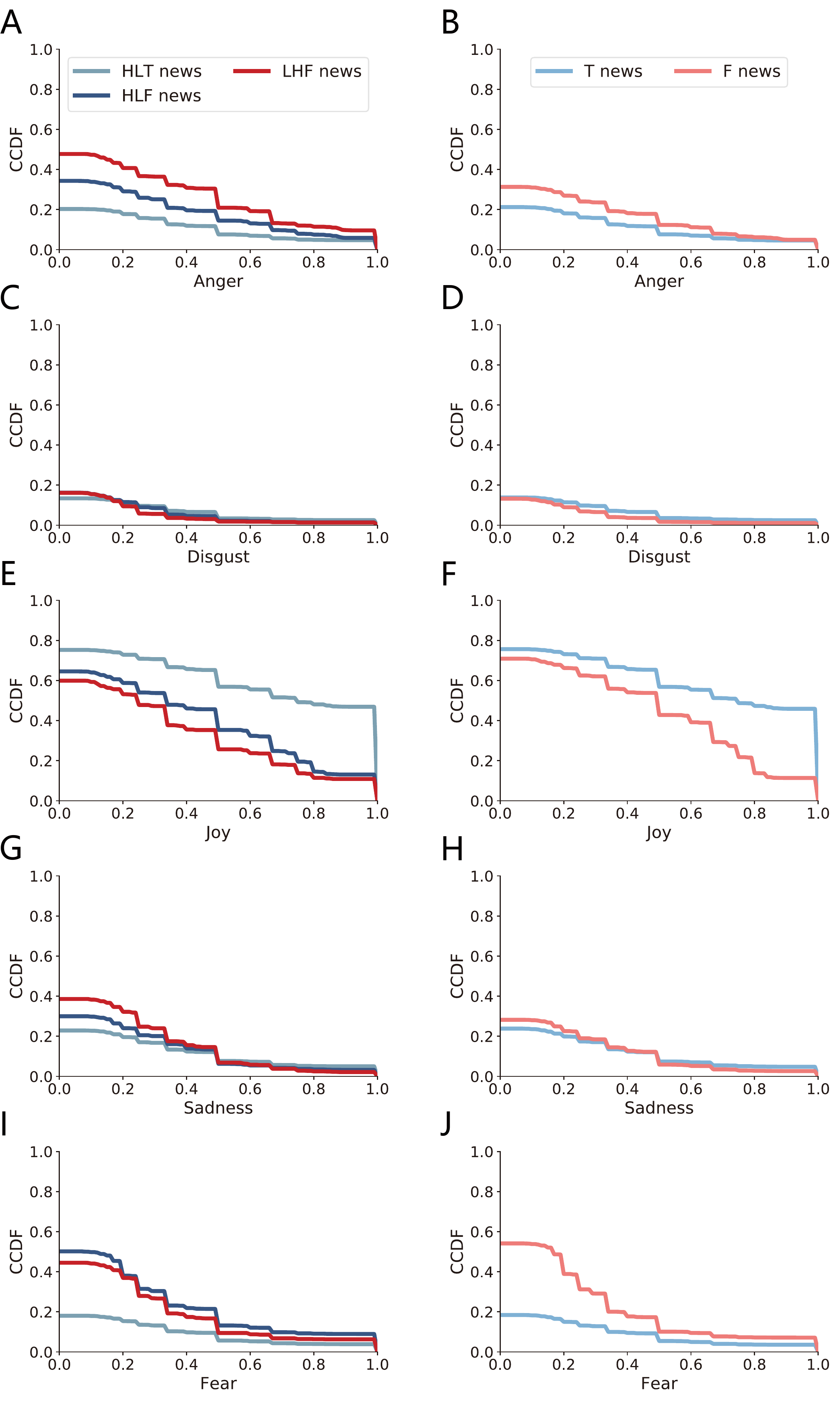}
	\caption{Complementary cumulative distribution functions (CCDFs) of emotions. (A and B) The proportion of anger. The proportion of anger greater than 0.5 in LHF news is nearly 3 times as much as that in HLT news (A). (C and D) The proportion of disgust. (E and F) The proportion of joy. The proportion of joy greater than 0.5 in HLT news is more than 2 times as much as that in LHF news (E). (G and H) The proportion of sadness. (I and J) The proportion of fear. The results of K-S tests are shown in SM S5, and consistent results from other methods can be seen in SM S7.}
	\label{fig:emo_ccdf}
\end{figure}

\begin{figure}	
	\centering
	\includegraphics[scale=0.4]{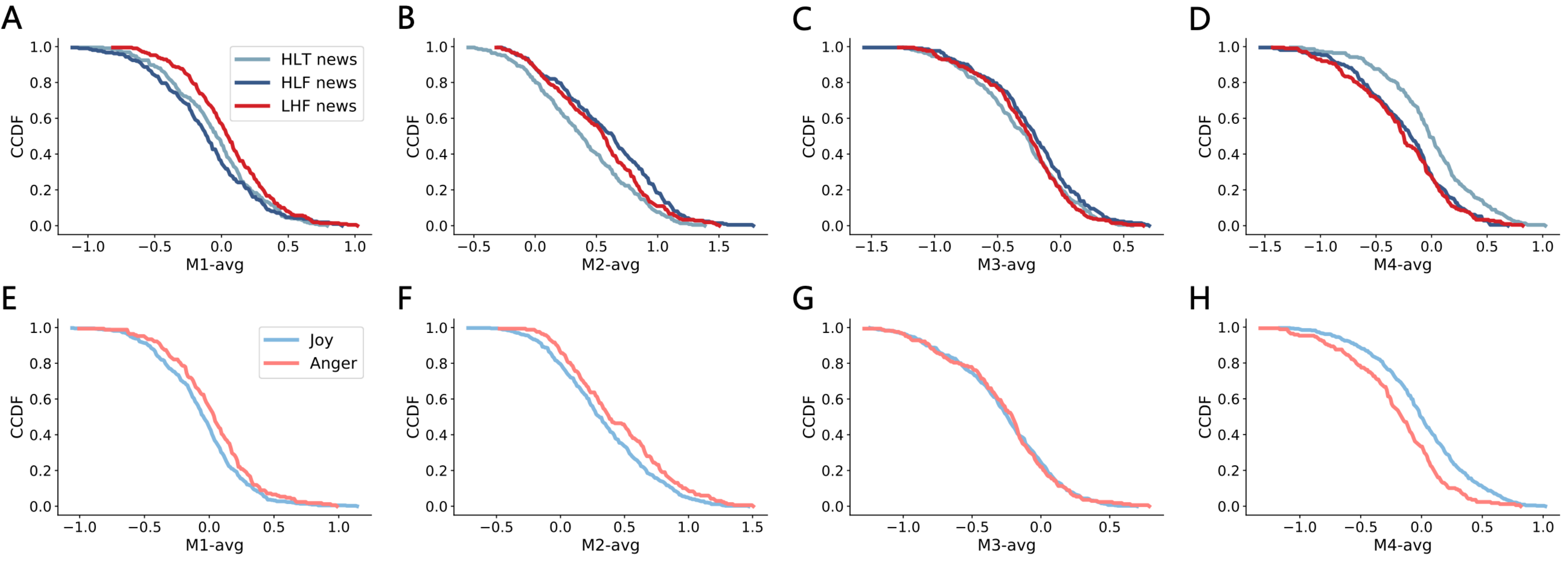}
	\caption{The CCDFs of motivations. (A and E) Anxiety management (M1-avg). (B and F) Information sharing (M2-avg). (C and G) Relationship management (M3-avg). (D and H) Self-enhancement (M4-avg). (A to D) The CCDFs of four motivations in HLT news, HLF news and LHF news. (E to H) The CCDFs of four motivations in anger-dominated news and joy-dominated news. The results of the K-S tests can be seen in SM S13.}
	\label{fig:moti_emo_ccdf_2}
\end{figure}

\begin{table}[ht]	
	\centering
	\includegraphics[scale=0.7]{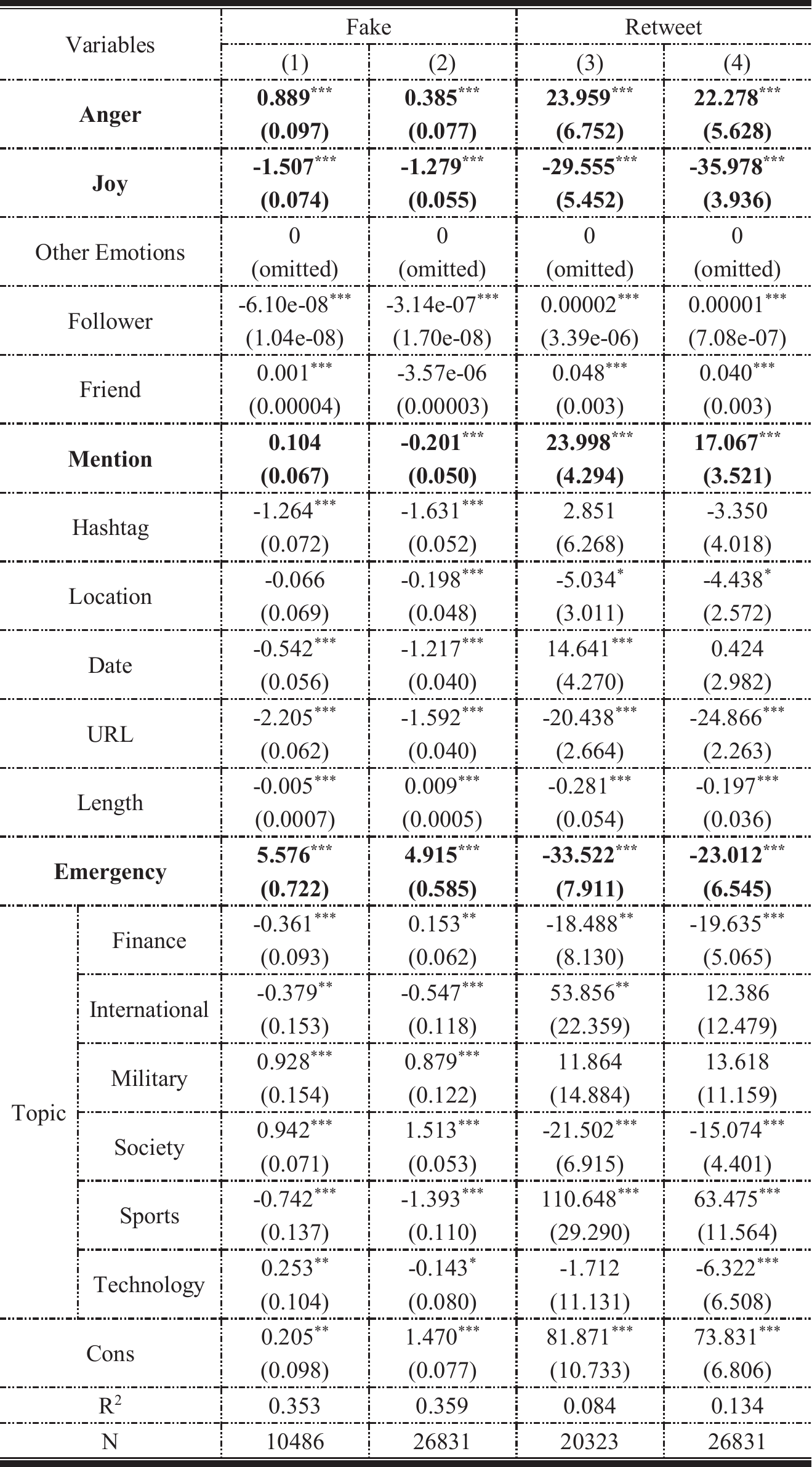}
	\caption{The results of logit and linear models in different groups. (1) The results of the logit model in LT news and HF news. (2) The results of the logit model in all true news and fake news. (3) The results of the linear model in LF news and HF news. (4) The results of the linear model in lowly retweeted (L) news and highly retweeted (H) news (see SM S9 for more details). The values in brackets are the robust standard errors. $^\ast P<0.1, ^{\ast\ast} P<0.05, ^{\ast\ast\ast} P<0.01$.}
	\label{table:table1}
\end{table}
\clearpage

\setcounter{table}{0}  
\renewcommand{\thetable}{S\arabic{table}}
\setcounter{figure}{0}  
\renewcommand{\figurename}{Fig.}
\renewcommand{\captionlabeldelim}{:}
\renewcommand{\thefigure}{S\arabic{figure}}

\vspace*{\fill}
\begin{center}
	{\huge Supplementary Materials for}\\	
	{\Large Anger makes fake news viral online}\\
	\author
	{Yuwei Chuai$^{1}$, Jichang Zhao$^{1,2,\ast}$\\
		\normalsize{$^{1}$School of Economics and Management, Beihang University, China}\\
		\normalsize{$^{2}$Beijing Advanced Innovation Center for Big Data and Brain Computing, China}\\
		\normalsize{$^\ast$Correspondence to: jichang@buaa.edu.cn}
	}\\	
\end{center}
\textbf{This PDF file includes:}\\
Materials and Methods\\
Figs. S1 to S25\\
Tables S1 to S27\\
\vspace*{\fill}
\clearpage

\section*{S1 Fake news and Real news}
The fake news and real news in this study were collected from Weibo, the most popular Twitter-like service in China, which had 200 million daily active users and generated over 100 million daily tweets (news) at the end of 2018 (\url{https://data.weibo.com/report/reportDetail?id=433}). Here, news refers to tweets including news-related content on Weibo. The users of Weibo are dominated by young people, and those aged between 18- and 30-years old account for 75\% of all users. There is also a distinctive verification mechanism in Weibo that ensures the reliability of the user demographics. Specifically, all users have to provide their IDs during registration because of the real-name certification regulation in China. Besides, influential users, including elites with a certain reputation and influence in specific domains, well-known enterprises and their executives, the mainstream media, and government agencies such as public authorities, can be further manually verified through documentary evidence \cite{2020a}. Weibo even presents red or blue badges on their online profiles. Weibo officially organizes a committee composed of professional fact-checkers outside Weibo to tag fake news authoritatively and publicly.

Through the open API of Weibo, we collected fake news rated and exposed by the official committee. Considering that fake news always draws attention from the committee after being widely disseminated, the digital traces of the spread of such news on Weibo can be completely traversed. Further probes on the timelines of all news items confirm this fact in S3. Real news, also termed true news in this study, refers to information that was not tagged as false by the committee and was posted by verified users, such as mainstream media, elites, or public authorities, with credibility. In total, we collected 22,479 fake news items (with 1,189,186 users) and 10,000 real news items (with 409,865 users) from 2011 to 2016. For each news item on Weibo, we also collected its attributes, namely, text, posting time, author profile (number of followers, number of reciprocal followers, etc.), retweets, and reposting time. A subset of the fake news and real news used in this study was employed in a previous study \cite{Zhao2020} on the structural uniqueness of fake news, in which equivalent results are derived from both Weibo and Twitter, implying the reliability and universality of our data. Additionally, authentic tweets from credible nonverified authors of Weibo further testified the representativeness of our real news data \cite{Zhao2020}. We have made the data publicly available at \url{https://doi.org/10.6084/m9.figshare.12163569.v2}. 
\clearpage

\section*{S2 News groups}
\subsection*{S2.1 Partition strategy}
The number of followers intuitively represents the influence of users on social media, i.e., more followers means the news will be broadcast to a larger audience and accordingly result in more retweets. Additionally, the number of retweets can represent the spreading capability of a given news item. Fake news might be widely retweeted because of the influence of its author; however, the broadcasting potential of authors does not sufficiently explain the fast spread of fake news \cite{Vosoughi2018}, e.g., fake news posted by lowly followed authors might be massively retweeted. To examine the causal impact of emotions on the circulation of fake news, treatment groups and control groups are established to control for variables and infer the significant roles of emotions underlying the spread. Considering that the role of emotions in information spreading might be subtle and easily interfered with by other variables, such as the influence of authors, we aim to split news, either fake or real, into a treatment group (e.g., highly retweeted news posted by authors with a low volume of followers) and a control group (e.g., lowly retweeted news posted by authors with a high volume of followers), through which the possible influence of authors can be controlled and the effects of emotions are amplified. Intuitively, for highly retweeted news posted by authors with a low volume of followers, promotion from the content, in particular, the emotions carried, would be more powerful and thus easier to detect. Therefore, we group the news according to the number of its author’s followers ($x$) and the number of retweets ($y$) \cite{Wang2019}. For example, based on real news with a high number of followers but a low number of retweets and fake news with a low number of followers but a high number of retweets, a division model of maximizing the difference between true and fake news is defined to determine the splitting interface, which is specified as
$$D\ =\ \left(\frac{N_{LHF}}{N_F}-\frac{N_{LHT}}{N_T}\right)+\left(\frac{N_{HLF}}{N_F}-\frac{N_{HLT}}{N_T}\right)-\left|\frac{N_{LLF}}{N_F}-\frac{N_{LLT}}{N_T}\right|-\left|\frac{N_{HHF}}{N_F}-\frac{N_{HHT}}{N_T}\right|,$$

\noindent where
\begin{itemize}
	\item $N_T$ is the number of true (T) news items.
	\item $N_F$ is the number of fake (F) news items.
	\item $N_{LLT}$ is the number of true news items with a low number of followers ($<x$) and a low number of retweets ($<y$).
	\item $N_{LHT}$ is the number of true news items with a low number of followers ($<x$) and a high number of retweets ($\geq y$).
	\item $N_{HHT}$ is the number of true news items with a high number of followers ($\geq x$) and a high number of retweets ($\geq y$).
	\item $N_{HLT}$ is the number of true news items with a high number of followers ($\geq x$) and a low number of retweets ($<y$).
	\item $N_{LLF}$ is the number of fake news items with a low number of followers ($<x$) and a low number of retweets ($<y$).
	\item $N_{LHF}$ is the number of fake news items with a low number of followers ($<x$) and a high number of retweets ($\geq y$).
	\item $N_{HHF}$ is the number of fake news items with a high number of followers ($\geq x$) and a high number of retweets ($\geq y$).
	\item $N_{HLF}$ is the number of fake news items with a high number of followers ($\geq x$) and a low number of retweets ($<y$). 
\end{itemize}

We let the number of followers (from 10 to $10^4$) and the number of retweets (from 10 to $10^8$) grow exponentially with a step size of 1 to maximize the value of $D$ and find the optimal partition line. As shown in Fig. S1, the best tuple is $\left(x^\ast,\ y^\ast\right)=\left(10,\ 1000\right)$.

According to the tuple $\left(10,\ 1000\right)$, we divide the news into low volume of followers and lowly retweeted true (LLT) news, low volume of followers and highly retweeted true (LHT) news, high volume of followers and highly retweeted true (HHT) news, high volume of followers and lowly retweeted true (HLT) news, low volume of followers and lowly retweeted fake (LLF) news, low volume of followers and highly retweeted fake (LHF) news, high volume of followers and highly retweeted fake (HHF) news and high volume of followers and lowly retweeted fake (HLF) news (Fig. S2). Lowly retweeted true (LT) news includes LLT news and HLT news, highly retweeted true (HT) news includes LHT news and HHT news, lowly retweeted fake (LF) news includes LLF news and HLF news and highly retweeted fake (HF) news includes LHF news and HHF news. Additionally, ignoring the label of fake or true, lowly retweeted news is categorized as L news, and highly retweeted news is categorized as H news. By pairing various groups, diverse assemblies of treatments and controls can be established to examine the causal impact of emotions on circulation. Specifically, HLT news accounts for the largest proportion of true news, and LLF news accounts for the largest proportion of fake news (Table S1).

\begin{figure}
	
	\centering
	\includegraphics[scale=0.8]{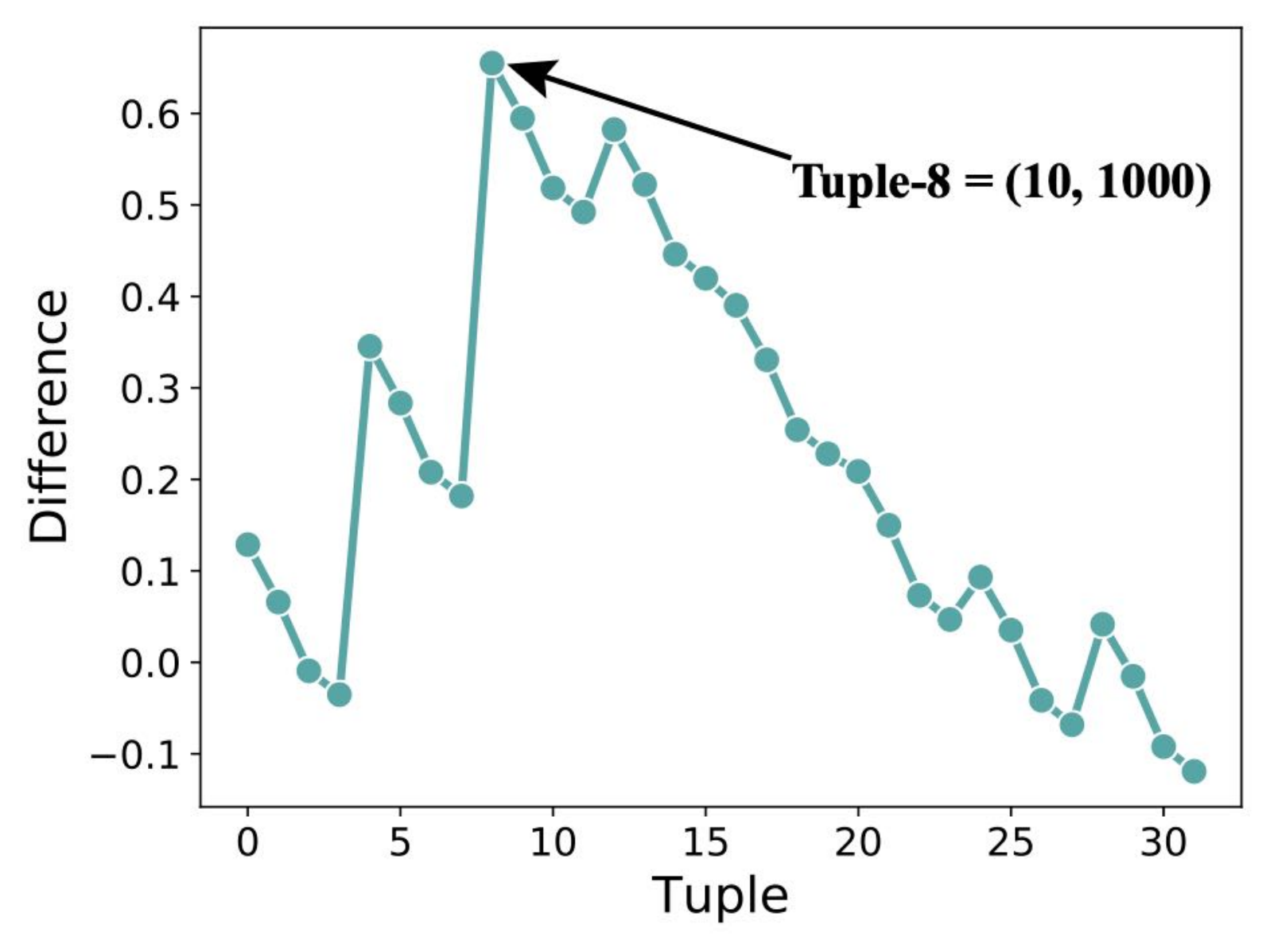}
	\caption{The difference ($D$) varies with the tuple $\left(x,y\right)$, where $x={10}^i\ \left(i=1,2,3,4\right)\ and\ y=\ {10}^j\ \left(j=1,2,\cdots,8\right)$.}
	\label{fig:best_tuple}
\end{figure}

\begin{figure}
	
	\centering
	\includegraphics[scale=0.5]{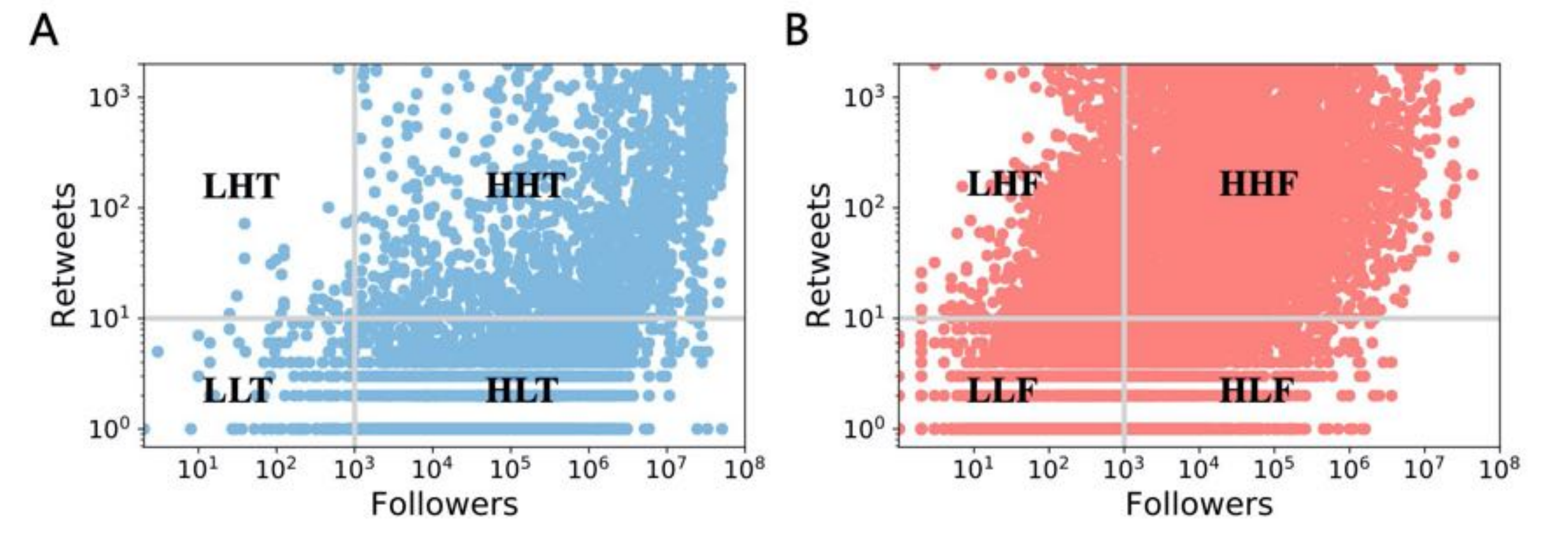}
	\caption{The scatter plots of news items. (A) Scatter plot of true news. (B) Scatter plot of fake news.}
	\label{fig:news_groups}
\end{figure}

\begin{table}
	
	\centering
	\includegraphics[]{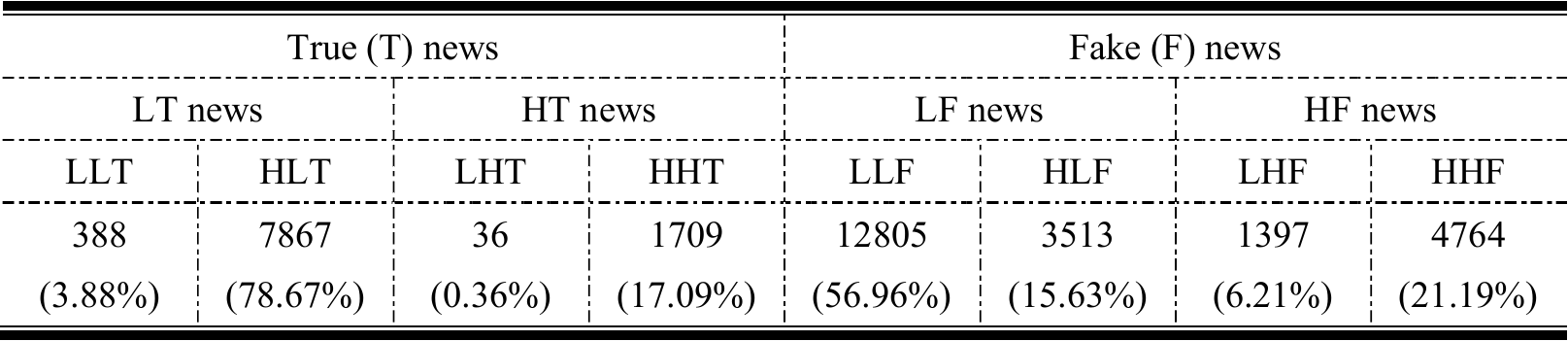}
	\caption{Numbers and proportions of all groups of both fake and real news items.}
	\label{table:news_groups_ratio}
\end{table}

\subsection*{S2.2 Information dominance}
To verify the rationality of the partition strategy in S2.1, we first examine the information dominance between different author groups. Here information dominance measures to which extent the authors of news items could dominate the spread in other spreader groups. According to their numbers of followers ($x$) , all users are divided into eight groups, including $G_0$ (users whose follower counts fall in the interval $\left[0,10\right)$), $G_1$ (users whose follower counts fall in the interval $\left[10,{10}^2\right)$), $G_2$ (users whose follower counts fall in the interval $\left[{10}^2,{10}^3\right)$), $G_3$ (users whose follower counts fall in the interval $\left[{10}^3,{10}^4\right)$), $G_4$ (users whose follower counts fall in the interval $\left[{10}^4,{10}^5\right)$), $G_5$ (users whose follower counts fall in the interval $\left[{10}^5,{10}^6\right)$), $G_6$ (users whose follower counts fall in the interval $\left[{10}^6,{10}^7\right)$), and $G_7$ (users whose follower counts fall in the interval $\left[{10}^7,\infty\right)$). The information transmitted from the news item $m$ in$G_i$ (if the author of $m$ belongs to $G_i$, $m$ is accordingly split to $G_i$) to $G_j$ is defined as

$$T_{i,m,j}=\frac{N_{i,m,j}}{\sum_{g=1}^{G}N_{i,m,g}},$$

\noindent where $N_{i,m,j}$ is the number of spreaders belonging to $G_j$ in the retweets of $m$ in $G_i$ and $G$ is the number of groups. Meanwhile, the coverage of $m$ to $G_j$ is defined as

$$C_{i,m,j}=\frac{N_{i,m,j}}{N_j},$$

\noindent where $N_j$ is the number of users belonging to $G_j$. According to $T_{i,m,j}$ and $C_{i,m,j}$, the transmission coverage of $G_i$ to $G_j$ is defined as

$${TC}_{i,j}=\frac{1}{M_i}\sum_{m=1}^{M_i}T_{i,m,j}C_{i,m,j},$$

\noindent where $M_i$ is the number of news items in $G_i$. Then, the information dominance of $G_i$ to $G_j$ is

$$ID\left(G_i,G_j\right)=\frac{{TC}_{i,j}-{TC}_{j,i}}{{TC}_{i,j}+{TC}_{j,i}}.$$
When the information dominance of $G_i$ ($G_{out}$) to $G_j$ ($G_{in}$) is positive, i.e., $ID\left(G_i,G_j\right)>0$, it is defined that $G_i$ has more information influence as compared to $G_j$. As shown in Fig. S3, since $G_2$, the information dominance of $G_{out}$ to $G_{in}$ is constantly larger than 0.5, implying authors with numbers of followers higher than ${10}^3$ indeed possess more information influence. Hence, it is reasonable to divide L users (with low influence) and H users (with high influence) by ${10}^3$ according to our partition strategy.

\begin{figure}
	\centering
	\includegraphics[]{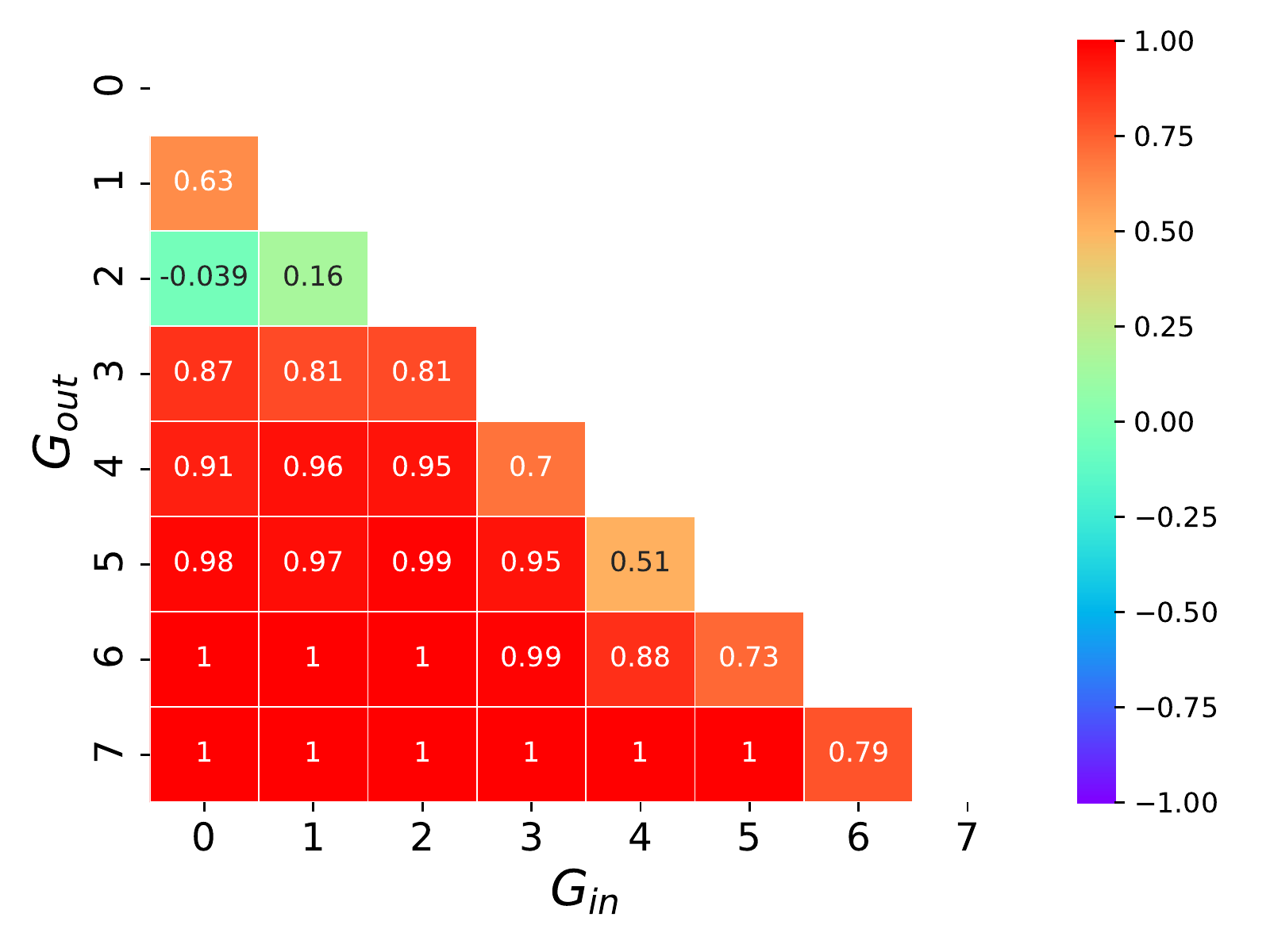}
	\caption{The information dominance of $G_{out}$ to $G_{in}$.}
	\label{fig:dominant_hmap}
\end{figure}

\subsection*{S2.3 Structural virality}
The spreading capability of news may not be comprehensively represented by the number of retweets, and the diffusion structure can also reflect the very viral nature of news. Therefore, we further examine the rationality of the partition strategy according to retweeting number ($y$) in S2.1 from the perspective of circulation structure. The structural virality is the average distance between all pairs of nodes in a diffusion \cite{Goel2016}, which can measure the diversity of diffusion structure. It is defined as

$$v\ =\ \frac{1}{n(n-1)}\sum_{i=1}^{n}\sum_{j=1}^{n}d_{i,j},$$

\noindent where $d_{i,j}$ denotes the length of the shortest path between nodes $i$ and $j$. When $v \sim 2$, it can be thought an approximately pure broadcast \cite{Goel2016}. The average structural virality of news diffusion with the number of retweets is shown in Fig. S4. For all true and fake news, approximately 97\% of the structural virality is lower than 2 when the number of retweets is less than 10, which is exactly same to the cutting point previously obtained, verifies the reliability of the division in S2.1 and again consolidates our partition strategy of news groups for treatment and control. Meanwhile, fake news is more viral (longer average path) than true news (K-S test $\sim$ 0.159, P $\sim$ 0) in terms of structural virality, which is consistent with previous results on Twitter \cite{Vosoughi2018}, implying the universality of our dataset from Weibo. Six typical diffusion networks of both fake and real news are also shown in Fig. S5 to further illustrate this point. 

\begin{figure}
	\centering
	\includegraphics[scale=0.6]{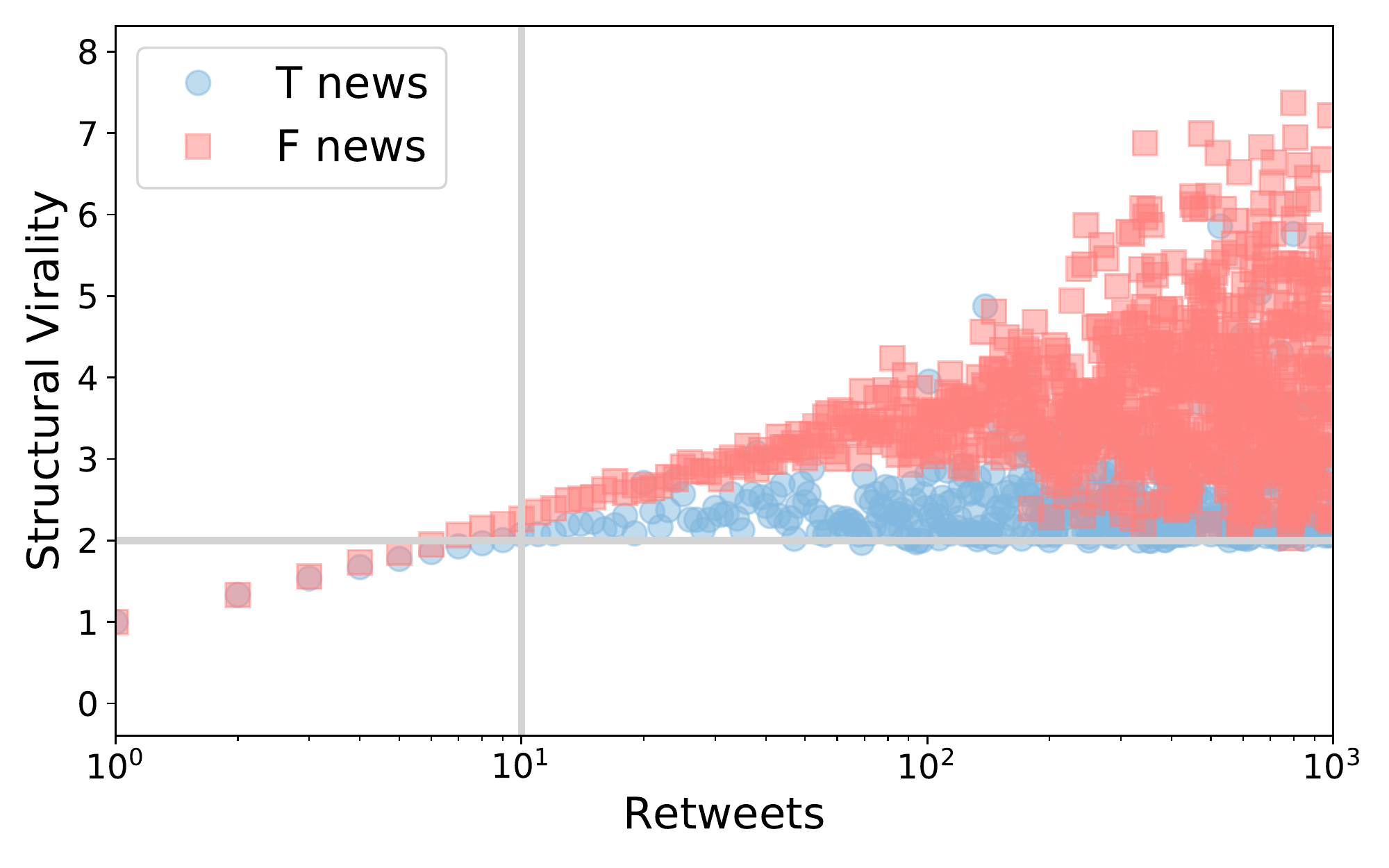}
	\caption{The average structural virality with the growing retweets.}
	\label{fig:StructuralVirality}
\end{figure}

\begin{figure}
	\centering
	\includegraphics[scale=0.35]{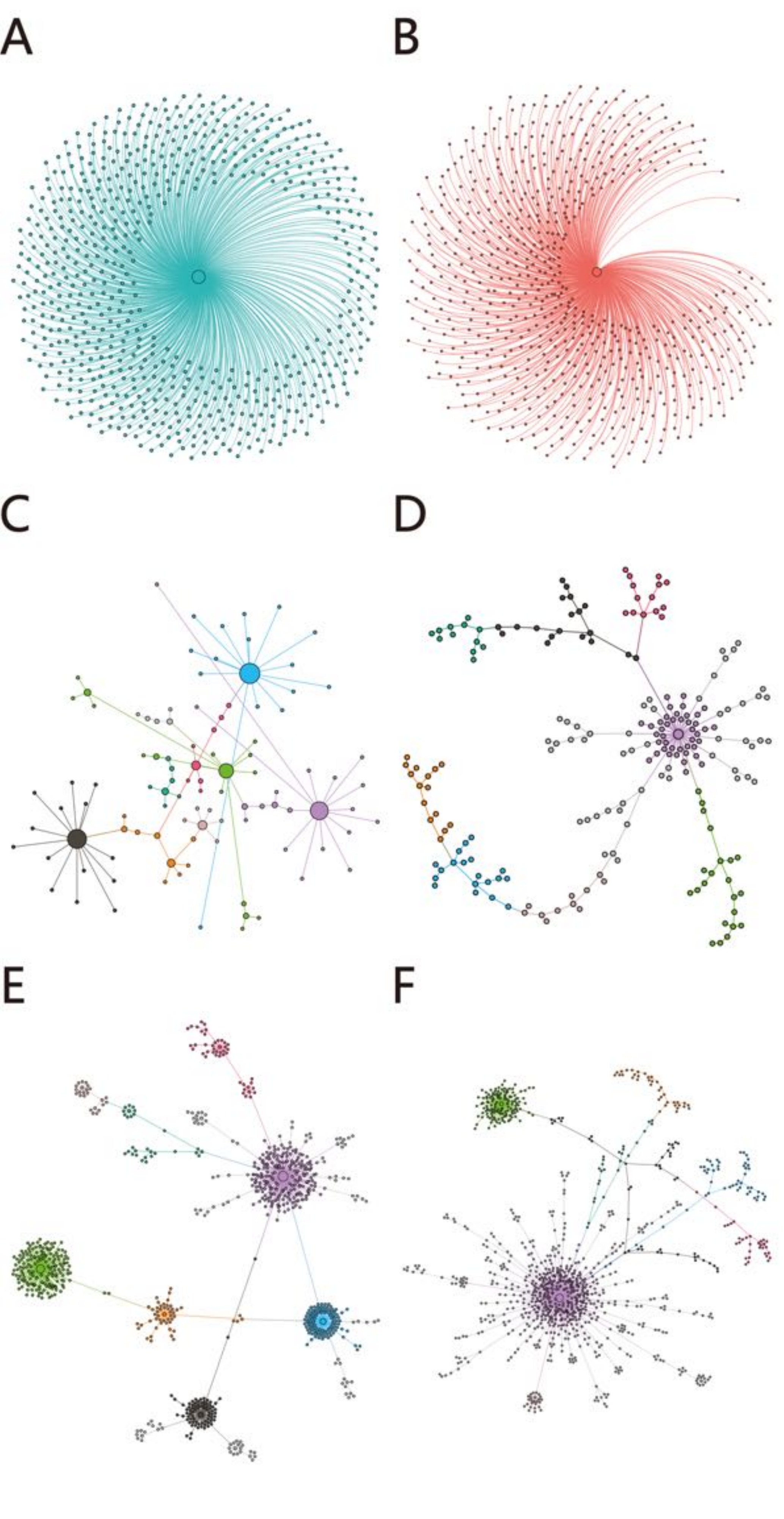}
	\caption{Typical examples of diffusion networks for true and fake news items. (A) A true news diffusion network with 630 nodes and $v \sim 2$ (pure broadcast). (B) A fake news diffusion network with 600 nodes and $v \sim 2$ (pure broadcast). (A) and (B) are both advertisements. (C) A true news diffusion network with 102 nodes, 9 communities, and $v \sim 7.142$. The content talks about the descendant of Confucius. (D) A fake news diffusion network with 207 nodes, 17 communities and $v \sim 9.895$. The content talks about Red Cross Society of China. (E) A true news diffusion network with 800 nodes, 21 communities, and $v \sim 5.763$. The content talks about a North Korean diplomat who joined South Korea. (F) A fake news diffusion network with 997 nodes, 63 communities, and $v \sim 7.748$. The content talks about some people using babies to make soup. Different colors represent different communities in the spread.}
	\label{fig:network_examples}
\end{figure}
\clearpage

\section*{S3 News timelines}
As mentioned in S1, both fake news and real news were collected before 2017 (our commercial access to Weibo API expired in 2017), and the news in our data set was posted from 2011 to 2016 (Fig. S6). The lifecycle of a news item starts from the posting time and ends with the final retweet in the sampling period. The timeline of each true or fake news item is analyzed by calculating the proportion of the number of new retweets within each hour of its lifecycle. For both true and fake news, retweets reach their peak within one hour after posting (Fig. S7A and S7B), illustrating the quick circulation on social media and, in particular, the explosive spread in the very early stage. Furthermore, we count the number of retweets every ten minutes and calculate the cumulative distribution functions (CCDFs) for different types of news. Fake (F) news demonstrates stronger vitality than true (T) news (K-S test $\sim$ 0.140, P $\sim$ 0.0) (Fig. S7C). Specifically, fake news still obtains 26\% of its retweets after 48 hours, while that proportion for true news is 20\%. More importantly, the stronger vitality of fake news than true news is consistently observed in groups of LT news vs. LF news (K-S test $\sim$ 0.114, P $\sim$ 0.0) (Fig. S7D) and HT news vs. HF news (K-S test $\sim$ 0.138, P $\sim$ 0.0) (Fig. S7E). Besides, we compared the distributions of the number of retweets within 48 hours of the posting and found that the propagation speed of fake news is significantly higher than that of true news (K-S test $\sim$ 0.195, P $\sim$ 0.0) (Fig. S7F). All this evidence suggests findings similar to those for Twitter \cite{Vosoughi2018}, that is, fake news is more viral than real news online. Compared to that of real news, its circulation lasts longer, has higher speed, and ultimately produces more retweets. 

\begin{figure}
	
	\centering
	\includegraphics[scale=0.8]{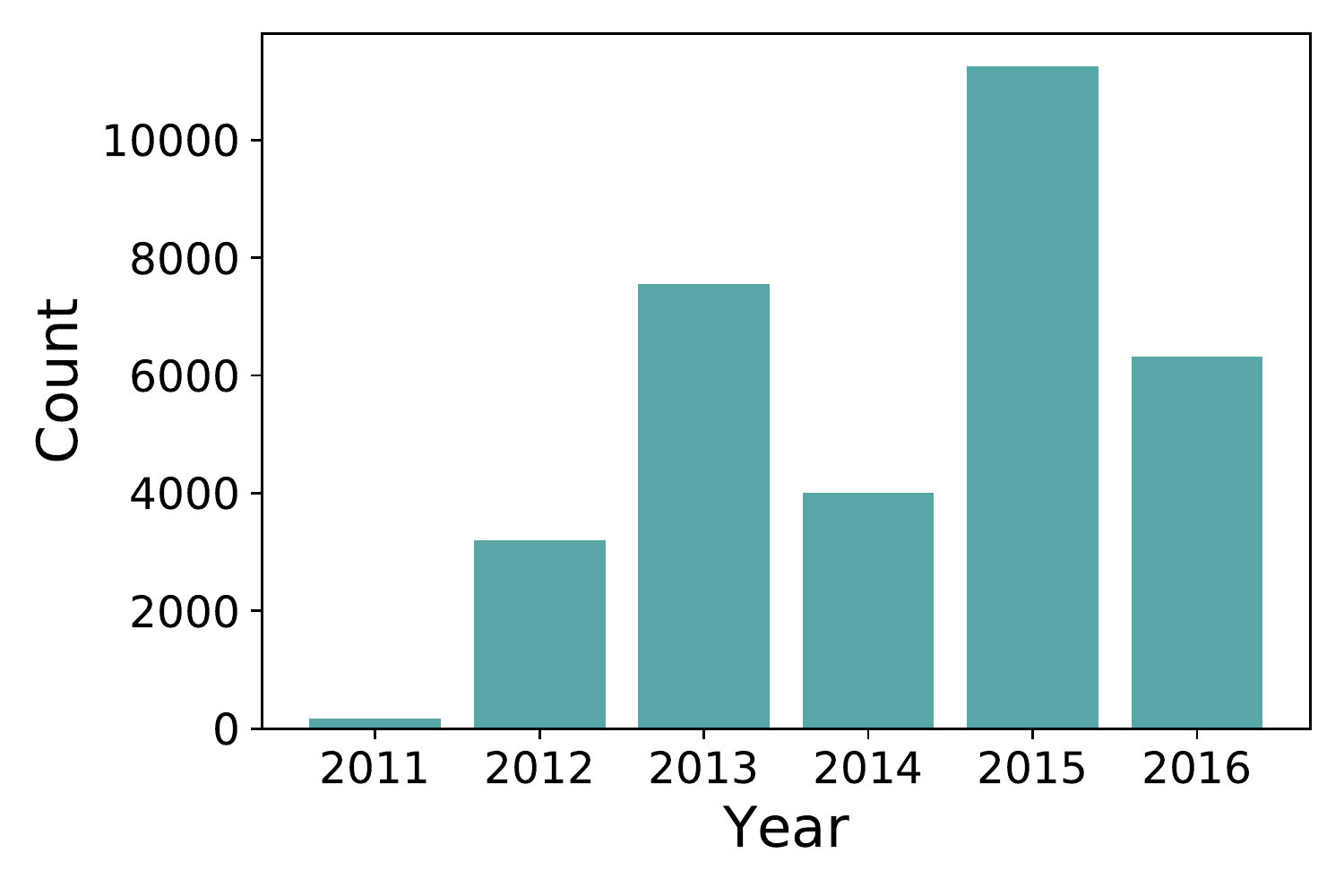}
	\caption{Yearly counts of news items.}
	\label{fig:yearly_count}
\end{figure}

\begin{figure}
	
	\centering
	\includegraphics[scale=0.4]{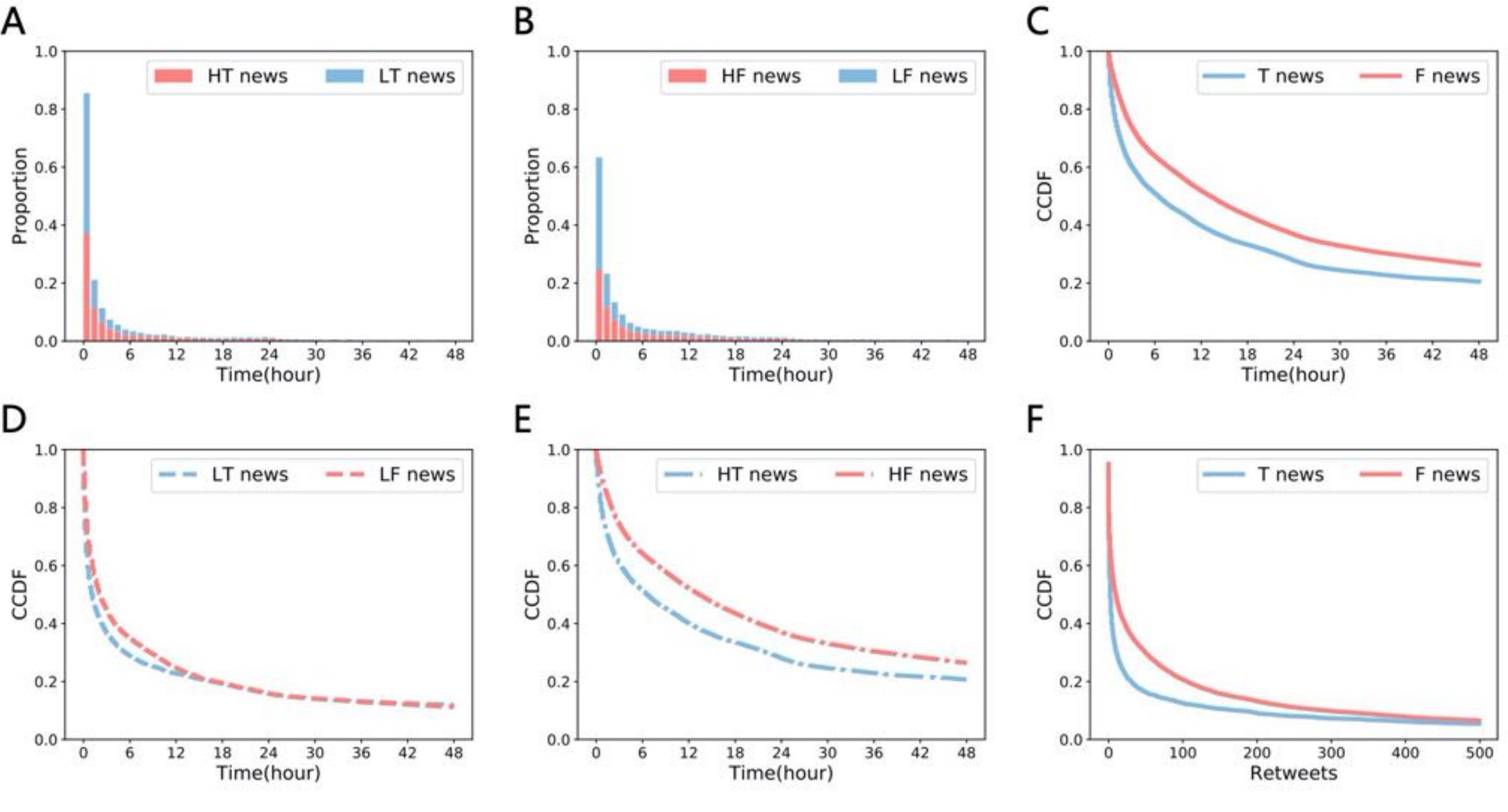}
	\caption{Timelines analysis. (A) The proportion of new retweets in each hour for both HT and LT news. (B) The proportion of new retweets in each hour for fake news. (C) CCDFs for retweeting time for true news and fake news. (D) CCDFs for retweeting time for LT news and LF news. (E) CCDFs for retweeting time for HT news and HF news. (F) CCDFs for the number of retweets within 48 hours for true news and fake news.}
	\label{fig:timeline}
\end{figure}
\clearpage

\section*{S4 Emotion lexicon }
In this study, the emotional texts of news in social media, both fake and true, are assumed to carry sophisticated signals that cannot be fully represented by binary values such as positive or negative. In contrast, emotions, in particular, negative emotions, are split into elementary compounds, including anger, disgust, sadness, and fear \cite{Sauter2010,LuminetIV2000}. Then, together with joy, which is used to reflect positive emotion, the distributions of the five emotions are derived to fully represent the emotional spectrum of each news item. An emotion lexicon must be established to obtain the emotional distribution of the text in both fake and true news intuitively and accurately; then, the occupation of a certain emotion can be calculated as the fraction of terms with this emotion in all emotional terms of the news text. We first segment all the texts into terms, filter by parts of speech, and keep nouns, verbs, adverbs, gerunds, adjectives, adjectives directly used as adverbials and adjectives with noun function to compose a candidate set. As a result, 34,227 preselected terms are obtained. Note that there might also be terms of nonemotion in the candidate set. We then hire human coders to manually label the terms: those without emotions are marked as neutral. A WeChat applet, named Word Emotion (Fig. S8), is built to make the labeling convenient. The whole labeling task was completed by nine well-instructed coders who are active users of Weibo with ages between 18 and 30 years old, and each term is labeled three times by randomly selected coders. Finally, terms with more than two identical emotional labels are screened out to build the lexicon. Ultimately, there are 6,155 emotional terms in total, including 1,323 anger terms, 710 disgust terms, 2,066 joy terms, 1,243 sad terms, and 813 fear terms. The emotion lexicon covers 87.1\% of the text of all fake and true news, and the remaining news items are labeled neutral, suggesting that the news in social media is indeed emotional. The emotion lexicon is publicly available and can be downloaded freely at \url{https://doi.org/10.6084/m9.figshare.12163569.v2}. 

\begin{figure}	
	\centering
	\includegraphics[scale=0.6]{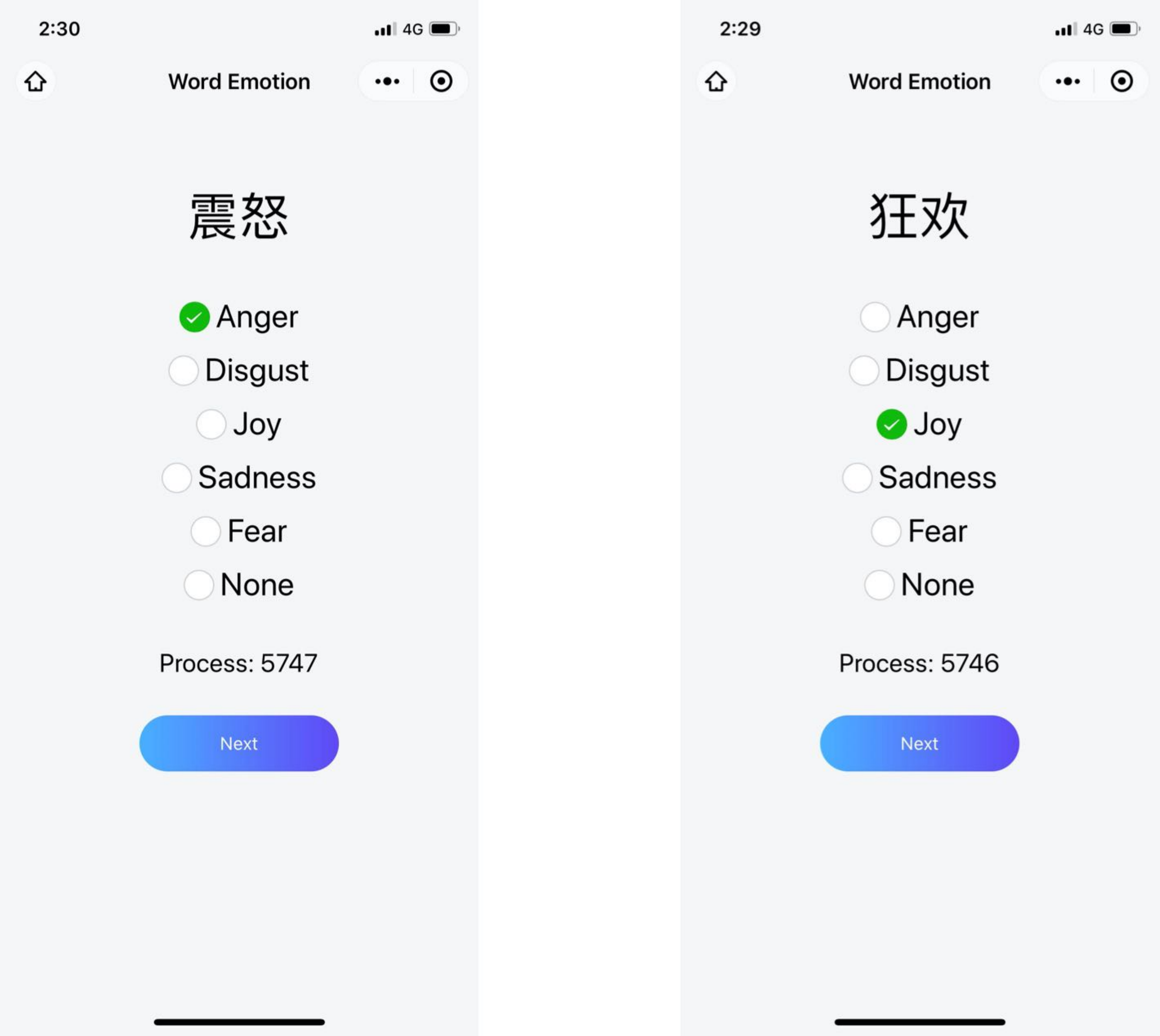}
	\caption{Main page of the WeChat applet ``Word Emotion". The Chinese word on the left describes a very angry state. The Chinese word on the right describes rejoice with wild excitement.}
	\label{fig:word_emotion}
\end{figure}
\clearpage

\section*{S5 Emotion distributions}
The emotional distributions of news items in the different groups are derived utilizing the established emotion lexicon. After the inference of emotion distributions, possible differences between treatment and control groups of news are comprehensively examined. These differences are expected to help reveal the mechanism underlying the circulation of fake news. In particular, more insights might be derived by splitting negative emotion into more elementary emotions.

In the main text, we discussed that the amount of anger in fake news is significantly higher than that in true news, and the amount of joy in true news is significantly higher than that in fake news. This phenomenon is more obvious in HLT news and LHF news after excluding the influence of the author. Moreover, to further examine the difference between anger and joy and its possible association with the fast spread of fake news, we compare the emotional differences between HLF news and LHF news. The results show that the amount of anger in LHF news is significantly higher than that in HLF news (Fig. 1A in the main text), and the amount of joy is significantly lower than that in HLF news (Fig. 1E in the main text), which is consistent with the comparison between L news and H news (Fig. S9A, S9C). That is, the amount of anger in widely circulated news is significantly higher than that in less widely circulated news. The statistics of the emotional distributions and the results of K-S tests are shown in Table S2-5. All these observations consistently suggest an association between anger and the virality of fake news and inspire later causal inference through regression models.

\begin{figure}	
	\centering
	\includegraphics[scale=0.4]{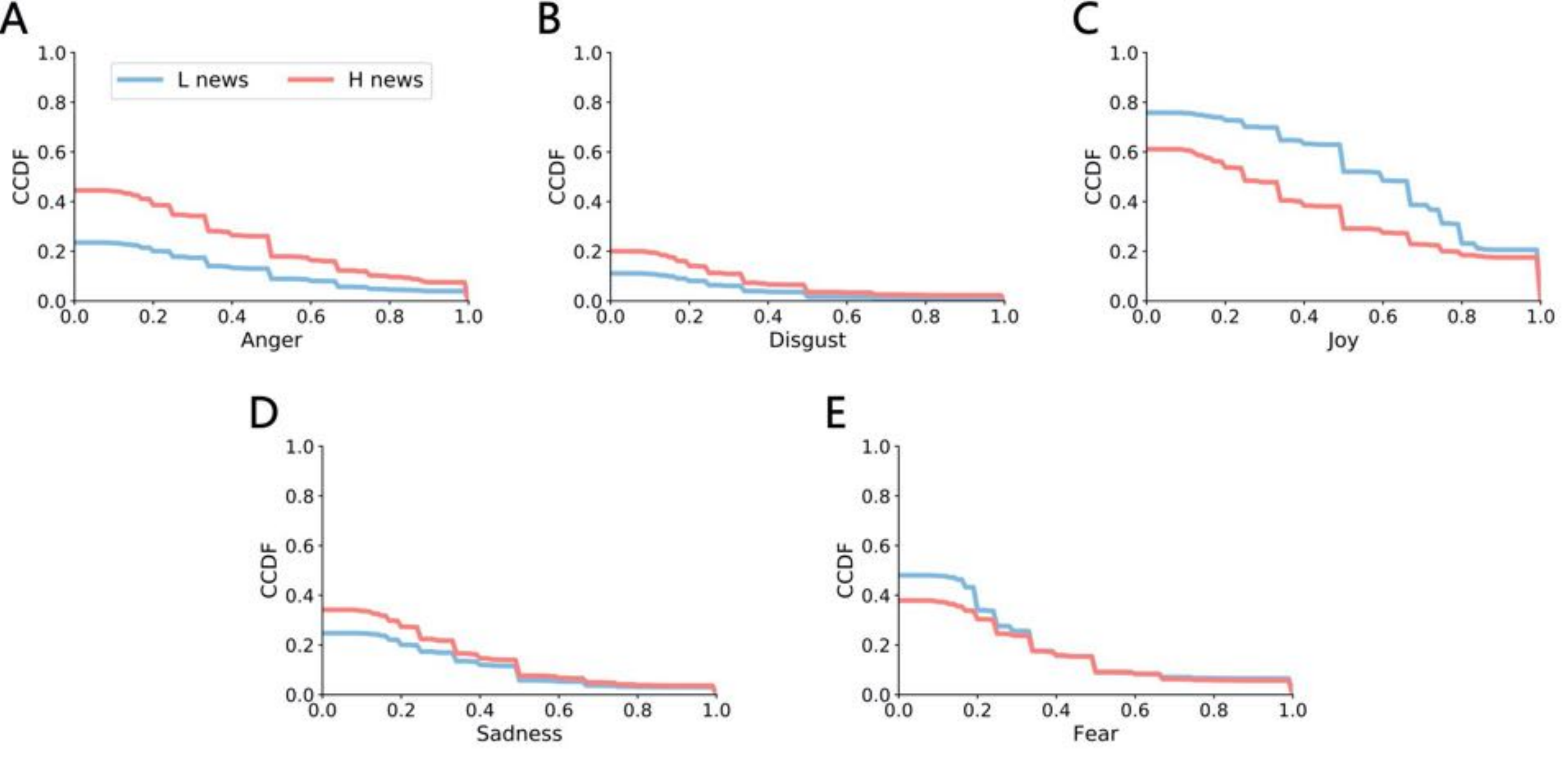}
	\caption{CCDFs of emotions in L news and H news items. (A) Anger, (B) Disgust, (C) Joy, (D) Sadness, (E) Fear. The results of the K-S tests can be seen in Table S5.}
	\label{fig:lh_news_ccdf}
\end{figure}

\begin{table}	
	\centering
	\includegraphics[]{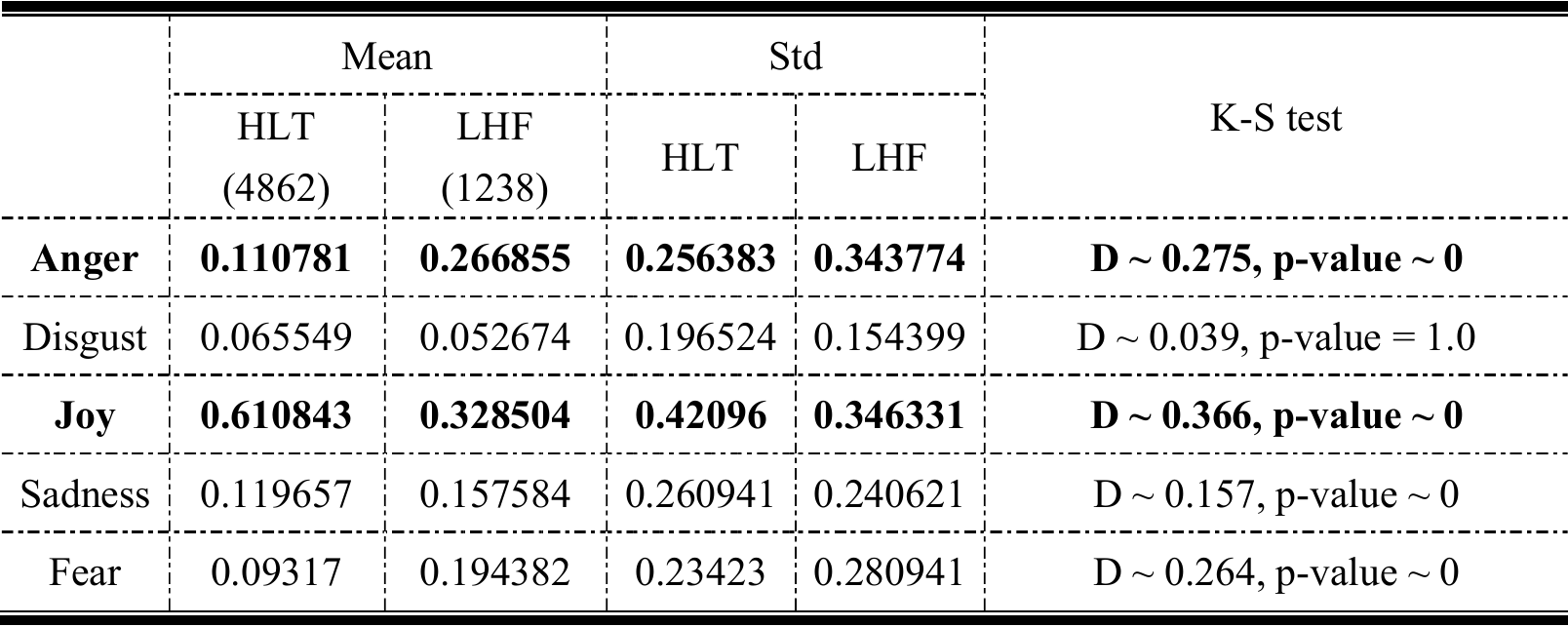}
	\caption{Statistics and K-S tests for HLT news and LHF news items.}
	\label{table:hlt_lhf_ks}
\end{table}

\begin{table}	
	\centering
	\includegraphics[]{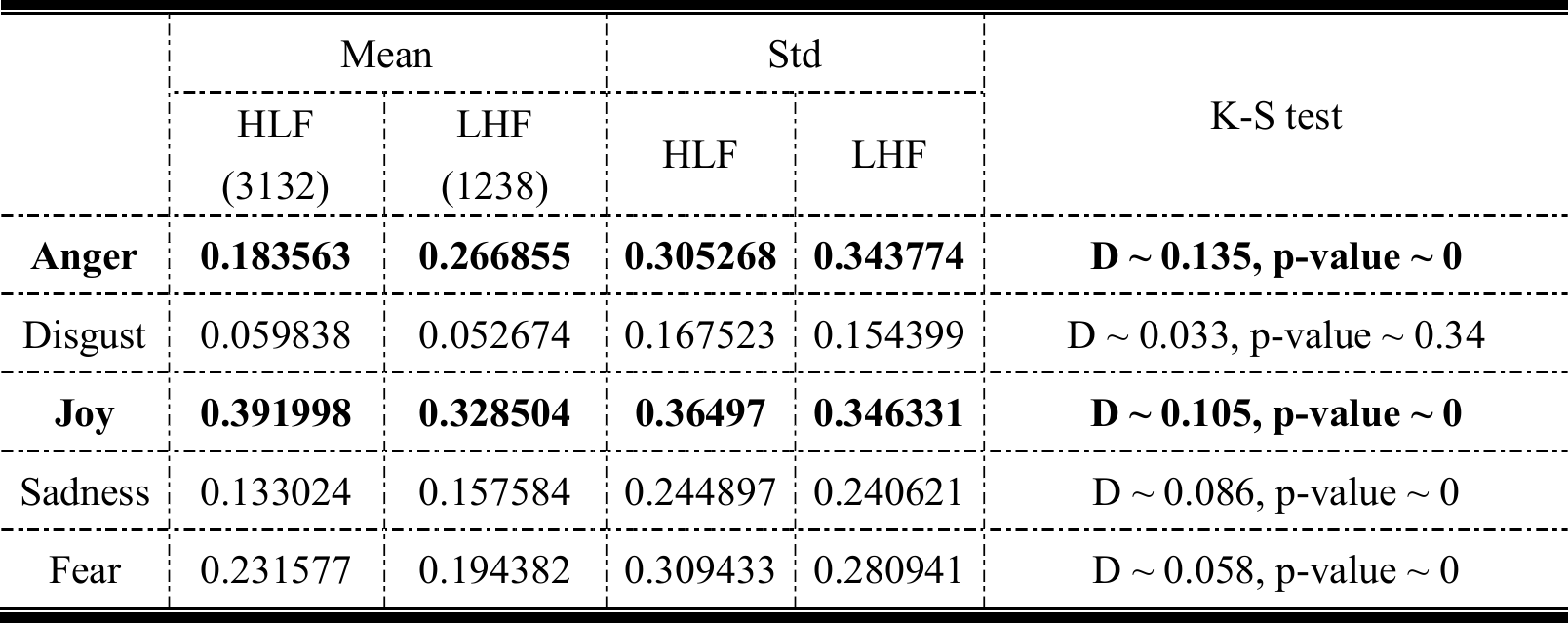}
	\caption{Statistics and K-S tests for HLF news and LHF news items.}
	\label{table:hlf_lhf_ks}
\end{table}

\begin{table}	
	\centering
	\includegraphics[]{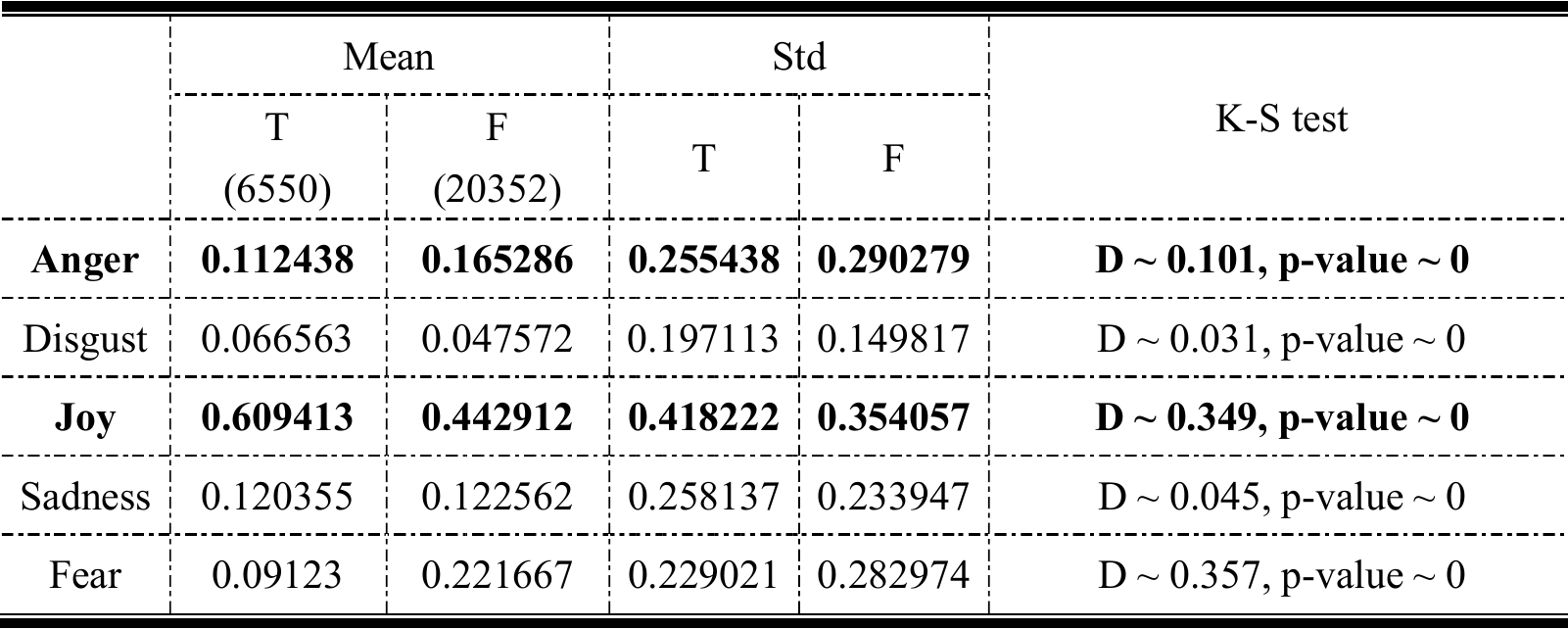}
	\caption{Statistics and K-S tests for true news and fake news items.}
	\label{table:t_f_ks}
\end{table}

\begin{table}	
	\centering
	\includegraphics[]{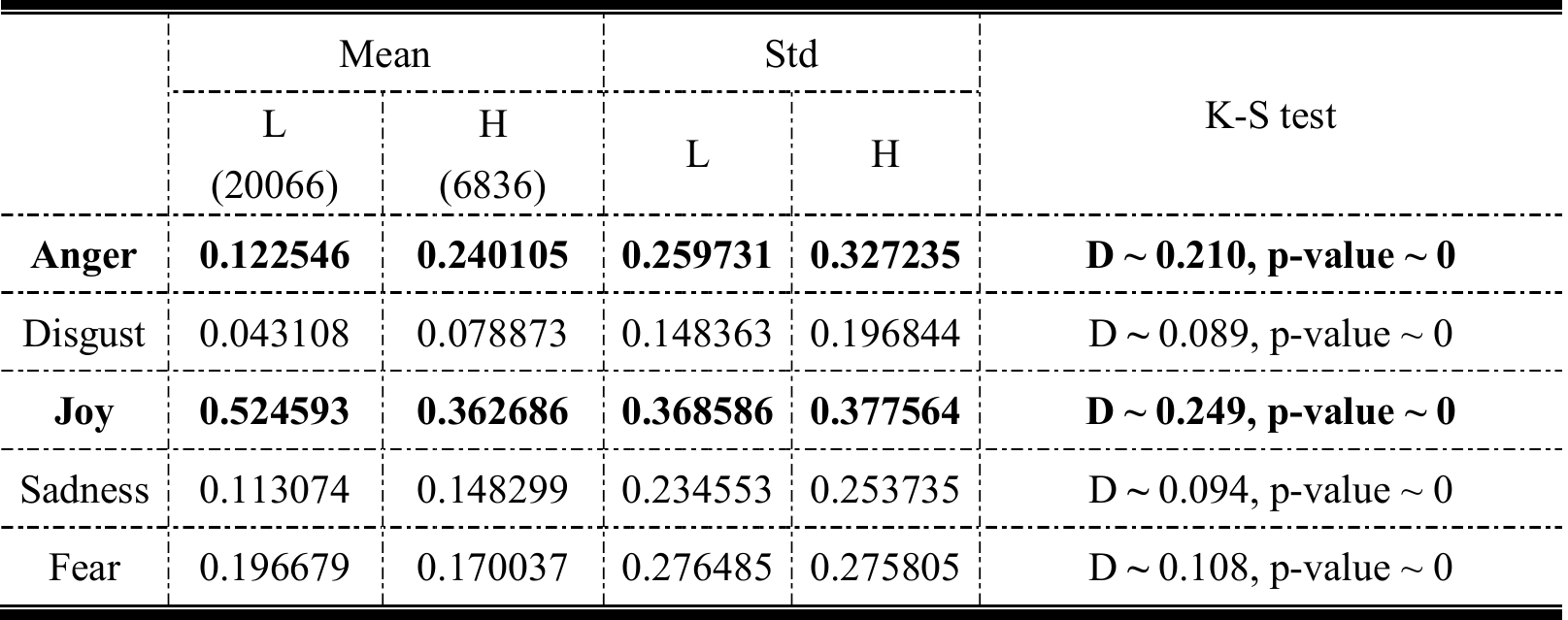}
	\caption{Statistics and K-S tests for L news and H news items.}
	\label{table:l_h_ks}
\end{table}
\clearpage

\section*{S6 Keywords in separating news groups}
The existence of highly retweeted tweets posted by authors with a low volume of followers in both fake news and real news implies the potential influence of content on circulation. Besides, emotions are carried by words in the text. The distinguishing distributions of emotions, in particular, anger and joy, between fake news and real news inspire us to pinpoint keywords that could split news groups. Additionally, these keywords could help in later offline questionnaires to strengthen the stimuli of anger and joy on the reposting incentives of the audience (see S11-13).

Specifically, for LHF news, HLT news, and HLF news items, we train an SVM and a logistic regression model, which are commonly employed to weigh words in text mining tasks, to evaluate the separability of the text and extract keywords that influence the separation. These groups of news are further split into two corpora to learn binary classification models, i.e., one corpus is composed of LHF news (positive class) and HLT news (negative class) and the other corpus is composed of LHF news (positive class) and HLF news (negative class). Words are used as text features to calculate the TF-IDF matrix for classification. After 5-fold cross-validation, the average accuracies are 0.94 (SVM) and 0.98 (logistic regression) in the corpus of LHF-HLT and 0.75 (SVM) and 0.81 (logistic regression) in the corpus of LHF-HLF, implying that using words as features results in good separation of LHF news from HLT news and HLF news. Moreover, content carrying emotions such as anger and joy could be an influential driver of news circulation. In particular, the better separability between LHF news and HLT news suggests the feasibility of keywords in strengthening the divergence of different news items in reposting stimuli. On this basis, we combine the chi-square test, mutual information, AdaBoost, and extra-trees for feature selection, and 150 influential keywords with the greatest weight in the classification are selected from each group of news items (Fig. S10A, C, and E) (These methods are implemented with the scikit-learn package in Python.). By analyzing the emotional distributions of keywords in each type of news, we found that the emotional keywords in HLT news are all joyful (Fig. S10B), and those in HLF news are mainly joyful (Fig. S10F), followed by fearful. However, negative emotions, especially anger, dominate the keywords in LHF news (Fig. S10D). These observations support the initial assumption that emotions carried by news, in particular, the dominant emotions of anger and joy, can be reflected by keywords that effectively separate different groups of news; therefore, these keywords will affect the incentives underlying retweets. Meanwhile, the exact same difference in the emotion distribution at the keyword level further confirms the consistency and robustness of the emotional divergence between fake news and true news revealed at the collective level (see S5).

\begin{figure}	
	\centering
	\includegraphics[scale=0.45]{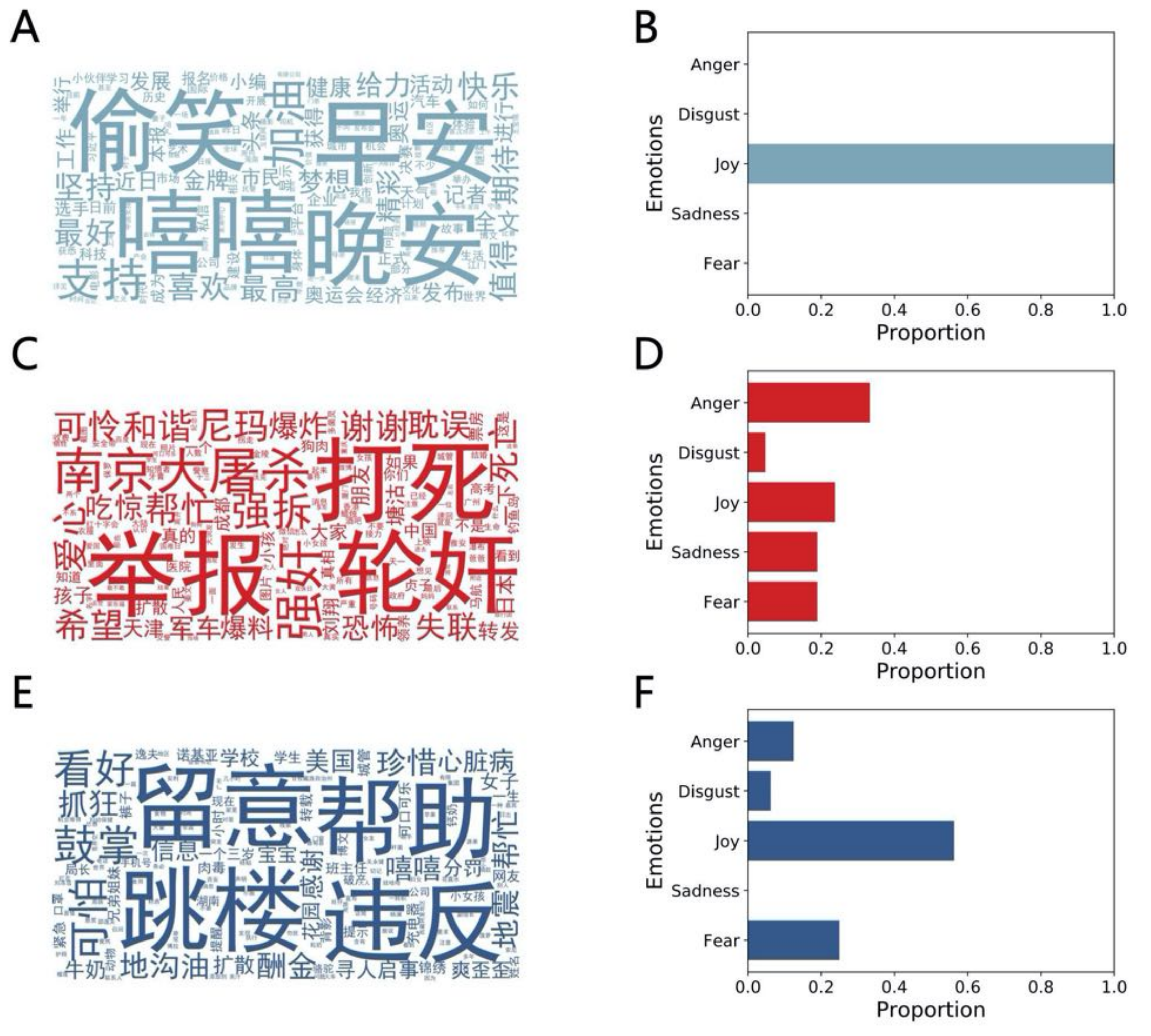}
	\caption{Word cloud and emotional distributions of keywords. (A) Word cloud of keywords in HLT news. (B) Emotional distribution of keywords in HLT news. (C) Word cloud of keywords in LHF news. (D) Emotional distribution of keywords. in LHF news. (E) Word cloud of keywords in HLF news. (F) Emotional distribution of keywords in HLF news. All the keywords in the word cloud are translated into English and can be found in the publicly available data at \url{https://doi.org/10.6084/m9.figshare.12163569.v2}.}
	\label{fig:emo_keywords}
\end{figure}
\clearpage

\section*{S7 Additional tests for emotion inference and divergence }
\subsection*{S7.1 Alternative approaches of emotion inference}
In addition to the emotion lexicon, which offers an intuitive measure of emotion occupation, machine learning models, in particular, state-of-the-art solutions such as deep neural networks, are alternative models to infer the emotion distributions of both fake and true news. In this study, to ensure the consistency and accuracy of our results on emotion distributions, we also considered classic machine learning and deep learning models. Specifically, two classifiers built for emotion detection in Chinese tweets from Weibo are employed to perform the additional tests, namely, a na\"{i}ve Bayesian classifier (termed Bayes, with an accuracy of 0.64) \cite{Zhao} and a backpropagation neural network based on an emotional dictionary (termed BP1, with an accuracy of 0.69, which was built with Keras), to calculate the emotion distributions of the texts in terms of probabilities of belonging to certain emotions. Then, the occupations of different emotions are further compared across groups, and the results are shown in Table S6-11. All the results support our conclusions obtained from the emotion lexicon, in particular, the difference in emotion distributions between anger and joy, suggesting the robustness of our understanding of emotion divergence between fake news and real news.

\begin{table}
	\centering
	\includegraphics[]{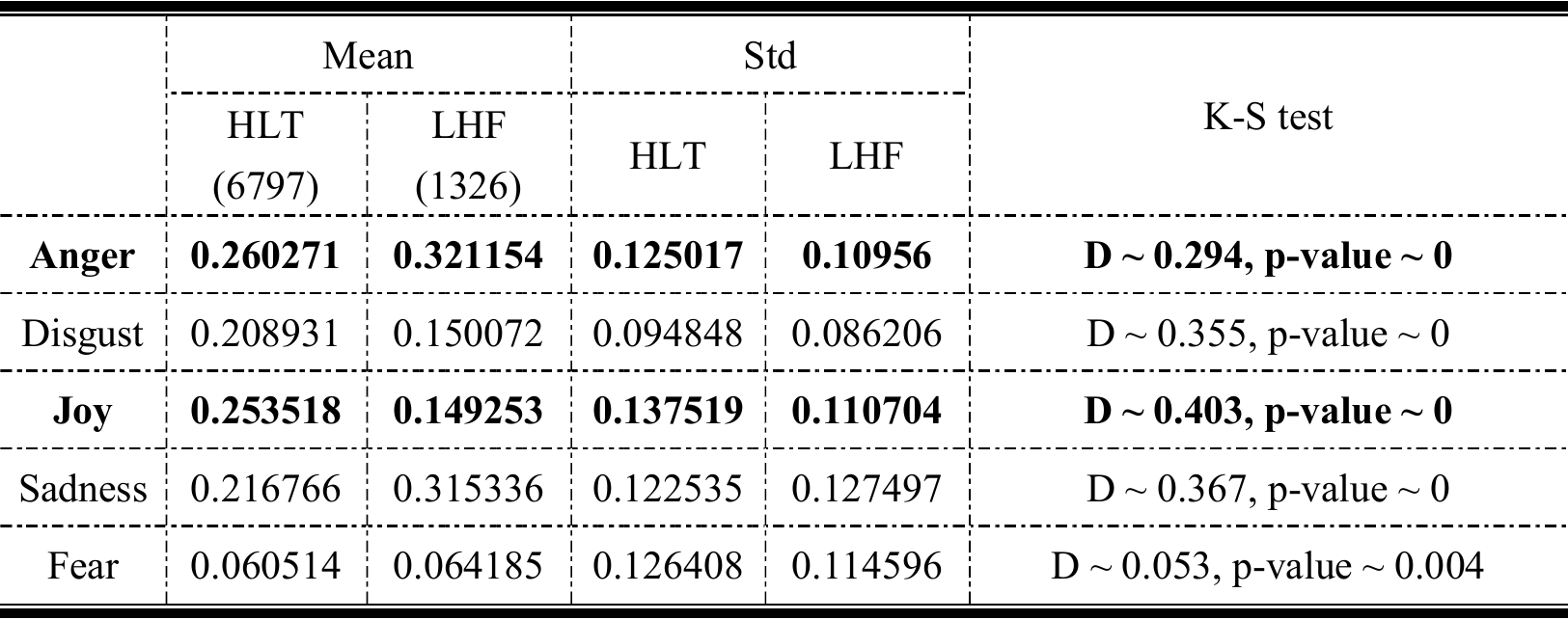}
	\caption{Statistics and K-S tests for HLT news and LHF news items based on Bayes.}
	\label{table:hlt_lhf_ks_bayes}
\end{table}

\begin{table}
	
	\centering
	\includegraphics[]{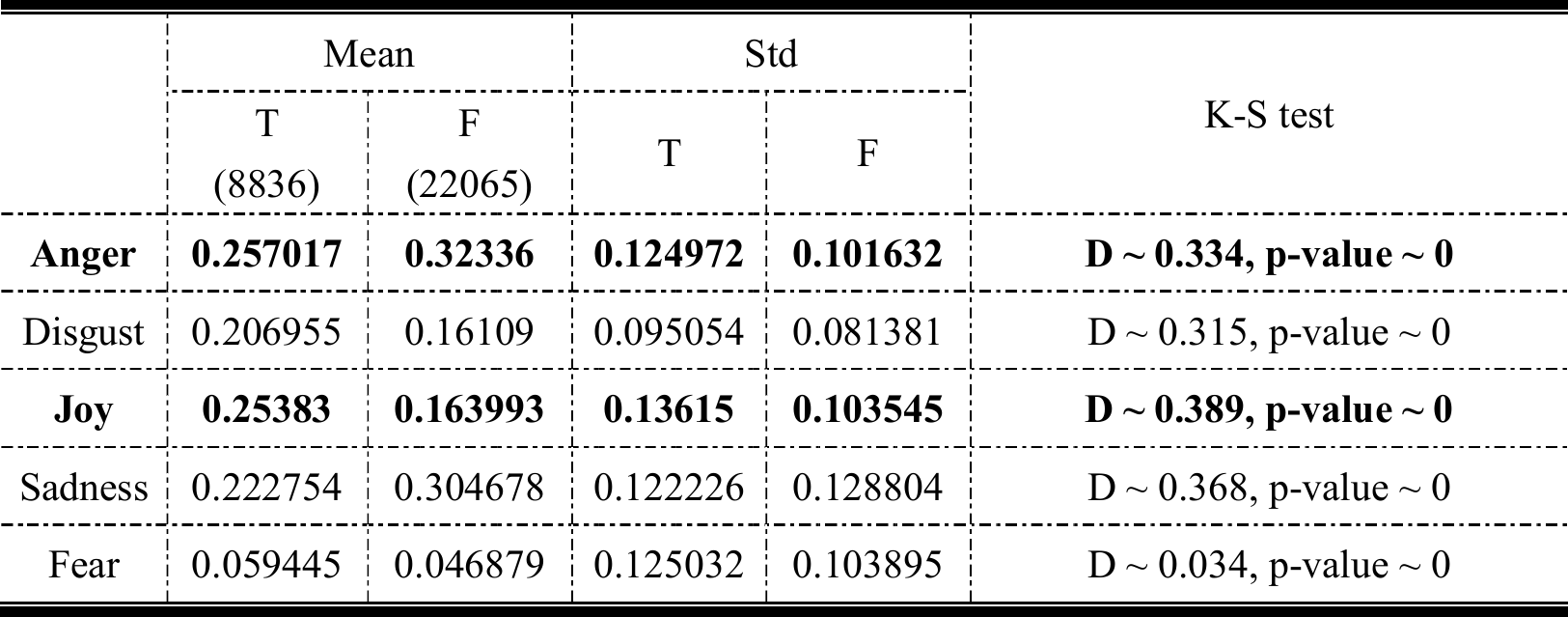}
	\caption{ Statistics and K-S tests for T news and F news items based on Bayes.}
	\label{table:t_f_ks_bayes}
\end{table}

\begin{table}
	
	\centering
	\includegraphics[]{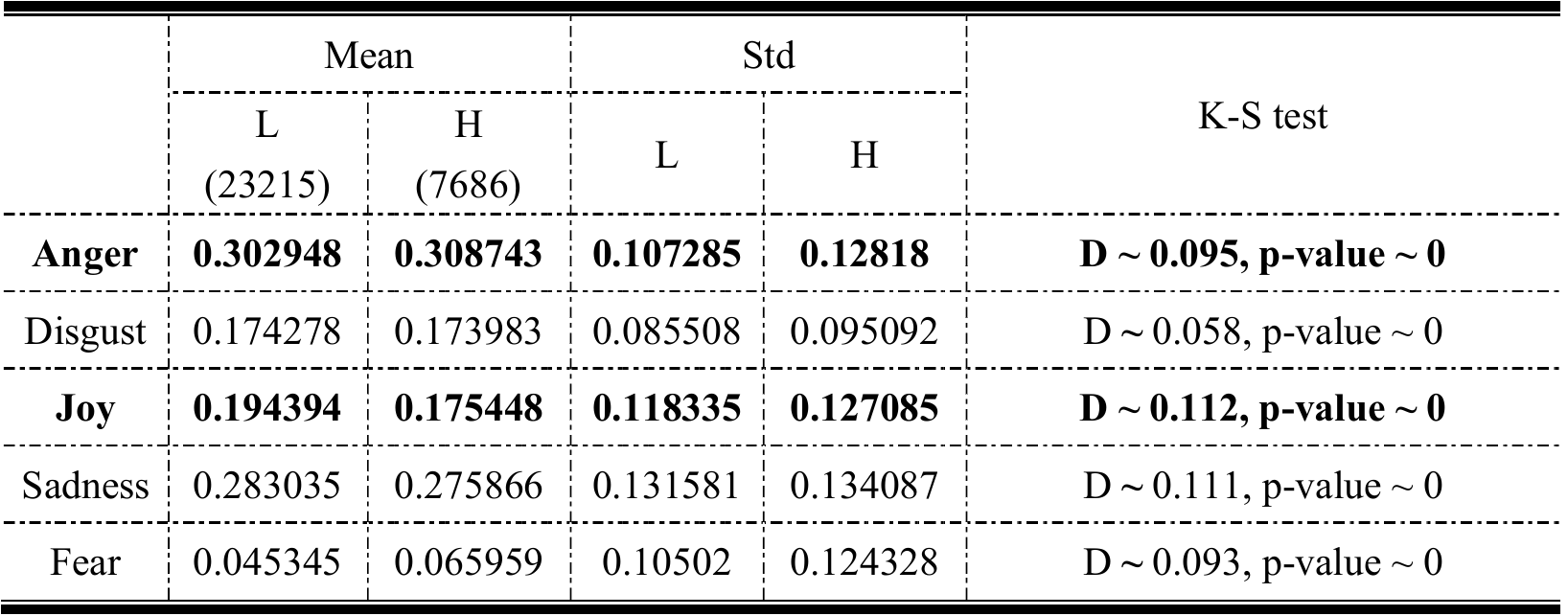}
	\caption{Statistics and K-S tests for L news and H news items based on Bayes.}
	\label{table:l_h_ks_bayes}
\end{table}

\begin{table}
	
	\centering
	\includegraphics[]{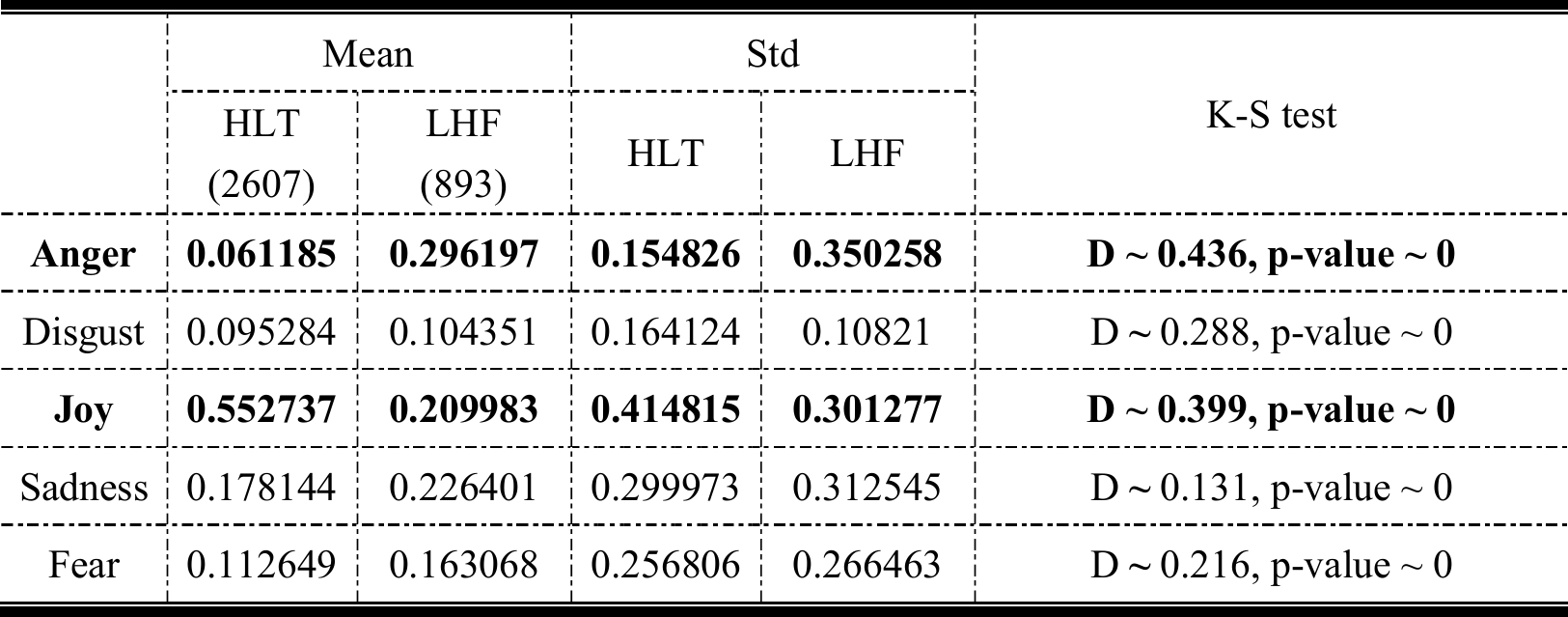}
	\caption{Statistics and K-S tests for HLT news and LHF news items based on BP1.}
	\label{table:hlt_lhf_bp1}
\end{table}

\begin{table}
	
	\centering
	\includegraphics[]{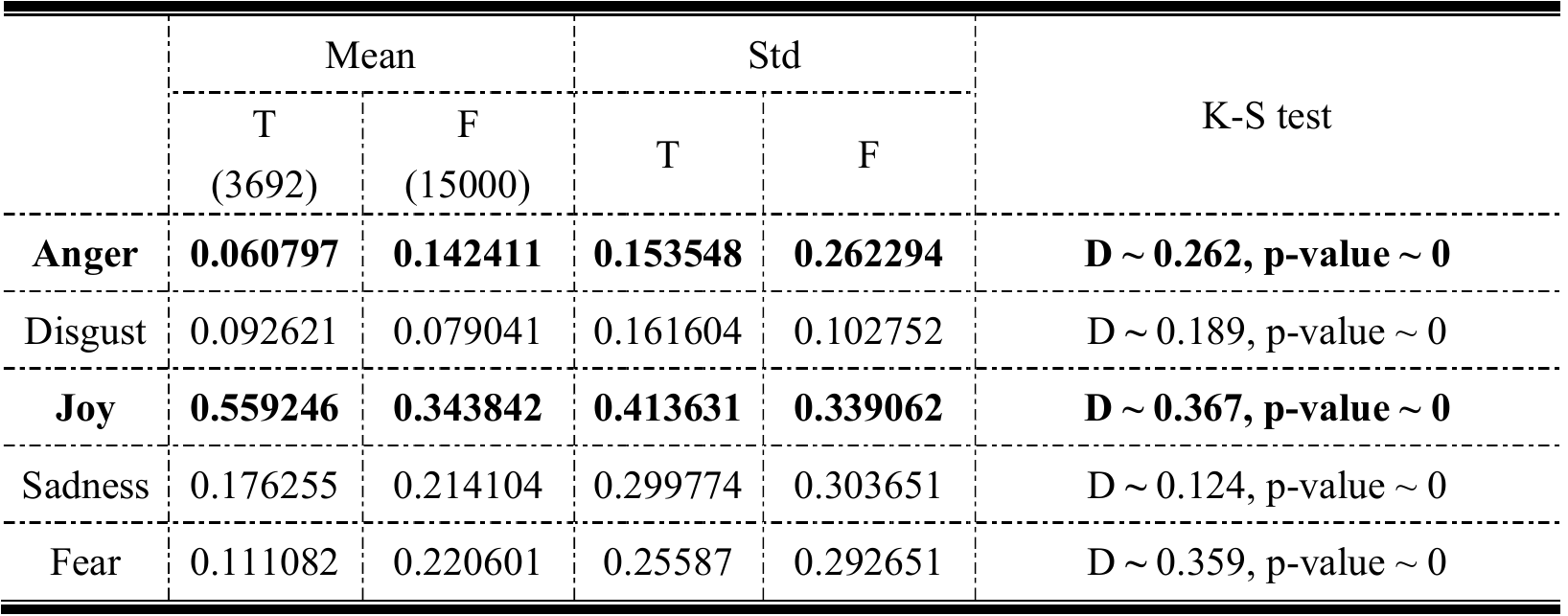}
	\caption{Statistics and K-S tests for T news and F news items based on BP1.}
	\label{table:t_f_bp1}
\end{table}

\begin{table}
	
	\centering
	\includegraphics[]{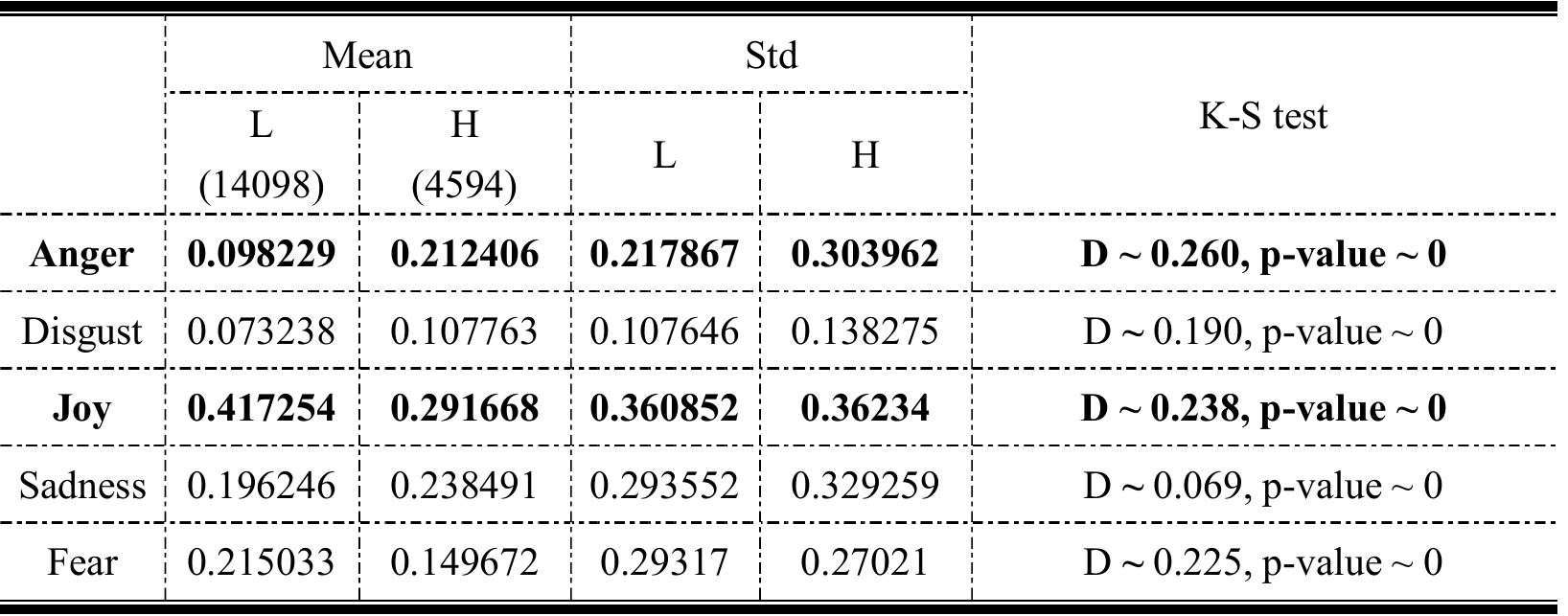}
	\caption{Statistics and K-S tests for L news and H news items based on BP1.}
	\label{table:l_h_bp1}
\end{table}

\subsection*{S7.2 Alternative measure of emotion distribution}
In the previous analysis and the additional test on emotion divergence, the emotion distribution of each news item is inferred exclusively by one method, i.e., lexicon-based, Bayes or BP1, and is simply represented as the occupations of the emotions in each text. However, it is possible that different methods could result in different inferences on the same text, which might undermine the consistency of emotion divergence we previously observed at the text level. To further assess the robustness of our conclusions about the different occupations of anger and joy in fake news and true news, a new text-level measure is presented to represent the emotion distribution by ranks. Specifically, for each news item text, a batch of models is employed separately to infer the probability of belonging to the five emotions, which are then ranked according to these probabilities: lower ranking values represent higher probabilities of the texts belonging to the corresponding emotions. Note that emotions with the same probability are ranked randomly. By aggregating the ranks of a certain emotion over all models, a distribution of rank can be obtained for the emotion in each text. Then, for each group of news items, the distributions of the five emotions can be obtained by averaging the rank distributions of the corresponding emotions in all texts.

First, employing a word2vec \cite{Shi} model that inferred over 560 million tweets of Weibo, each term is embedded into a vector of 200 dimensions. Then, the text of a news item is converted into a vector of 200 dimensions by averaging the embeddings of all terms in the text. To increase the number of inference models of emotions, six additional emotion classifiers are constructed on the emotion lexicon: including AdaBoost, decision tree, logistic regression, ridge classifier, SVM, and backpropagation neural network (BP2) (The classic machine learning models are built with scikit-learn and BP2 is built with PyTorch.). Specifically, terms with emotional labels in the emotion lexicon are first embedded to train these models; then, the emotions of news item text in the same embedding space are inferred. The accuracies of these models in 5-fold cross-validations are 0.67, 0.73, 0.79, 0.76, 0.75 and 0.86. From the results of the rank distributions, ranks of anger in LHF news, F news and H news items are significantly lower than those in HLT news, T news, and L news items (Fig. S11A, B, C), while the ranks of joy show the opposite trends (Fig. S11G, H, I). Note that a lower rank represents a higher probability of belonging to the corresponding emotion. This result is consistent with all previous results, indicating that the divergence in anger and joy between fake news and real news is robust and independent of emotion inference model and emotion distribution measure. However, the differences in other negative emotions across news groups, though significant, are inconsistent and varying. The ranks of sadness in LHF news, F news, and H news items are significantly higher than those in HLT news, T news, and L news items (Fig. S11J, K, L), which is inconsistent with the previous results (see Fig. 1 in the main text). The ranks of disgust fluctuate inconsistently across different assemblies of news groups. Although the rank of fear in LHF news is significantly lower than that in HLT news, as the rank is smaller than 4, it becomes higher than that of HLT, as the rank is 5. (Fig. S11M). Therefore, in the following causal inference on the impact of emotions on circulation, negative emotions other than anger are not considered separately.

\begin{figure}	
	\centering
	\includegraphics[scale=0.4]{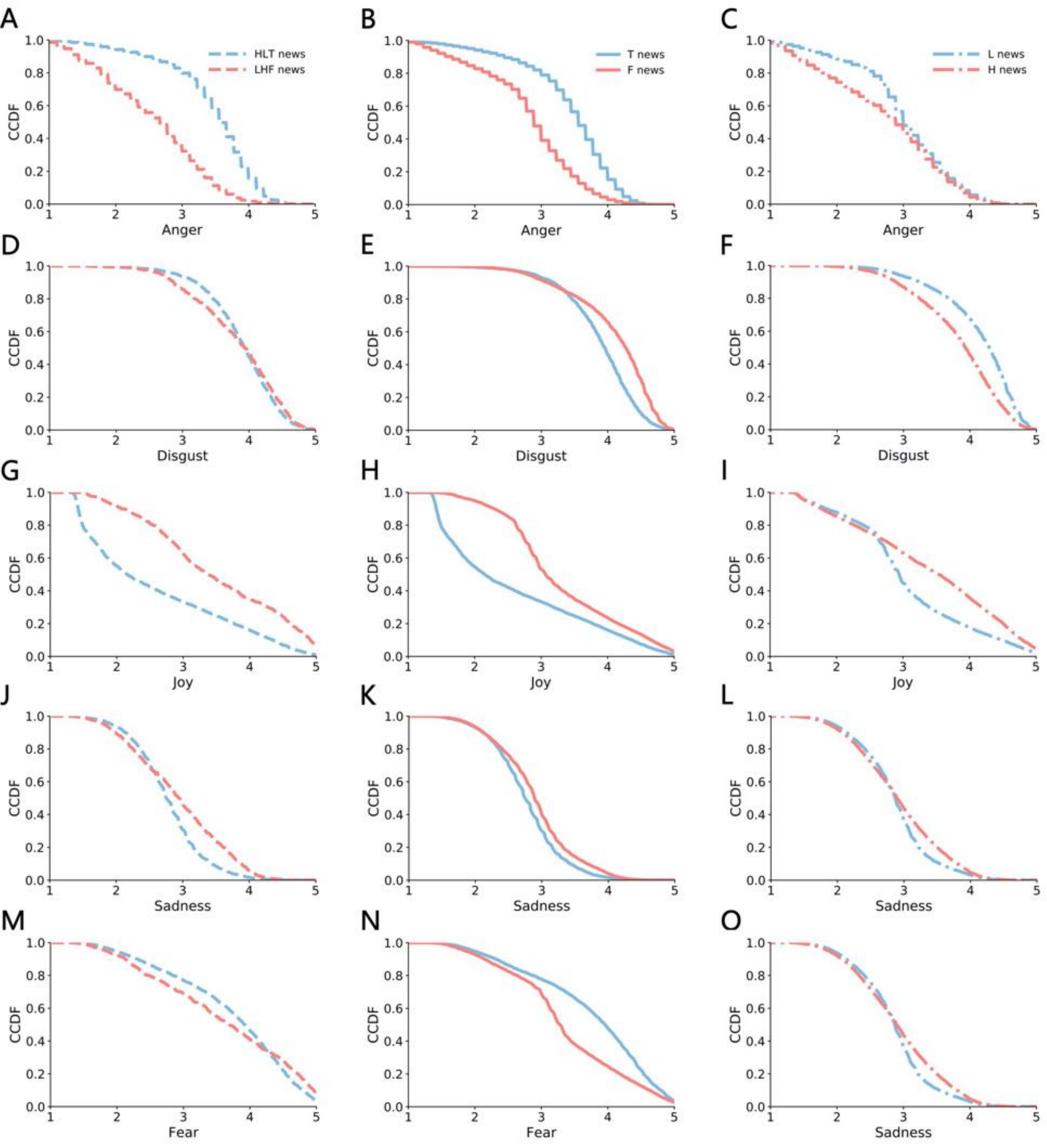}
	\caption{CCDFs of emotion ranks in HLT news and LHF news items, T news and F news items, L news and H news items. (A, B, C) Anger, (D, E, F) Disgust, (G, H, I) Joy, (J, K, L) Sadness, (M, N, O) Fear. The results of the K-S tests are shown in Table S12-14.}
	\label{fig:emo_ccdf_rank}
\end{figure}

\begin{table}
	
	\centering
	\includegraphics[]{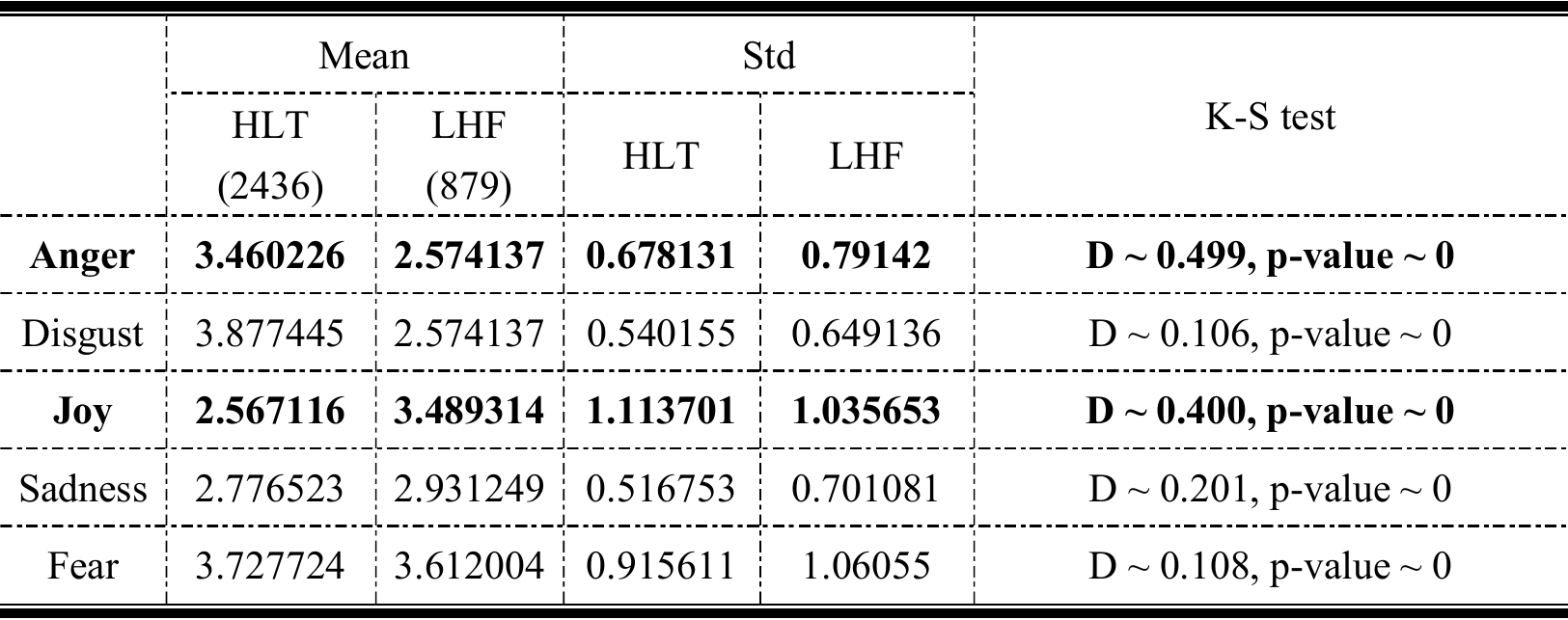}
	\caption{Statistics and K-S tests for the rank distributions of HLT news and LHF news items.}
	\label{table:hlt_lhf_ks_rank}
\end{table}

\begin{table}
	
	\centering
	\includegraphics[]{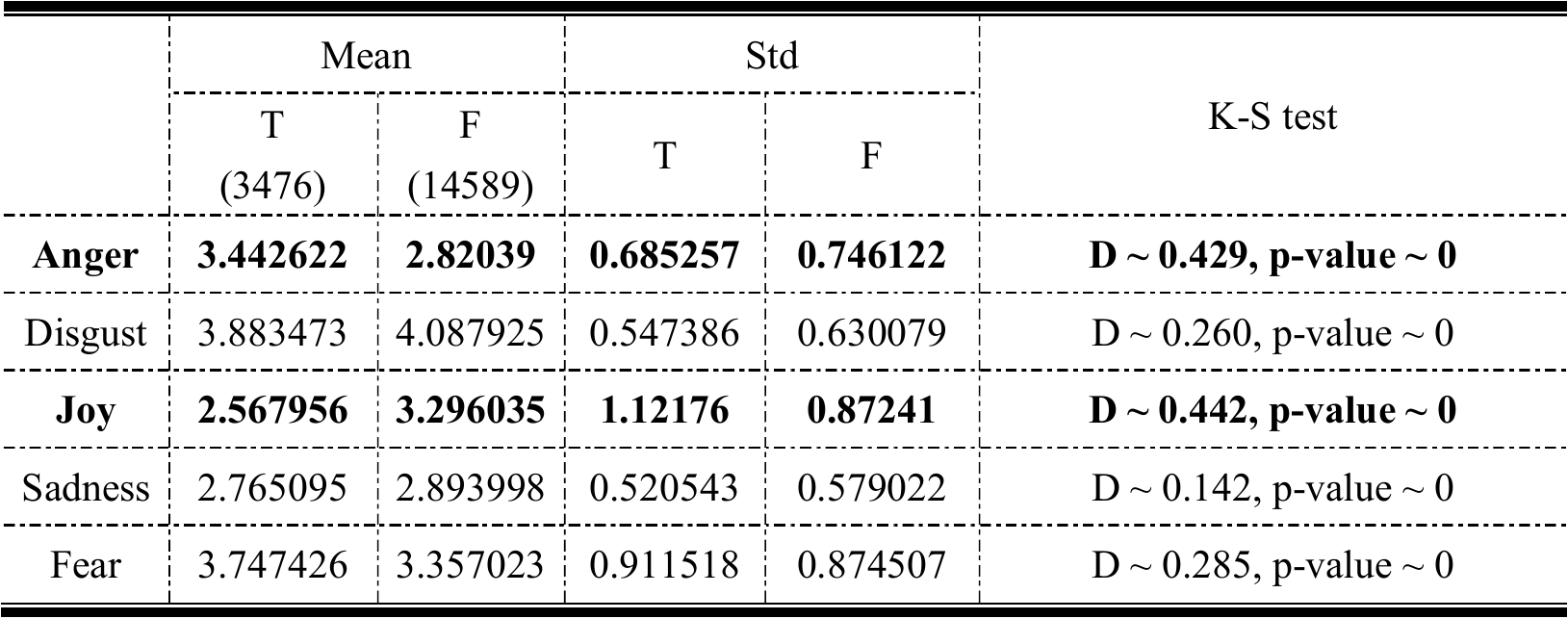}
	\caption{ Statistics and K-S tests for the rank distributions of T news and F news items.}
	\label{table:t_f_ks_rank}
\end{table}

\begin{table}
	
	\centering
	\includegraphics[]{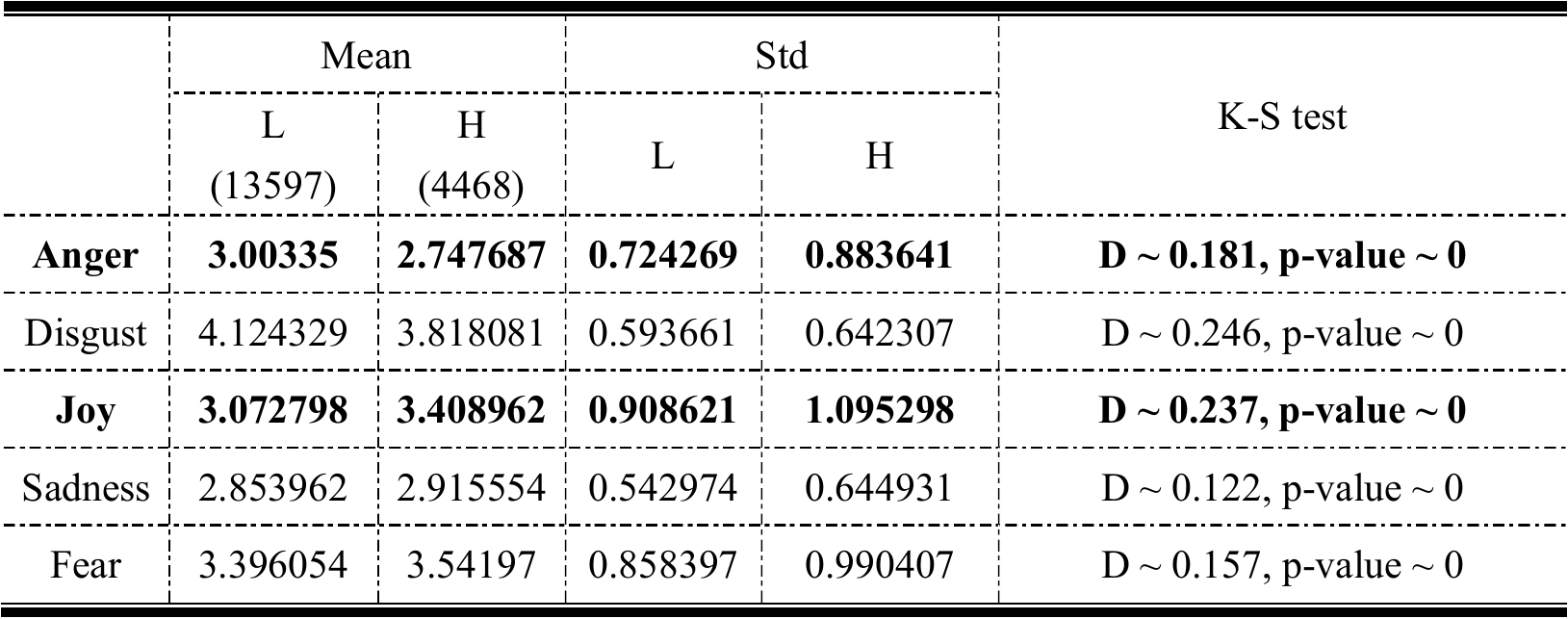}
	\caption{Statistics and K-S tests for the rank distributions of L news and H news items.}
	\label{table:l_h_ks_rank}
\end{table}

\subsection*{S7.3 A case study of fake news in COVID-19}
Emergent events, particularly disasters, always spur fake news items and social media can be fertile ground for their fast spread. With the sudden outbreak of COVID-19, epidemic-related fake news flooded the Internet, disseminating false information and resulting in collective panic. Here, we further collected 324 fake news items (including 31,284 retweets) related to the epidemic from January 22 to March 1, 2020 (\url{https://covid19.thunlp.org/archives/5/}), and examined the divergence between anger and joy in their emotional distributions to validate our findings in the specific circumstance of an emergency event. Using the emotion lexicon built in this paper, the emotional distributions of 200 fake news items were inferred. The results consistently found that HF news items carried more anger and less joy than LF news items. The dominance of anger to joy (the occupation of anger minus that of joy) was significantly larger in the HF news group (T $\sim$ 2.851, P $\sim$ 0.006) (Fig. S12 and Table S15). However, it should be noted that here we only support a case study on fake news caused by a specific event such as COVID-19 epidemic. Due to the very small sample size (e.g., 200) and lack of a control real news group, further explorations such as regression models (see S9) were not performed on this data set.

\begin{figure}	
	\centering
	\includegraphics[]{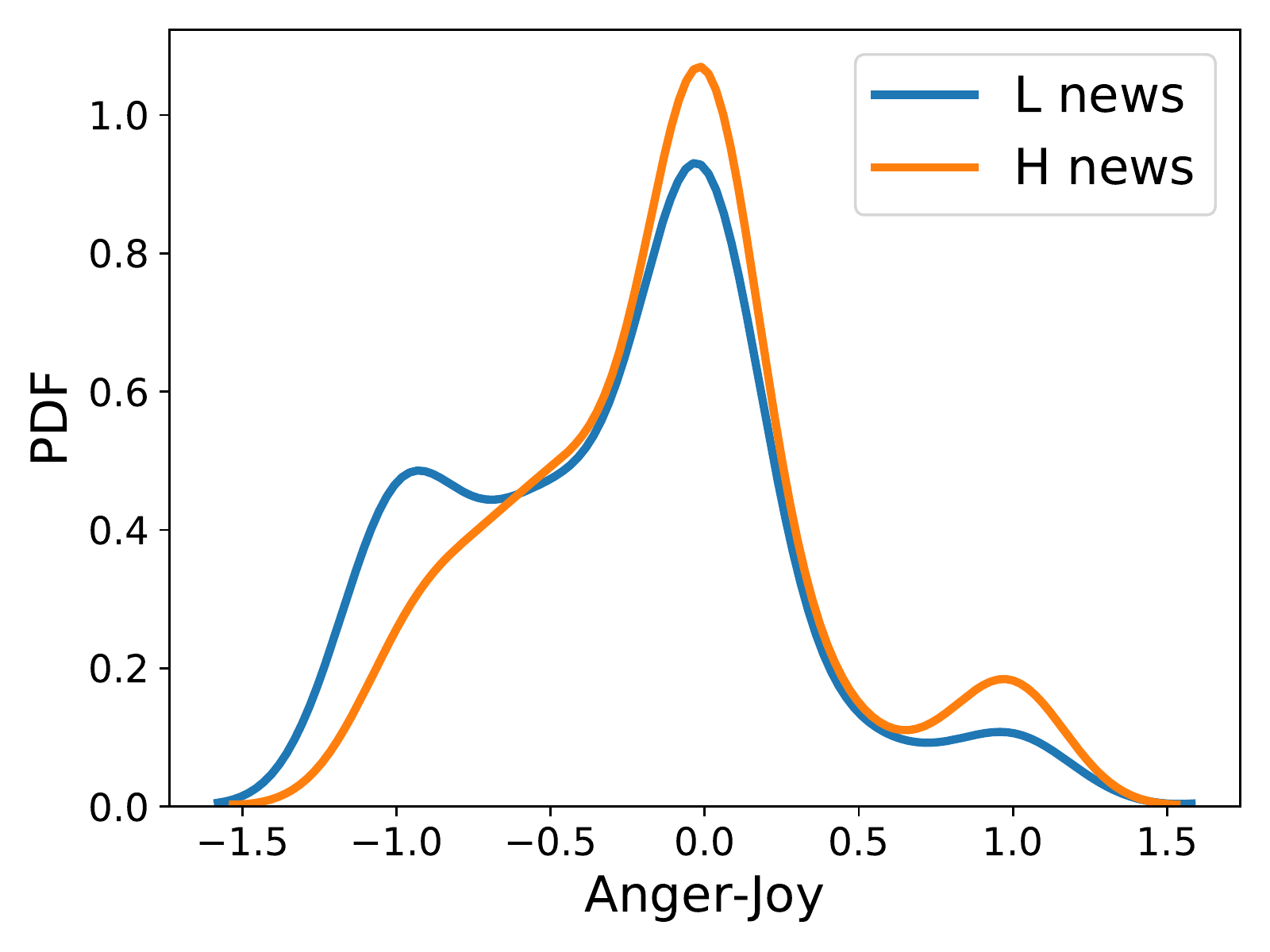}
	\caption{Probability density function (PDF) of Anger-Joy in LF news and HF news items.}
	\label{fig:covid-pdf-diff}
\end{figure}

\begin{table}
	
	\centering
	\includegraphics[]{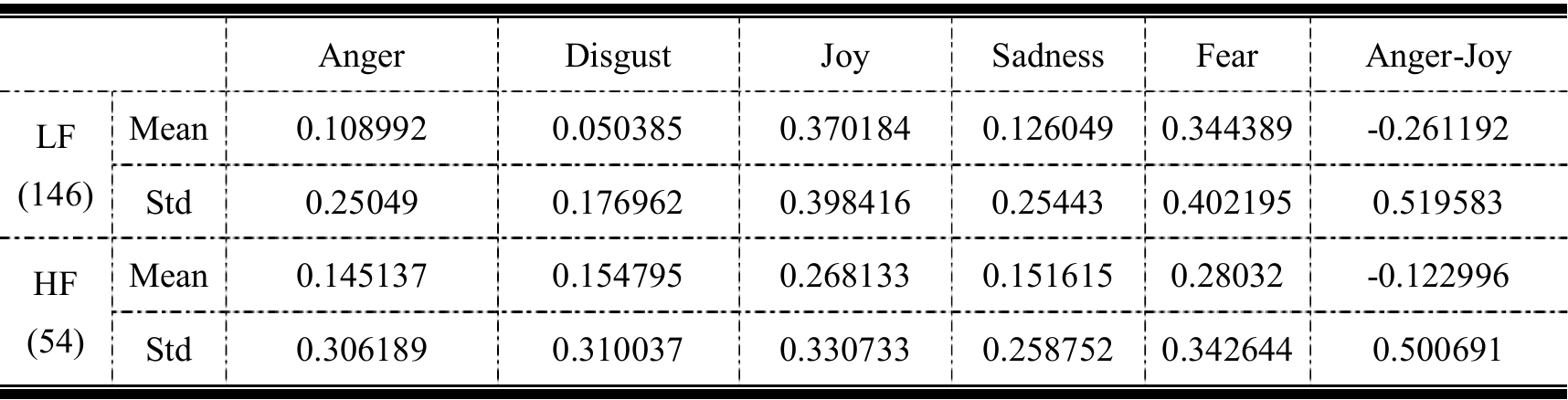}
	\caption{Statistics for LF news and HF news items in COVID-19. }
	\label{table:lf_hf_mean_covid}
\end{table}
\clearpage

\section*{S8 Control variables }
Carrying more anger but less joy is significantly associated with the fast spread of fake news. To further examine the causal impact of anger and joy on the circulation of news online, variables that might be correlated with the spread should be comprehensively considered and controlled. In addition to emotions inferred from texts, other factors such as content \cite{Suh}, user profiles \cite{Vosoughi2018}, and external shocks such as disaster events \cite{Spinney2019} that could be obtained from the content are considered and controlled. Note that considering the fast spread of fake news (see S3) and, in particular, that most people do not critically question its credibility \cite{Lazer2018}, only variables that can be derived at the very beginning of the posting are considered, while those related to spreading structures that are usually employed in the detection of fake news \cite{Zhao2020} are not considered due to the ex post facto inference. In addition to variables derived from content at the source, we introduce the number of followers and the number of friends, i.e., reciprocal followers, in Weibo as control variables to further consider the possible impact from user profiles. Notably, the ages of the authors are missing from the user profiles returned by Weibo’s open API. However, evidence from previous efforts of the impact of age on spread is inconsistent \cite{Vosoughi2018,Guess2019}. In the meantime, according to the annual report (\url{https://data.weibo.com/report/index}), most Weibo user ages are concentrated in a narrow range between 18 and 30 years old, so the impact of age could be trivial because of context dependence. Additionally, according to recent results in \cite{Guess2019}, the users’ ages are associated with the content topics, e.g., those aged 60 or greater are more likely to post/repost political tweets. Hence, in our model, the users’ age could be indirectly controlled through the considered topics. Thus, age can be omitted without significant disturbance to the results.  

In total, the following variables are derived and controlled:
\begin{itemize}
\item	Mention: Whether the text contains @.
\item	Hashtag: Whether the text contains a hashtag.
\item	Location: Whether the text contains location information.
\item	Date: Whether the text contains date information.
\item	URL: Whether the text contains a URL.
\item	Length: The length of the text.
\item	Emergency: Whether the text content is related to a disaster event. The emergency event in this study refers to the explosion accident in the Tianjin Binhai New Area on August 12, 2015, which occurred within the sampling period.
\item	Topic: The topic discussed in the text.
\item	Follower: The number of followers of the author.
\item	Friend: The number of friends of the author.
\end{itemize}

\subsection*{S8.1 Analysis of binary factors}
Table S16 shows the statistics of binary factors including mention, hashtag, location, date, URL, and emergency. From the perspective of the proportions of all binary factors, mention, and emergency have high proportions in LHF news, followed by H news, suggesting that both promote the spread of fake news. Hashtag, date, and URL have higher proportions in true news than in fake news, implying that they contribute little to the spread of fake news. Meanwhile, although the proportion of location is relatively high in fake news, it is concentrated mainly in L news, so its impact on spread might be trivial. These preliminary analyses offer directions for examining the causal impact of these factors on the spread of news.

\begin{table}	
	\centering
	\includegraphics[]{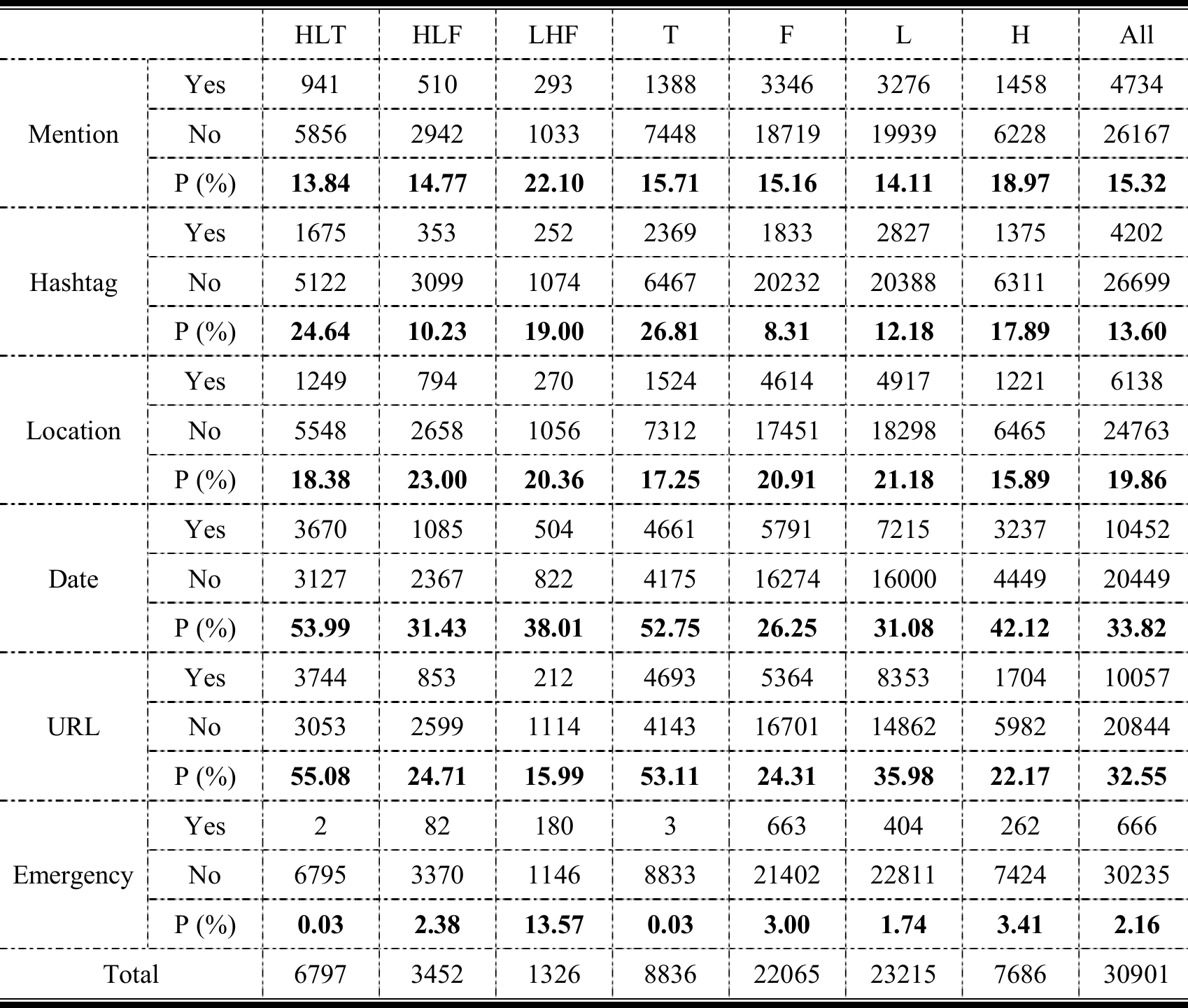}
	\caption{Statistics of binary factors. }
	\label{table:binary_factors}
\end{table}

\subsection*{S8.2 Analysis of Length}
We calculated the length distribution of the text as the number of characters and letters. The length of LHF news has a more concentrated distributed than that of HLT news (K-S test $\sim$ 0.145, P $\sim$ 0) (Fig. S13A), and the difference is also significant in fake news and true news (K-S test $\sim$ 0.134, P $\sim$ 0) (Fig. S13B). Therefore, fake news may be more deliberate and planned in terms of linguistic organization, while real news is more casually narrated. However, the text length is more concentrated in HLF news (compared with LHF news, K-S test $\sim$ 0.073, P $\sim$ 0) (Fig. S13A) and L news (compared with H news, K-S test $\sim$ 0.095, P $\sim$ 0) (Fig. S13C), indicating that this factor might have little effect on promoting the spread of false news.

\begin{figure}
	
	\centering
	\includegraphics[scale=0.4]{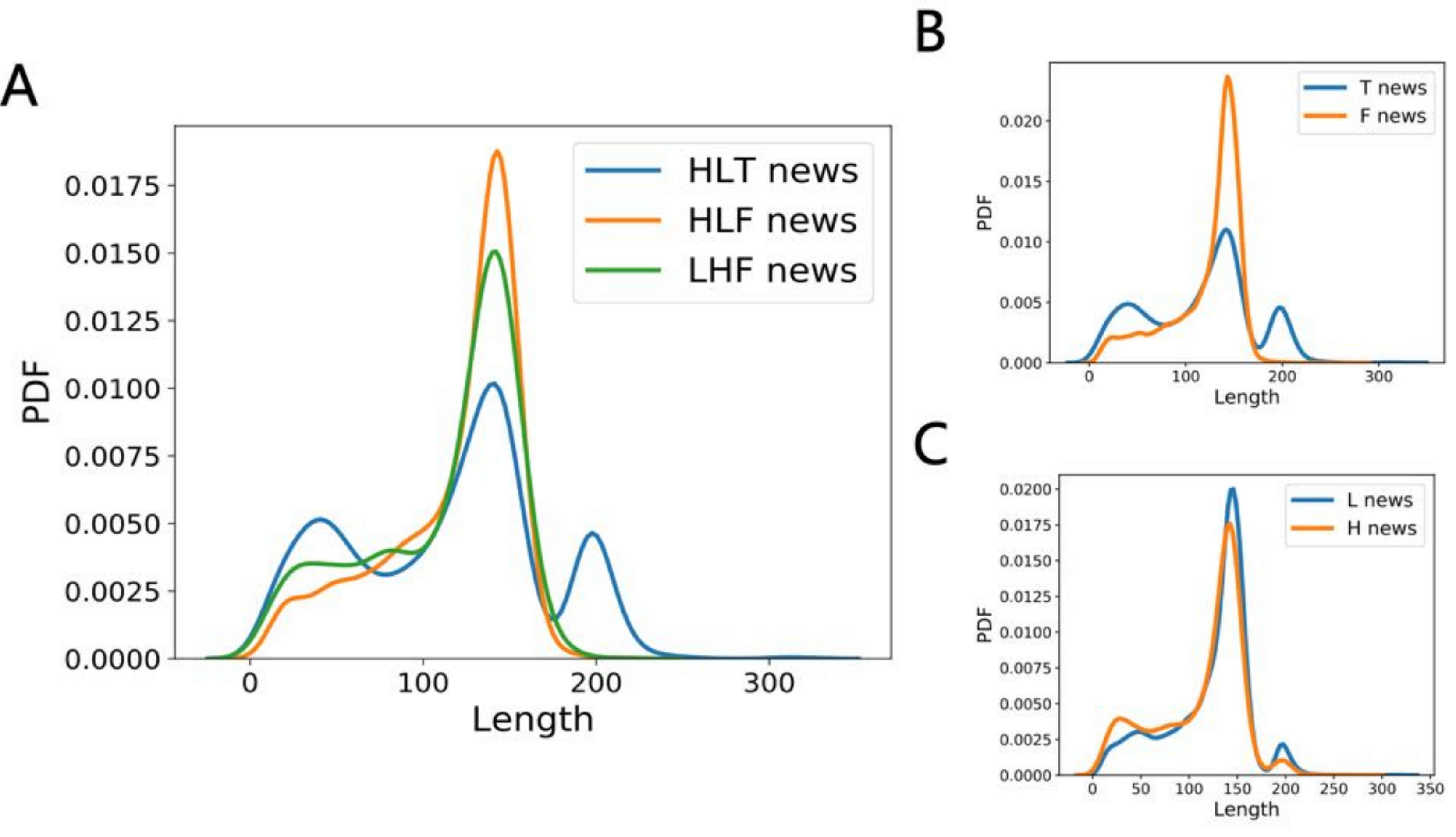}
	\caption{Probability density functions (PDFs) of length. }
	\label{fig:text_len_pdf}
\end{figure}

\subsection*{S8.3 Analysis of Topic}
The topics discussed in the news are also important features of the text. We used a na\"{i}ve Bayesian topic classifier \cite{Fan2015} to analyze the topic distributions of different types of news. The classifier was trained on more than 410,000 Weibo tweets, which were grouped into seven categories that fit the news taxonomy of Weibo: entertainment, finance, international, military, society, sports, and technology. The accuracy and F-measure are greater than 0.84, indicating good performance in topic classification. Besides, incremental training in this classifier can help solve the problem of new words. News that cannot be classified into the above seven categories is omitted in the analysis. As shown in Fig. S14, significant differences are observed in the distribution of topics among different groups of news. Specifically, the topic of society accounts for the largest proportion in HLF news, LHF news, and F news, suggesting that fake news focuses on social issues that are closely related to people’s daily lives. Hot social topics would make fake news more likely to spread but do not necessarily make fake news widely spread because H news’s proportion of society topic is lower than that of L news.

Through the analysis of the above eight variables derived from content, the differences between true and fake news are examined, but many do not promote the spread of fake news. Two factors, mention and emergency, may play promoting roles in the spread of fake news; however, they only occupy small proportions of all news items, which might undermine their effect on fast circulation. 

\begin{figure}
	
	\centering
	\includegraphics[scale=0.6]{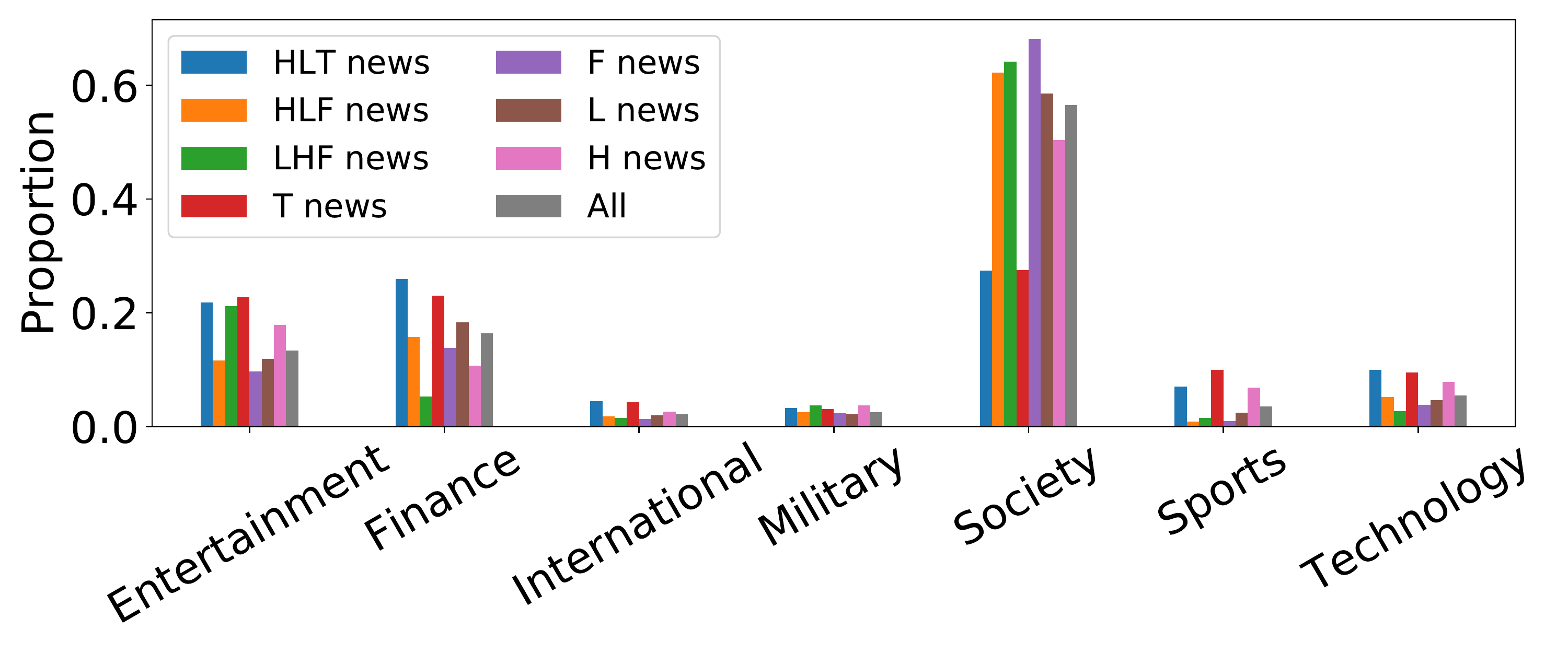}
	\caption{Topic distributions of different groups of news items.}
	\label{fig:topics}
\end{figure}

\subsection*{S8.4 Analysis of variables from authors}
We also examine the variables from the author profiles. Interestingly, whether true or fake, news with more retweets was posted by authors with more followers (Fig. S15) and friends (Fig. S16). However, the greater numbers of followers and friends associated with true news (as compared to fake news, and is consistent with the Twitter findings \cite{Vosoughi2018}) suggest that these factors might not be the key factors making fake news more viral than true news online.
By controlling all these variables, we establish both logit and linear models to examine the causal impact of anger and joy on the spread of fake news.

\begin{figure}	
	\centering
	\includegraphics[scale=0.6]{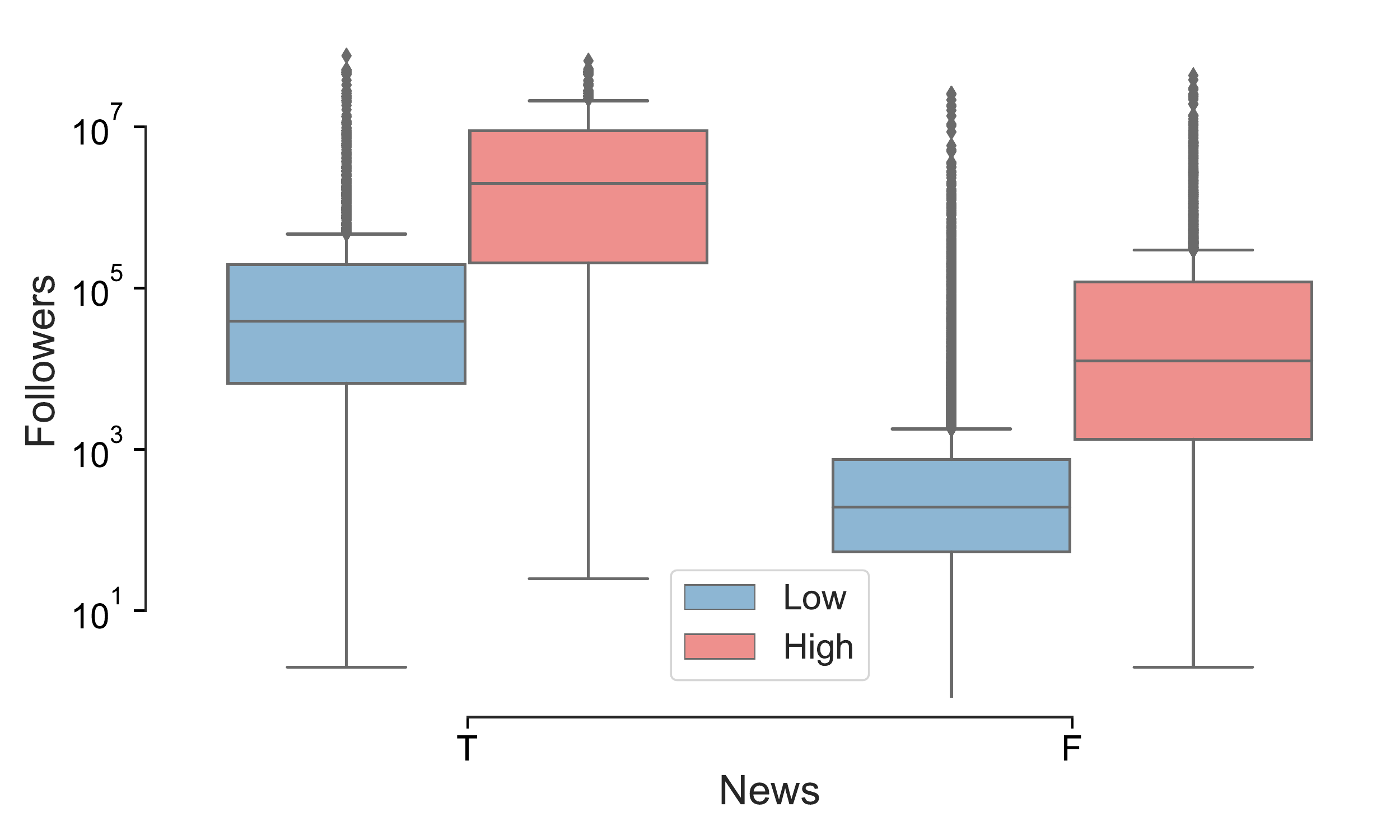}
	\caption{The boxplots of followers in true (LT and HT) news and fake (LF and HF) news items.}
	\label{fig:box_followers}
\end{figure}

\begin{figure}
	
	\centering
	\includegraphics[scale=0.6]{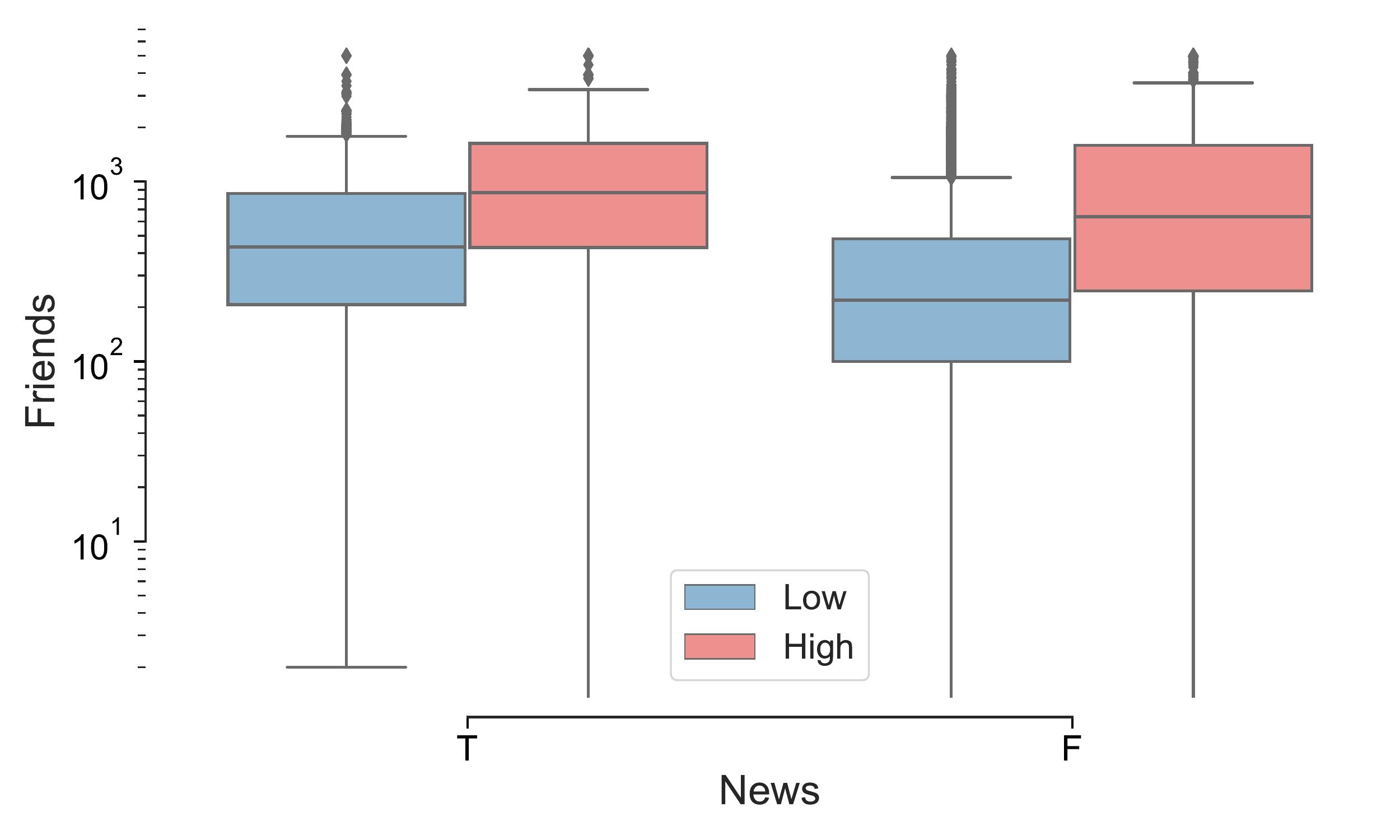}
	\caption{ The boxplots of friends in true (LT and HT) news and fake (LF and HF) news items. }
	\label{fig:box_friends}
\end{figure}
\clearpage

\section*{S9 Logit and linear regression models }
Logit and linear regression models are established to causally examine the impact of anger and joy on the spread of fake news. Note that for emotion variables, we focus primarily on anger and joy and combine the other emotions into other emotions. Note that there is a linear relationship between emotion-related variables because the ratios of the five emotions sum to 1. All the control variables from content, user profiles, and the external shock, as presented above, are comprehensively introduced into both models.
The logit model is defined as

$logit\left(p_{fake}\right)\ =\ \beta_0\ +\ \beta_1v_1\ +\ \beta_2v_2\ +\ \beta_3v_3\ +\ \beta_4v_4\ +\ \beta_5v_5\ +\beta_6v_6\ +\beta_7v_7\ +\ \beta_8v_8\ +\ \beta_9v_9\ +\ \beta_{10}v_{10}\ +\ \beta_{11}v_{11}\ +\ \beta_{12}v_{12}\ + \bm{\alpha^{'}X},$

\noindent where
\begin{itemize}	
\item $p_{fake}$ is the probability of fake news.
\item $\beta_0$ is the intercept.
\item $\beta_1,\ \beta_2,\ ...,\ \beta_{12}$ and $\bm{\alpha}$ are the coefficients of variables.
\item $v_1,\ v_2,\ ...,\ v_{12}$ represent anger, joy, other emotions, follower, friend, mention, hashtag, location, date, URL, length, and emergency.
\item $\bm{X}$ represents topic control dummy variables.
\item Mention, hashtag, location, date, URL, and emergency are dummy variables.
\end{itemize}

Emotion variables derived from emotion distributions in the logit model are calculated for all methods, namely, emotion lexicon, Bayes, and BP1. The results of the model based on the emotion lexicon are shown in Table 1 of the main text. We hereby supplement the estimation results for the remaining two methods (Table S17). In all the results, the coefficients of anger are uniformly and significantly positive after controlling for all other variables, indicating that anger is causally associated with fake news, particularly news that is highly retweeted. By contrast, the coefficients of joy are significantly negative in all results, especially for HF news and H news, indicating its prevention on the spread or news, particularly fake news. The coefficients of emergency and military and the topic of society are significantly positive, while the coefficients of mention are positive but nonsignificant (Table 1 in the main text and Table S17), which is consistent with our analysis in S8.

Then, a linear regression model is established to further qualify the influence of anger and joy on the spread of fake news. The model is defined as

$reg\left(Num_{retweet}\right)\ =\ \beta_0\ +\ \beta_1v_1\ +\ \beta_2v_2\ +\ \beta_3v_3\ +\ \beta_4v_4\ +\ \beta_5v_5\ +\beta_6v_6\ +\beta_7v_7\ +\ \beta_8v_8\ +\ \beta_9v_9\ +\ \beta_{10}v_{10}\ +\ \beta_{11}v_{11}\ +\ \beta_{12}v_{12}\ + \bm{\alpha^{'}X},$

\noindent where
\begin{itemize}
	\item The dependent variable $Num_{retweet}$ is the number of retweets within 48 hours of news release. Note that over 70\% of retweets of fake news and 80\% of retweets of real news occurred within 48 hours after posting (see S3). Other settings, e.g., longer than 48 hours, do not influence the results.
	\item The independent variables are consistent with the explanatory variables of the logit model.
\end{itemize}

We first estimate the linear model on fake news and then for all news, neglecting the labels of true or fake; the results can be found in Table 1 (3, 4) of the main text, in which the emotion distributions are inferred through the method based on the emotion lexicon. We also apply the linear model on emotion distributions from the other two methods, and consistent results are obtained, as shown in Table S17 (3, 6). Specifically, the positive coefficient of anger indicates its causal promotion on the spread, while the negative coefficient of joy indicates its preventive effect on the circulation of fake news. Furthermore, the coefficients of emergency, military topic, and social topic are significantly positive, implying their roles in enhancing the spread of information. 

\begin{table}	
	\centering
	\includegraphics[scale=0.6]{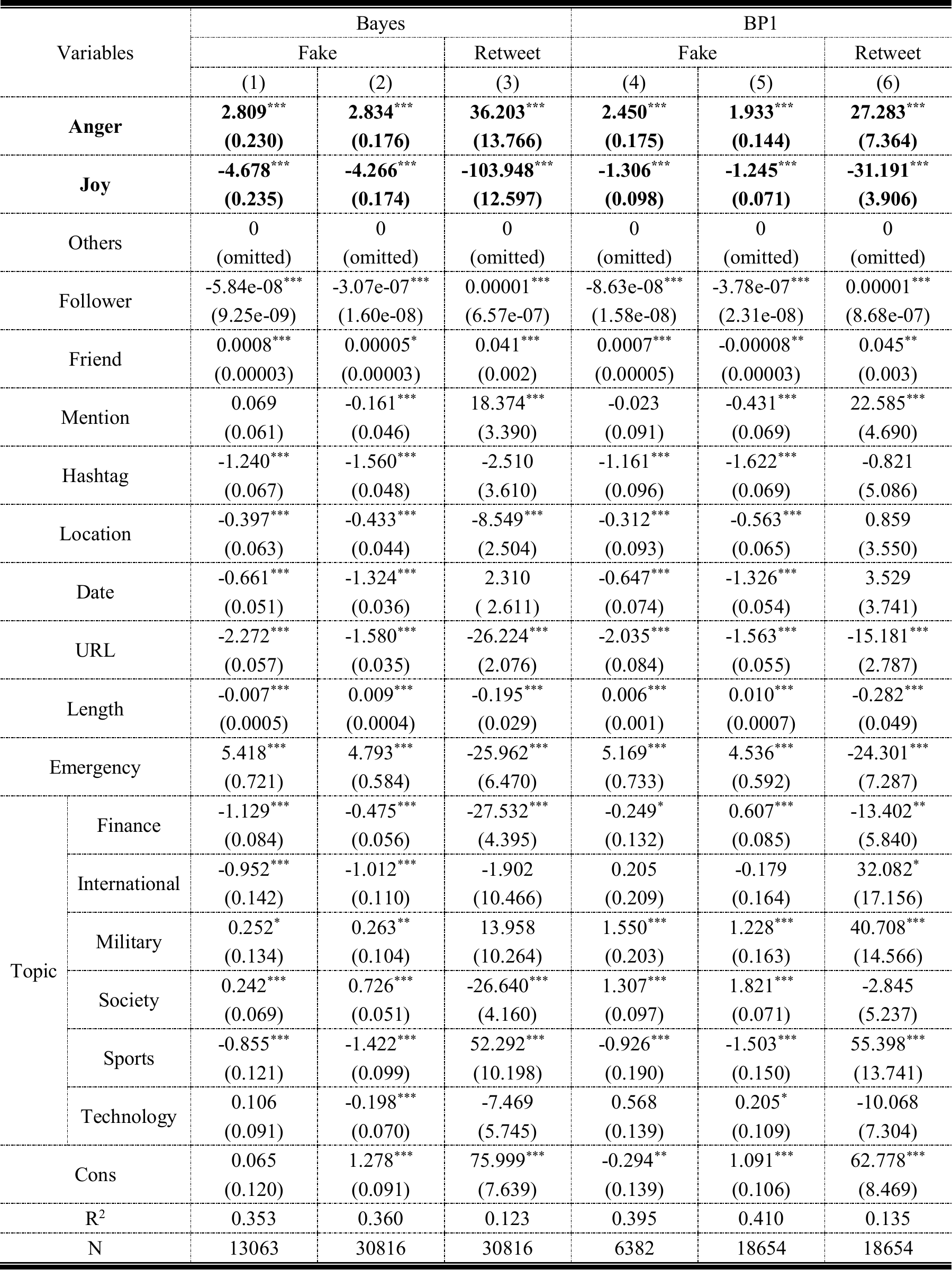}
	\caption{The validation results of the logit and linear models in different groups. (1,4) The logit model for LT news and HF news. (2,5) The logit model for T news and F news items. (3,6) The linear model for L news and H news. $^\ast P<0.1, ^{\ast\ast} P<0.05, ^{\ast\ast\ast} P<0.01$.}
	\label{table:model_others}
\end{table}

\section*{S10 Additional validations on English news}
It has been stated that emotional expression is culture dependent \cite{Bagozzi1999}. Though previous results on diffusion networks (see S2) and timeline analyses (see S3) demonstrated consistency with English tweets in Twitter and suggest the universality of our data from Weibo, more evidence on the roles of anger and joy in circulation through regression models of causal inference are still necessary. Here, six publicly available online datasets are accordingly utilized to ensure that our results can be applied to English news items (tweets) from Twitter and even other mainstream news media like WASHINGTON (Reuters). These datasets include:

\begin{enumerate}[(1)]
	\item Dataset S1: 12,247,065 coronavirus (COVID-19) tweets posted from 4 March 2020 to 28 March 2020 in Twitter.\footnote{\url{https://www.kaggle.com/smid80/coronavirus-covid19-tweets\#2020-03-00\%20Coronavirus\%20Tweets\%20(pre\%202020-03-12).CSV}}
	\item Dataset S2: 8,642,360 coronavirus (COVID-19) tweets posted from 29 March 2020 to 15 April 2020 in Twitter.\footnote{\url{https://www.kaggle.com/smid80/coronavirus-covid19-tweets-early-april}}	
	\item Dataset S3: 3,835,546 coronavirus (COVID-19) tweets posted from 16 April 2020 to 24 April 2020 in Twitter.\footnote{\url{https://www.kaggle.com/smid80/coronavirus-covid19-tweets-late-april}}
	\item Dataset S4: 397,629 election day tweets scraped on the day of 2016 United States election in Twitter.\footnote{\url{https://www.kaggle.com/kinguistics/election-day-tweet\#selection_day_tweets.csv}}
	\item Dataset S5: 23,481 fake news and 21,417 real news posted from 31 March 2015 to 19 February 2018 and miss retweets.\footnote{\url{https://www.kaggle.com/clmentbisaillon/fake-and-real-news-dataset}}
	\item Dataset S6: 478 fake news (tweets) posted during breaking news related to the events including Prince Toronto, Charlie Hebdo, Germanwings-crash, Sydney siege and etc. in Twitter.\footnote{\url{https://figshare.com/articles/PHEME_dataset_for_Rumour_Detection_and_Veracity_Classification/6392078}}
\end{enumerate}

Each English tweet in Datasets S1-4 contains the text, retweet count, follower count, friend count, etc. Though there are no labels of whether these tweets are fake or real, the promoting effect of anger on retweeting can still be verified. We randomly extracted 2,000 news items from each file (one file per day) in Datasets S1-3 and obtained 90,000 news items (57,508 news items with retweets) related to COVID-19. Besides, there are 72,182 politically-related news items with retweets in Dataset S4. News items with retweets extracted from Datasets S1-3 (COVID-19) and Dataset S4 (Politics) are thus combined to examine the effects of emotions on information spread. We divided news items into L news and H news items according to the number of retweets and built the logit model as follows: 

$logit\left(p_{h-news}\right)\ =\ \beta_0\ +\ \beta_1w_1\ +\ \beta_2w_2\ +\ \beta_3w_3\ +\ \beta_4w_4\ +\ \beta_5w_5\ +\beta_6w_6\ +\beta_7w_7\ +\ \beta_8w_8\ +\ \beta_9w\ +\ \beta_{10}w_{10}\ +\ \beta_{11}w_{11}\ +\ \beta_{12}w_{12}\ +\ \ \beta_{13}w_{13}+\ \ \beta_{14}w_{14}+\ \ \beta_{15}w_{15}+\ \ \beta_{16}w_{16}+\ \ \beta_{17}w_{17},$

\noindent where
\begin{itemize}
	\item $p_{h-news}$ is the probability of H news (tweets with more than 10 retweets in the dataset).
	\item $\beta_0$ is the intercept.
	\item $\beta_1,\ \beta_2,\ ...,\ \beta_{17}$ are the coefficients of variables.
	\item $w_1,\ w_2,\ ...,\ w_{17}$ represent variables of anger, disgust, joy, sadness, fear, surprise, anticipation, trust, follower, friend, mention, hashtag, location, date, URL, length, and topic.
	\item Topic indicates politics or COVID-19.
\end{itemize}

The emotion lexicon from the National Research Council of Canada (NRC) was employed to infer the emotional distributions of all of the English news items. It contains 14,182 words with eight emotions: anger, disgust, joy, sadness, fear, surprise, anticipation, and trust \cite{Mohammad,Mohammad2013}. The coverage of this emotion lexicon is 73.3\% on the dataset used in the logit model. Though emotions carried by the English news items here are expanded to eight emotions, the promoting effect of anger is still significant and that of joy is opposite as expected (Table S18). These results suggest that the promoting effect of anger for information spread is independent of cultural differences and our results can be confidently extended to English news items. Other emotions such as disgust and anticipation were also found to be significant but with negative coefficients, implying they prevent the spread of information. It should also be noted that here the linear model was not examined due to the missing of retweeting time in these datasets, i.e., whether the spread of news was sufficiently sampled cannot be assured. Consequently, it is problematic to treat the number of retweets directly as a dependable variable.

Since whether the news in Dataset S1-4 is true or fake is not labeled, Dataset S5, containing 21,417 true news items (with 11,264 political news and 10,133 world news items) and 23,481 fake news items (with 9,050 news, 6,718 political news, 1,548 government news, 4,415 left-news, 781 U.S. news, and 776 Middle-east news items) was further utilized to verify the divergence of anger between true and fake news. Note that true news may be from sources such as WASHINGTON (Reuters) and Twitter. Hence, the title and body texts of the news items are joined together for the emotion inference analysis (the coverage of the emotion lexicon is nearly 100\%). As expected, anger occupation in fake news items is higher than that in true news items (true news items $\sim$ 0.110, fake news items $\sim$ 0.123, K-S test $\sim$ 0.108, P $\sim$ 0). There is also a very small dataset (Dataset S6) of fake news items containing 117 LF news items (tweets) with emotions and 361 HF news items (tweets) with emotions. It is also consistent with the Weibo results (see Table 1) in that HF news items in this Twitter dataset carry more anger than their counterpart LF news items (LF news items $\sim$ 0.020, HF news items $\sim$ 0.142, K-S test $\sim$ 0.416, P $\sim$ 0). 

In summary, the results from these supplementary datasets of English news items confirm our conclusions derived from Weibo and support that the finding is independent of the culture and platform, fake news items carry more anger than real news items, and anger promotes the circulation of news online.

\begin{table}	
	\centering
	\includegraphics[scale=0.6]{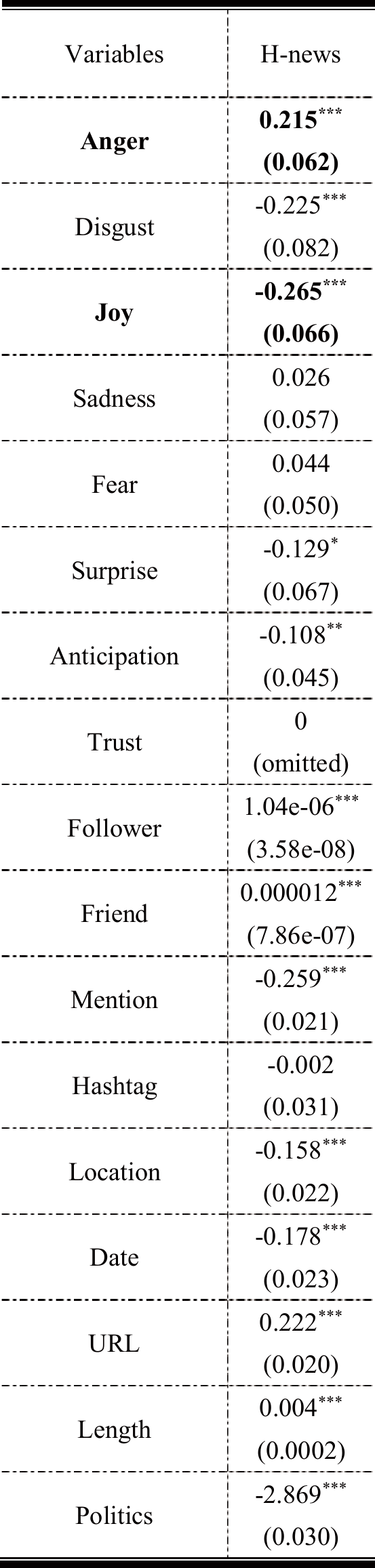}
	\caption{The logit model for the English news items about COVID-19 and politics. $^\ast P<0.1, ^{\ast\ast} P<0.05, ^{\ast\ast\ast} P<0.01$.}
	\label{table:logit_english}
\end{table}
\clearpage

\section*{S11 Selecting typical news for questionnaires }
Emotions of high arousal, such as anger and joy, are associated with information diffusion, particularly information sharing \cite{Stieglitz2014}. To further investigate how anger and joy carried in news influence incentives underlying retweeting, which reignites the circulation of news on social media, offline questionnaires are conducted to bind the emotion divergence between fake news and real news with retweeting incentives. Due to the time consuming and intensive labor costs, it is challenging for questionnaires to cover all the fake news and true news in our data. Therefore, five typical news items from groups of HLT news, LHF news, and HLF news are selected to perform the surveys. Similarly, in terms of news in these groups, the possible stimuli from emotions such as anger and joy to the retweeting incentives are hoped to be amplified to ease the following detection. To guarantee that the selection of news samples from each group is representative, each group of news is clustered before sampling. First, we use the word2vec model to convert the words in each news item into vectors of 200 dimensions and take the mean of these word vectors to represent the news item, i.e., the news item is similarly embedded in a space of 200 dimensions. Then, K-means clustering is employed to cluster each group of news items into five clusters. Next, based on including keywords with high importance in each news item (see S6) and intrinsic factors such as mentions and hashtags in each group (see S8), representative texts are sampled from those near the cluster centers. Note that we do not deliberately consider emotion distributions in the selection to avoid the impact of subjective bias on subsequent incentive stimuli and to ensure the objectivity of the results. Finally, we select 15 typical news items (Table S19-22), and their positions in the group can be found in Fig. S17. The sampled texts and the keywords in these texts are distributed evenly in the embedding space of different groups of news, suggesting that they are indeed typical and representative. Notably, the selected keywords that help separate the groups of news in sampling the texts are anticipated to help strengthen the stimuli of reposting incentives, which would further enhance the impact of anger and joy.

\begin{figure}
	\centering
	\includegraphics[scale=0.4]{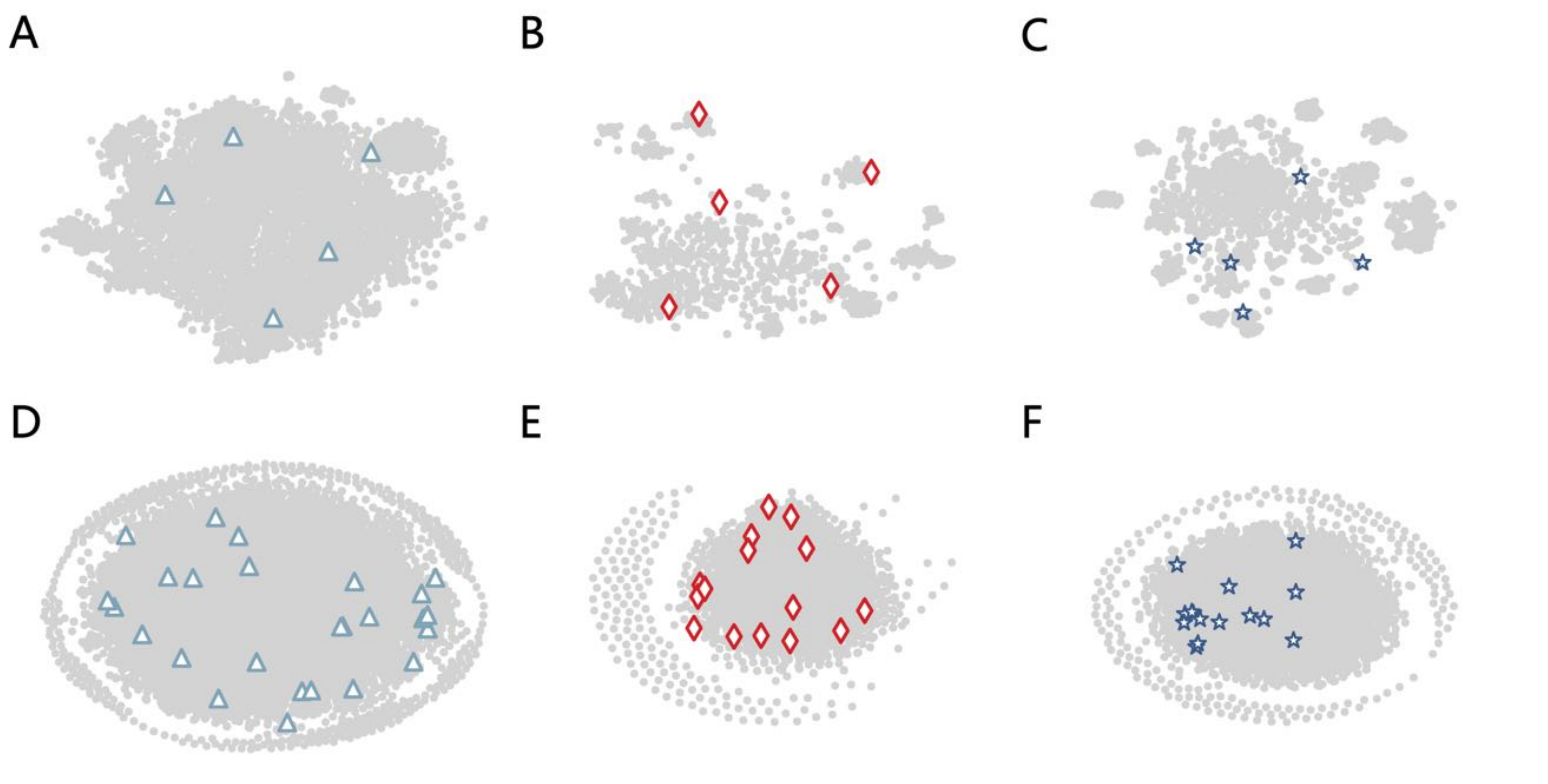}
	\caption{Positions of sampled texts and keywords in the embedding space. (A) Text in HLT news. (B) Text in LHF news. (C) Text in HLF news. (D) Keywords in HLT news. (E) Keywords in LHF news. (F) Keywords in HLF news.}
	\label{fig:space_point}
\end{figure}

\begin{table}
	
	\centering
	\includegraphics[scale=0.8]{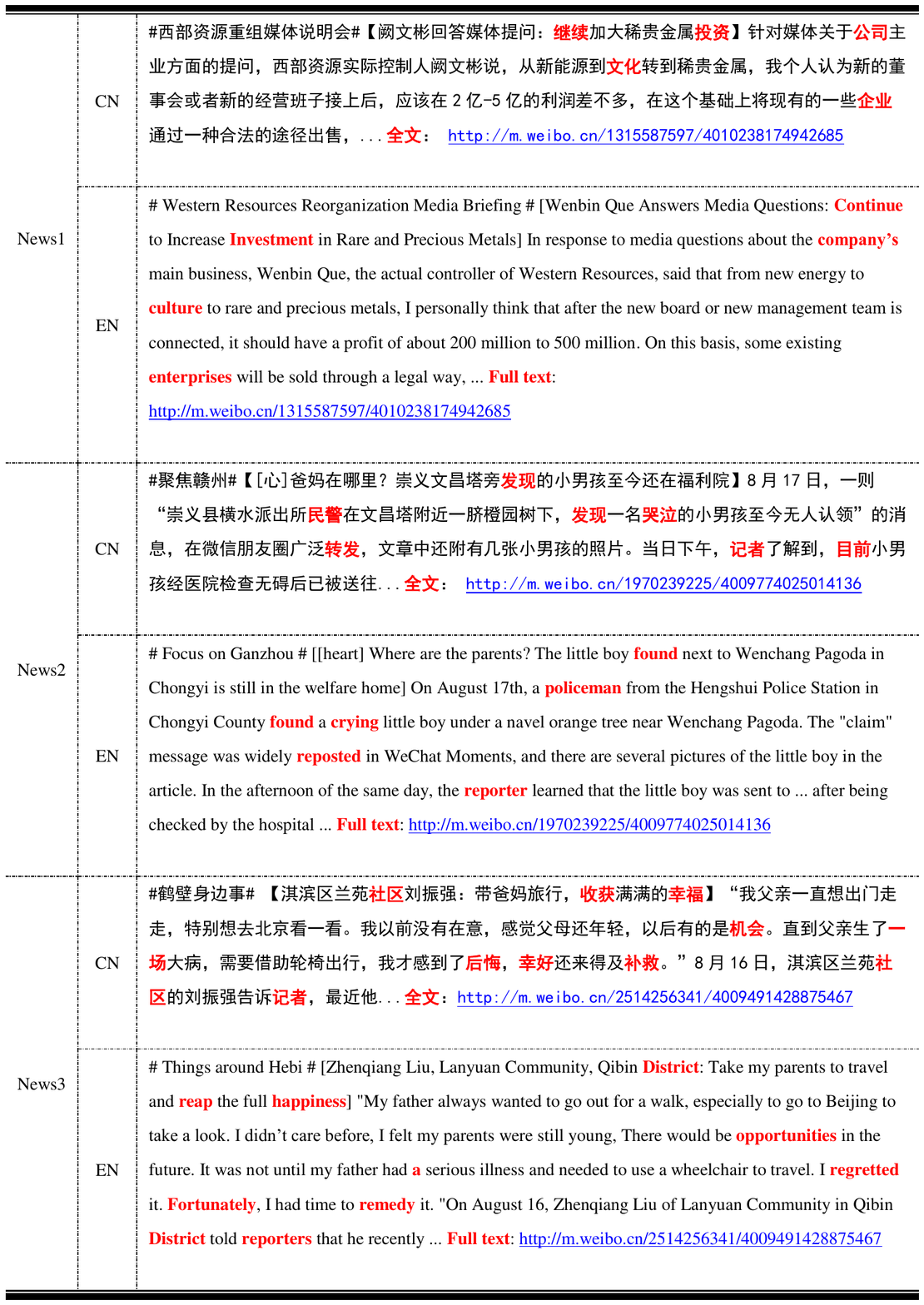}
	\caption{HLT-News1-3 selected in HLT news items. Keywords are highlighted in red.}
	\label{table:hlt_news_1-3}
\end{table}

\begin{table}
	
	\centering
	\includegraphics[scale=0.8]{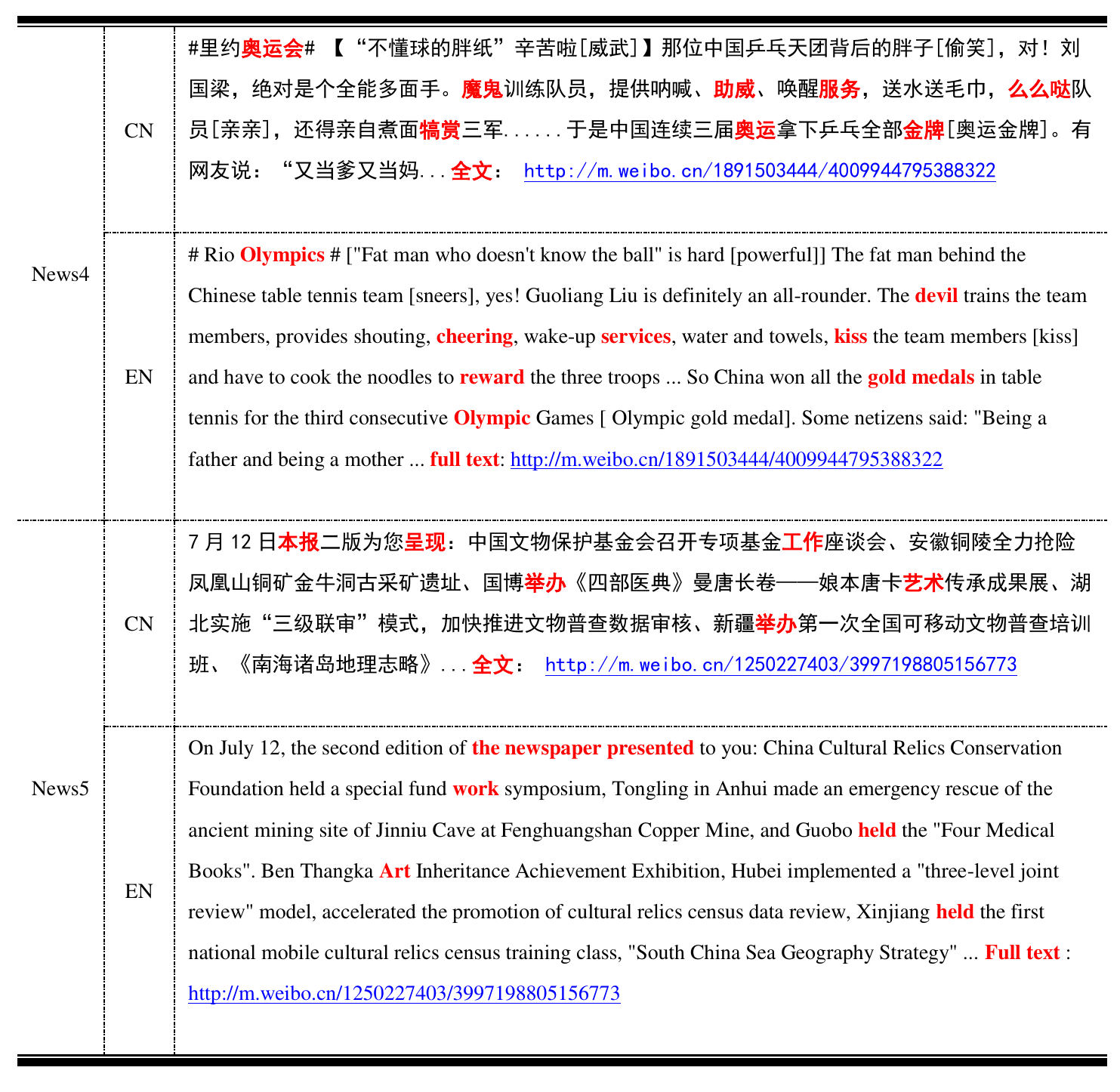}
	\caption{HLT-News4-5 selected in HLT news items. Keywords are highlighted in red.}
	\label{table:hlt_news_4-5}
\end{table}

\begin{table}	
	\centering
	\includegraphics[scale=0.8]{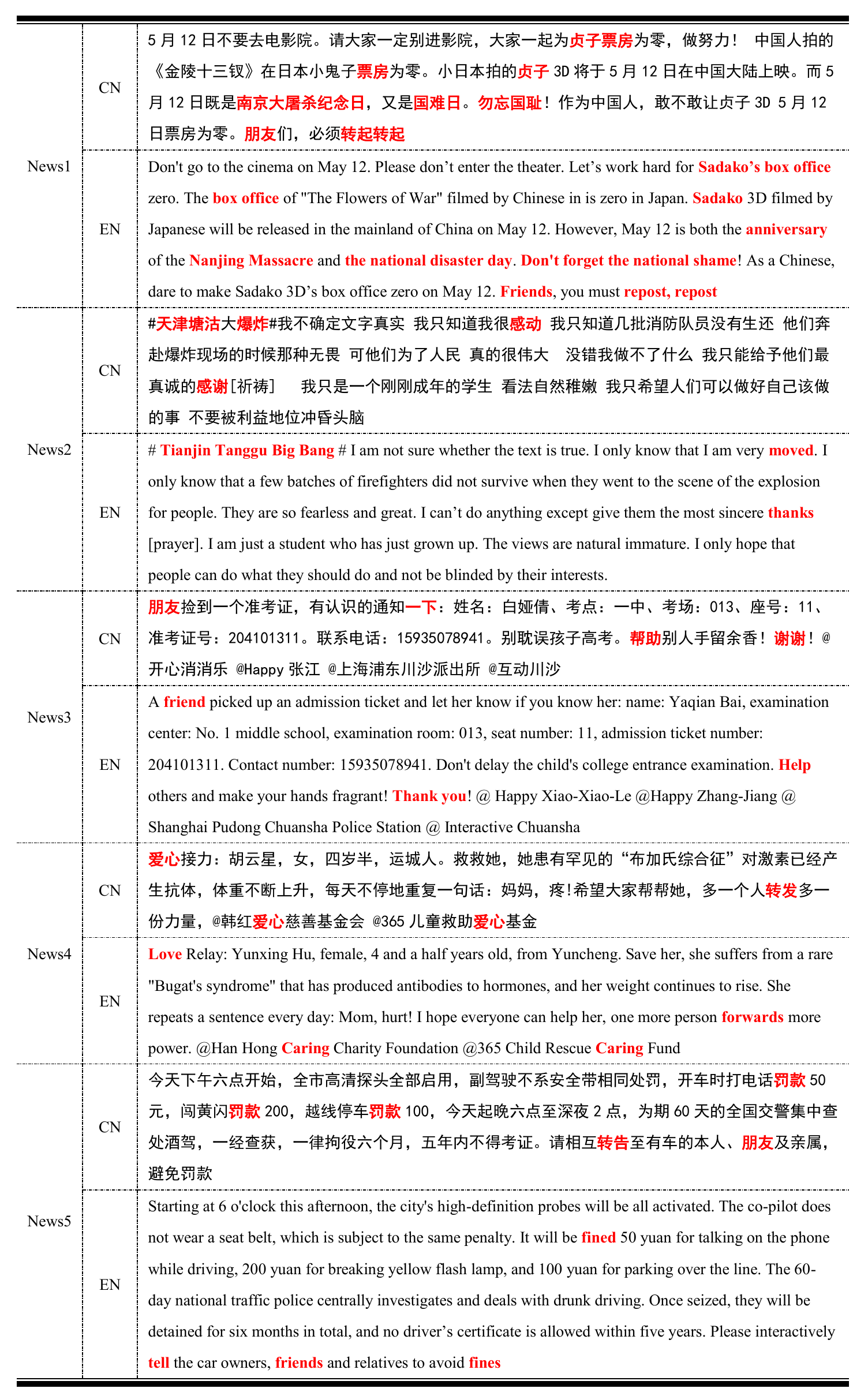}
	\caption{LHF-News1-5 selected in LHF news items. Keywords are highlighted in red.}
	\label{table:lhf_news_1-5}
\end{table}

\begin{table}
	
	\centering
	\includegraphics[scale=0.8]{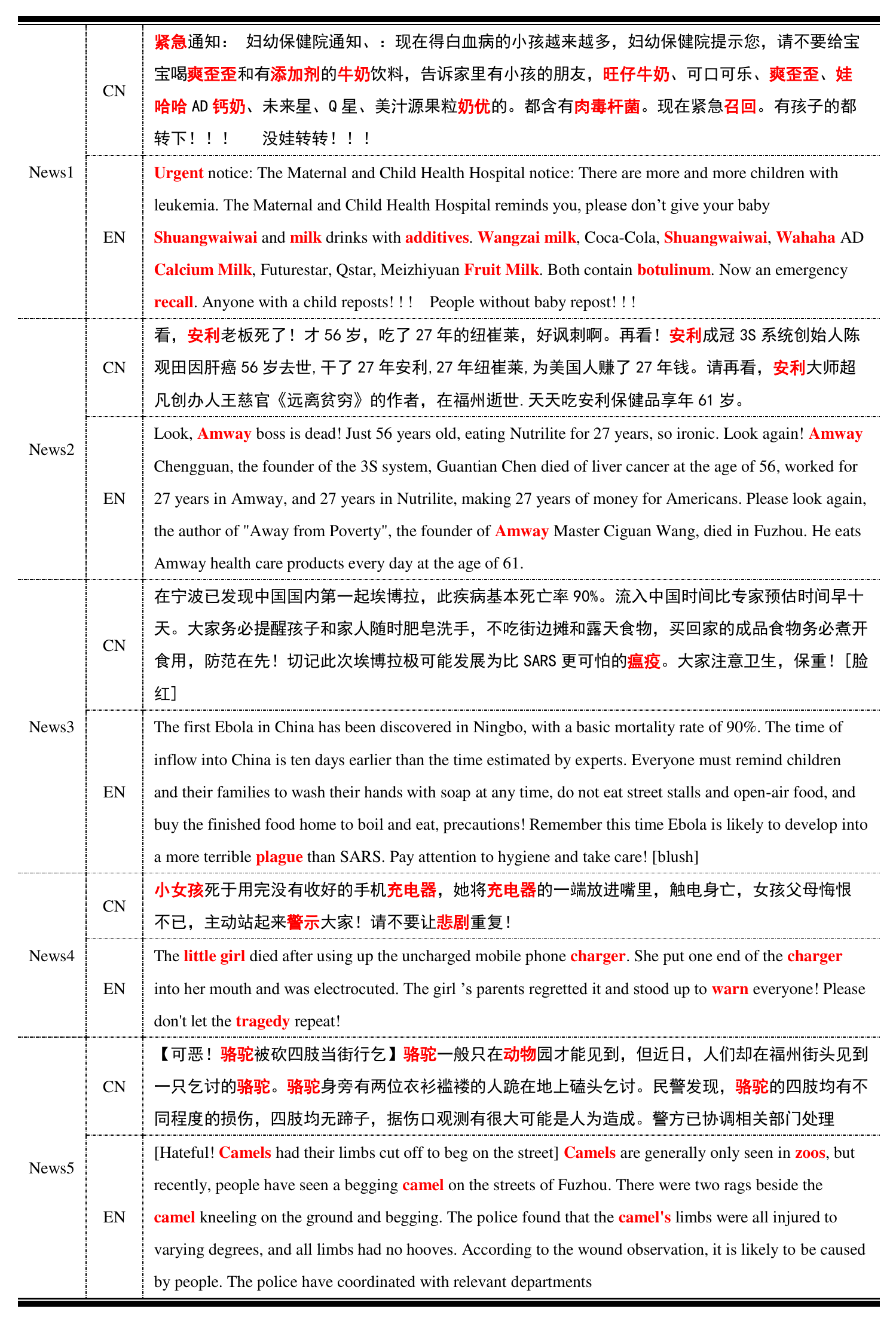}
	\caption{HLF-News1-5 selected in HLF news items. Keywords are highlighted in red.}
	\label{table:hlf_news_1-5}
\end{table}
\clearpage

\section*{S12 Questionnaires}
We employ a carefully designed questionnaire that is commonly used for rumor sharing motivation surveys on social media \cite{Sudhir2014}, which comprehensively measures four motivations of the subjects: anxiety management, information sharing, relationship management, and self-enhancement. There are six items for anxiety management (Fig. S18), six items for information sharing (Fig. S19), five items for relationship management (Fig. S20) and four items for self-enhancement (Fig. S21). Each item is measured on a four-point scale (1-strongly disagree, 2-disagree, 3-agree, 4-strongly agree). There are six questionnaires in total. For each group of news items, we implement two online questionnaires, one showing the original text and one showing the text with keywords marked in red squares (Fig. S22). Meanwhile, five news items from each group appear in each questionnaire randomly. Except for the news presented, all other circumstances in the questionnaires, e.g., author profile, posting time, and posting source, are carefully controlled to be consistent. Specifically, the difference in stimuli to the incentives of subjects is only the news itself. For the presentation of the text, we attempted to simulate the real Weibo interface by adding the background of the mobile version of the Weibo App to each news item (Fig. S22). For subjects who completed the questionnaires, we required them to be Weibo users aged between 18 and 30 years old (according to the 2018 Weibo user development report, this age group accounts for 75\% of all users), matching users in online data as much as possible (\url{https://data.weibo.com/report/index}). Note that subjects are not specifically targeted based on occupation or income level because we want to probe the general effect of emotion divergence on the retweeting incentives for the majority of Weibo users. More importantly, considering the widespread global impact of fake news online, revealing a mechanism that is independent of user demographics would be powerful in inspiring new cures.

\begin{figure}	
	\centering
	\includegraphics[scale=0.6]{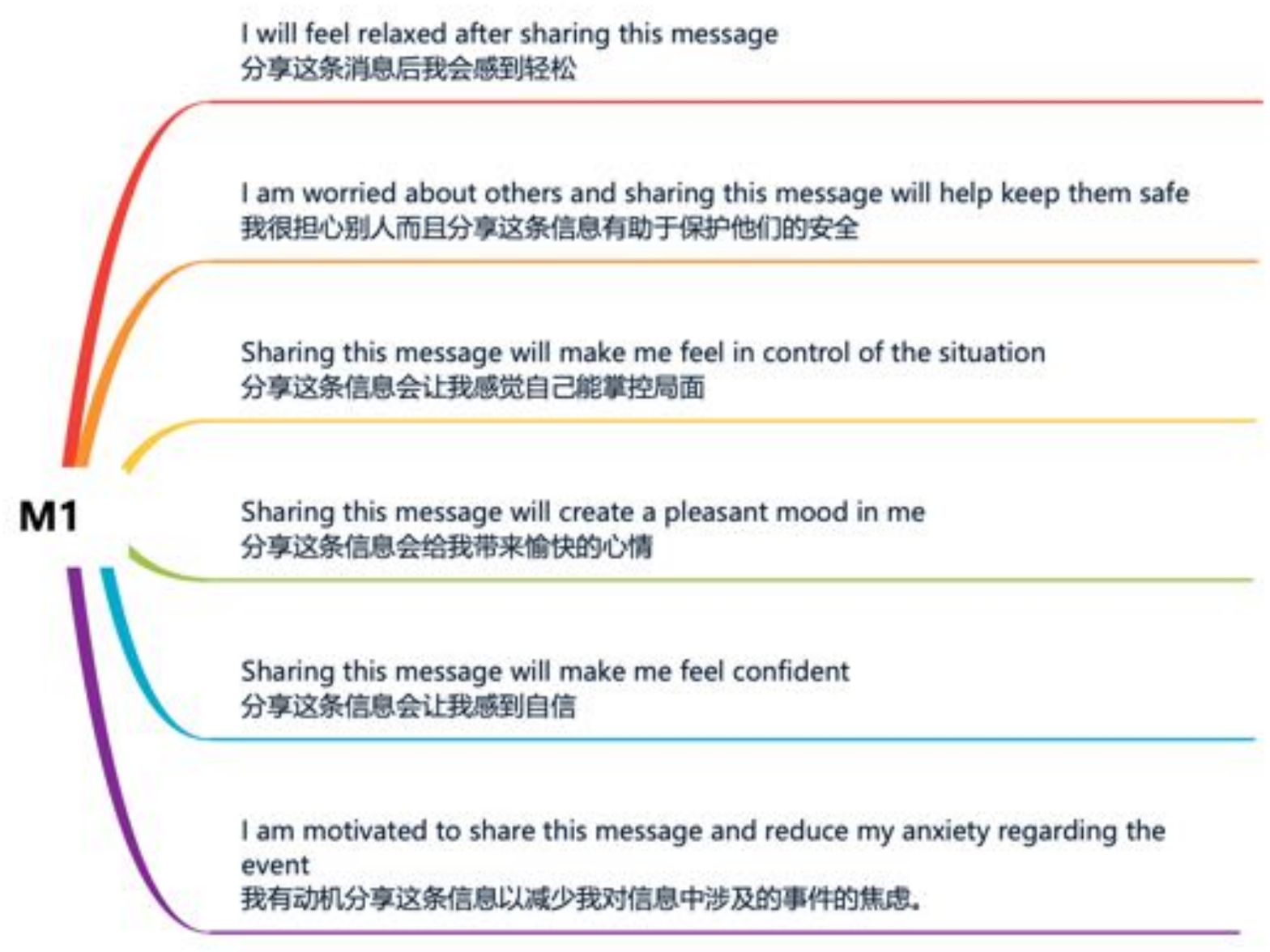}
	\caption{Anxiety Management Motivation (M1).}
	\label{fig:m1}
\end{figure}

\begin{figure}	
	\centering
	\includegraphics[scale=0.6]{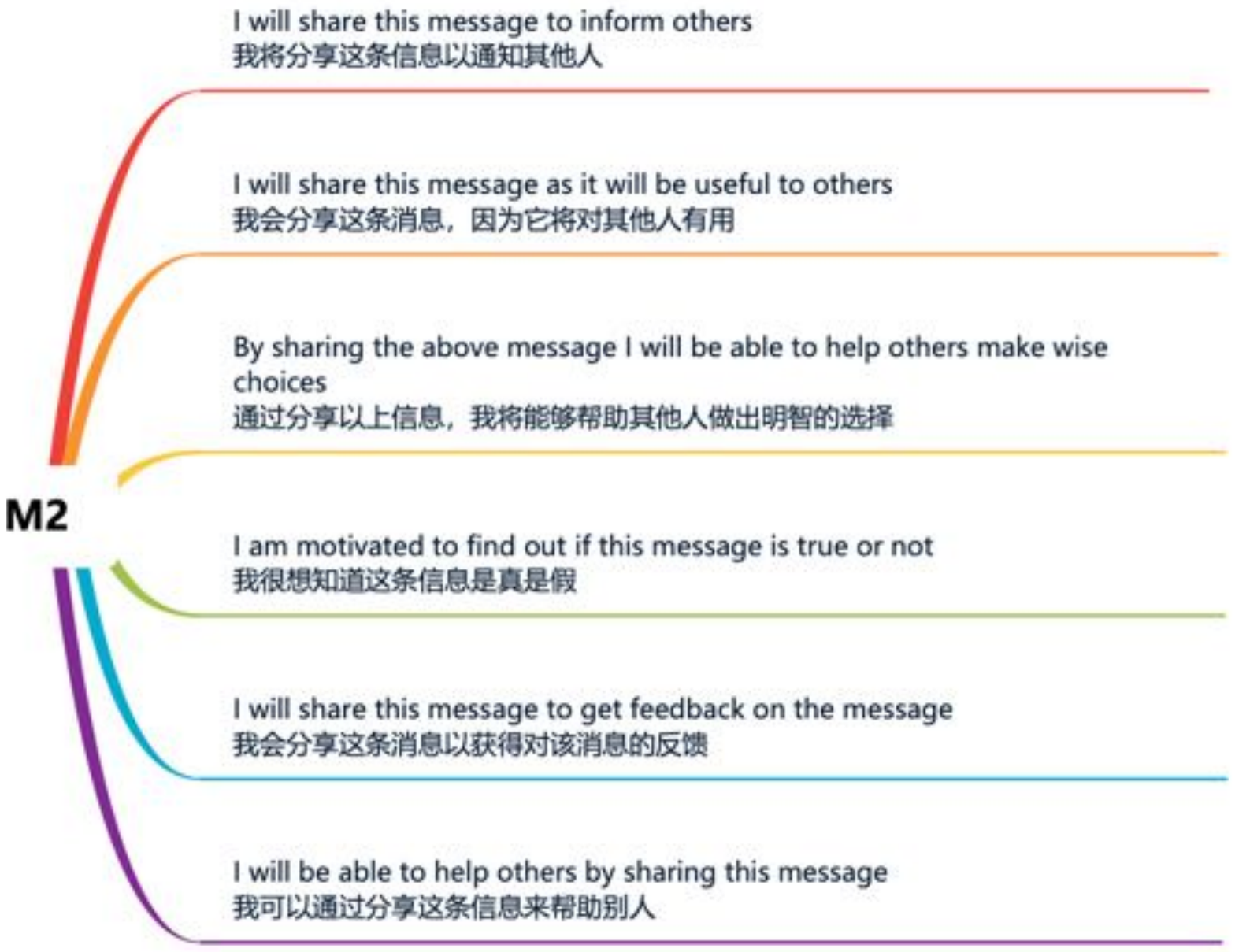}
	\caption{Information Sharing Motivation (M2).}
	\label{fig:m2}
\end{figure}

\begin{figure}
	\centering
	\includegraphics[scale=0.6]{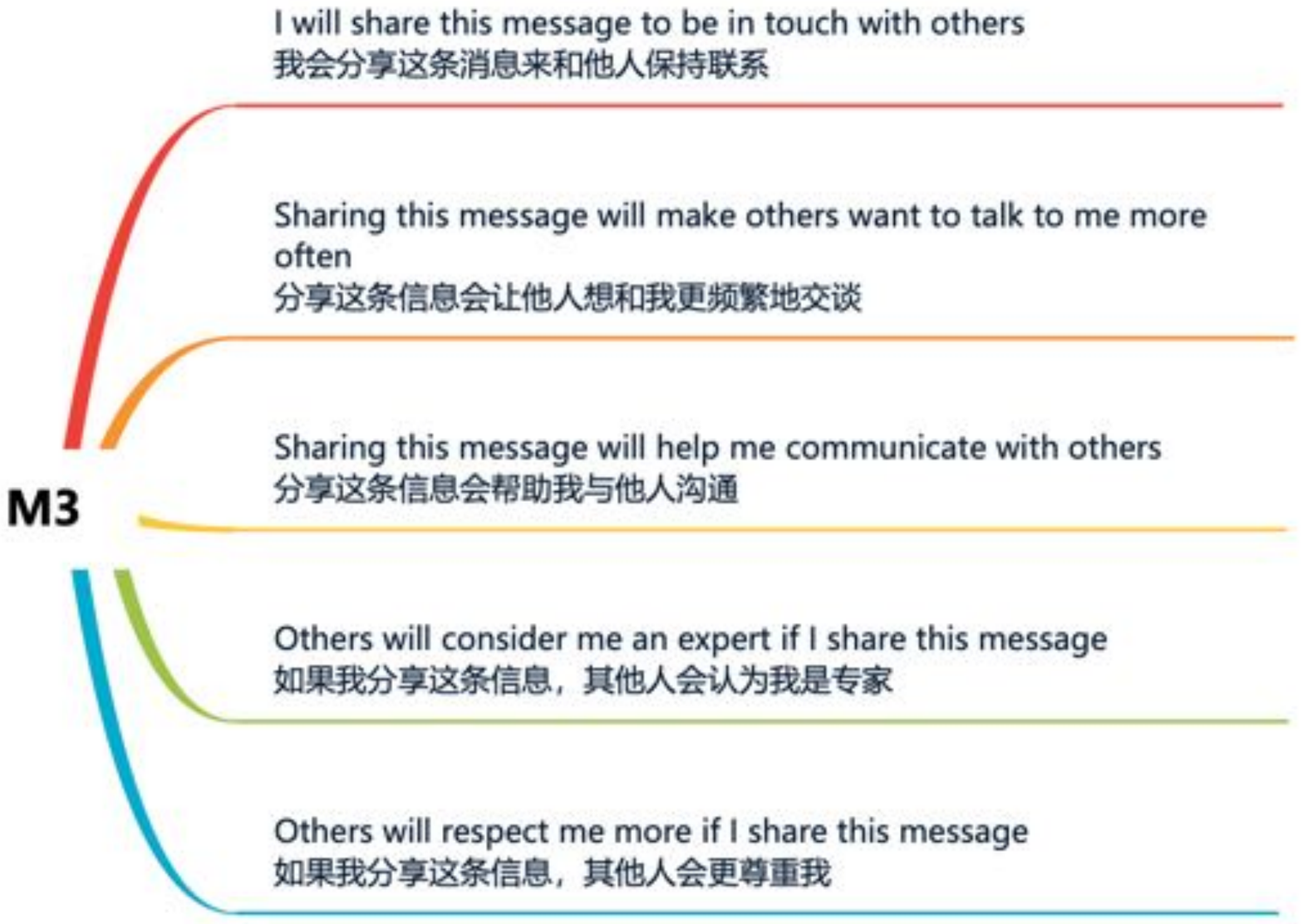}
	\caption{Relationship Management Motivation (M3).}
	\label{fig:m3}
\end{figure}

\begin{figure}	
	\centering
	\includegraphics[scale=0.6]{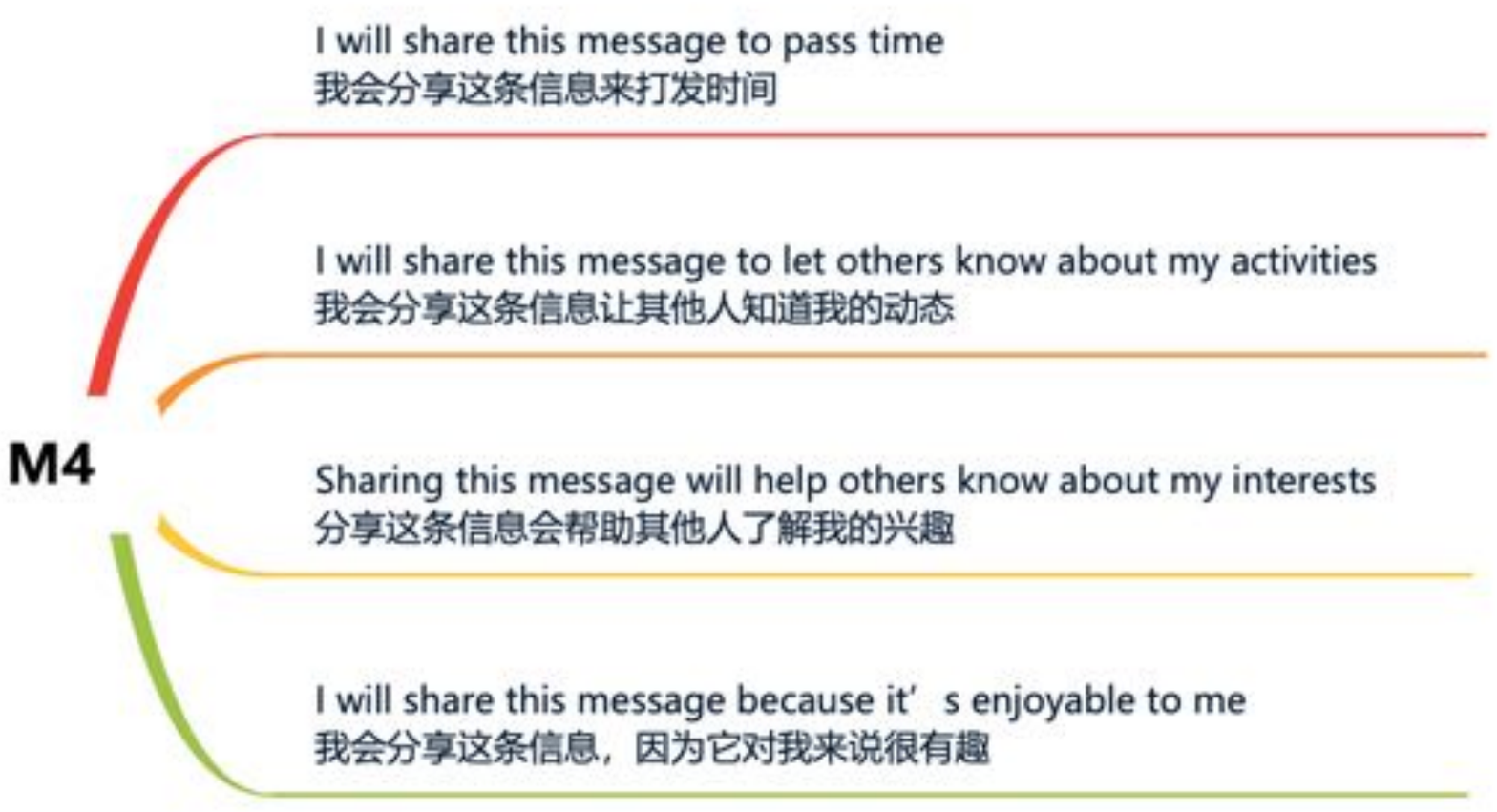}
	\caption{Self-enhancement motivation (M4).}
	\label{fig:m4}
\end{figure}

\begin{figure}	
	\centering
	\includegraphics[scale=0.4]{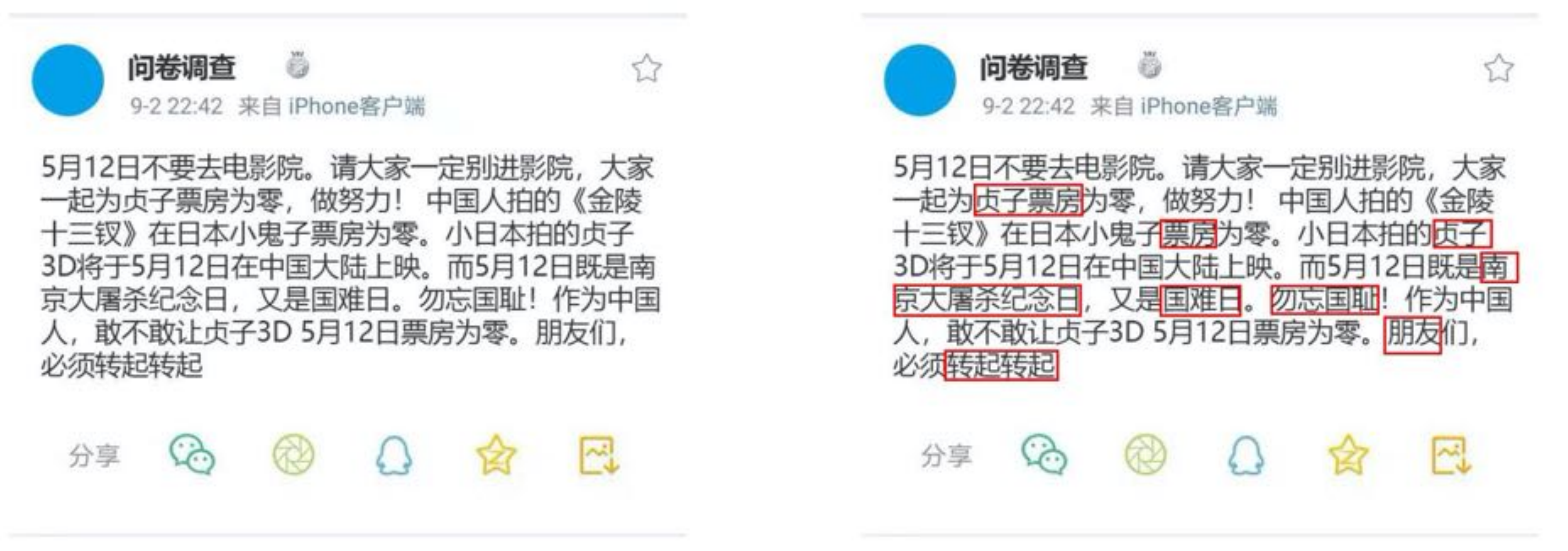}
	\caption{Questionnaire examples of original text (left) and text with marked keywords (right).}
	\label{fig:original_text}
\end{figure}
\clearpage

\section*{S13 Questionnaire results}
We hired a well-reputed online survey company (\url{https://www.wjx.cn/}) and collected a total of 1291 valid responses from 1316 subjects within China. Specifically, we obtained 224 responses to the unmarked HLT news questionnaire (HLT-Q1), 214 responses to the marked HLT news questionnaire (HLT-Q2), 210 responses to the unmarked LHF news questionnaire (LHF-Q1), 212 responses to the marked LHF news questionnaire (LHF-Q2), 211 responses to the unmarked HLF news questionnaire (HLF-Q1) and 210 responses to the marked HLF news questionnaire (HLF-Q2). All the responses are carefully validated, and the values of Cronbach’s alpha are provided in Table S23. The collected responses are also publicly available at \url{https://doi.org/10.6084/m9.figshare.12163569.v2}. Since subjective bias may exist, that is, the response degree might vary across different subjects, the following method is adopted to eliminate the subjective bias:

$$Mi-avg\ =m_i-\frac{m_1+m_2+m_3+m_4}{4},\ i=1,2,3,4$$

\noindent where $m_i$ is the average score of all the items in motivation $Mi$ and$Mi-avg$ is the debiased average score for $Mi$.

\begin{table}
	\centering
	\includegraphics[]{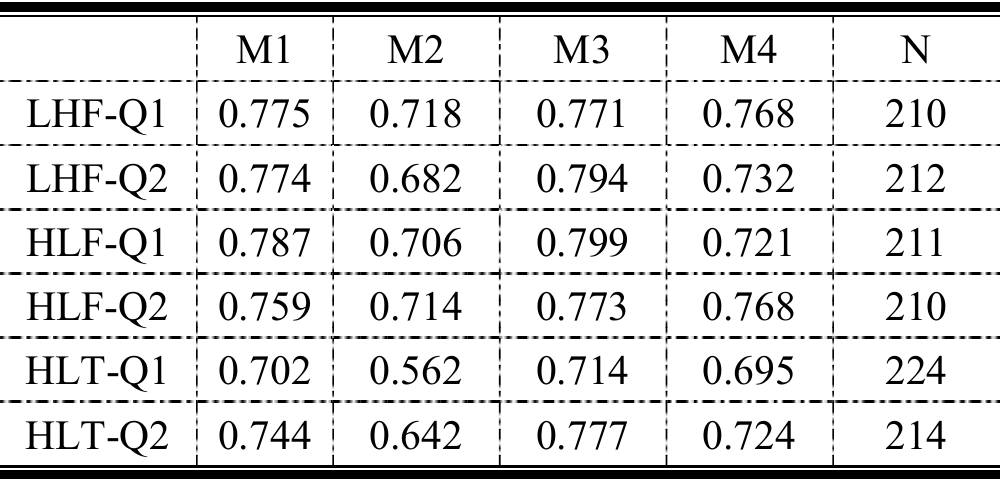}
	\caption{The values of Cronbach's alpha in different questionnaires.}
	\label{table:cronbach}
\end{table}

\subsection*{S13.1 Differences in motivations between different groups of news}
The main text showed that the motivation of information sharing of false news is stronger than that of real news, and the motivation of anxiety management of LHF news is significantly stronger than that of news in both HLF and HLT. For responses with keywords outlined, these differences are significant and even augmented, and interestingly, the differences between LHF news and the other two groups of news are more significant in M1 (Fig. S23A), implying audiences of highly retweeted fake news are more incentivized in terms of anxiety management. The statistics and K-S tests are shown in Table S24 and Table S25.

\begin{figure}
	\centering
	\includegraphics[scale=0.4]{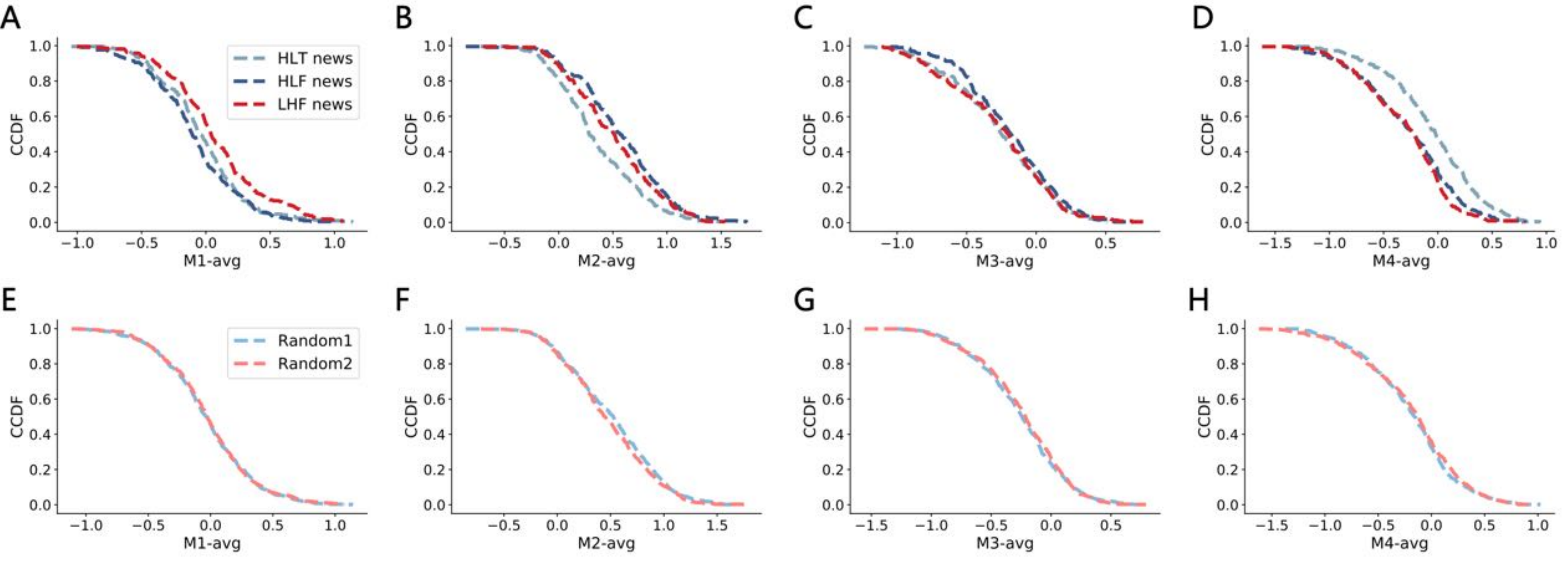}
	\caption{The CCDFs of motivations. (A to D) CCDFs for the motivations in different groups of news with marked keywords. (E to F) CCDFs for the motivations in the two groups separated randomly.}
	\label{fig:moti_emo_ccdf_1}
\end{figure}

\begin{table}
	\centering
	\includegraphics[]{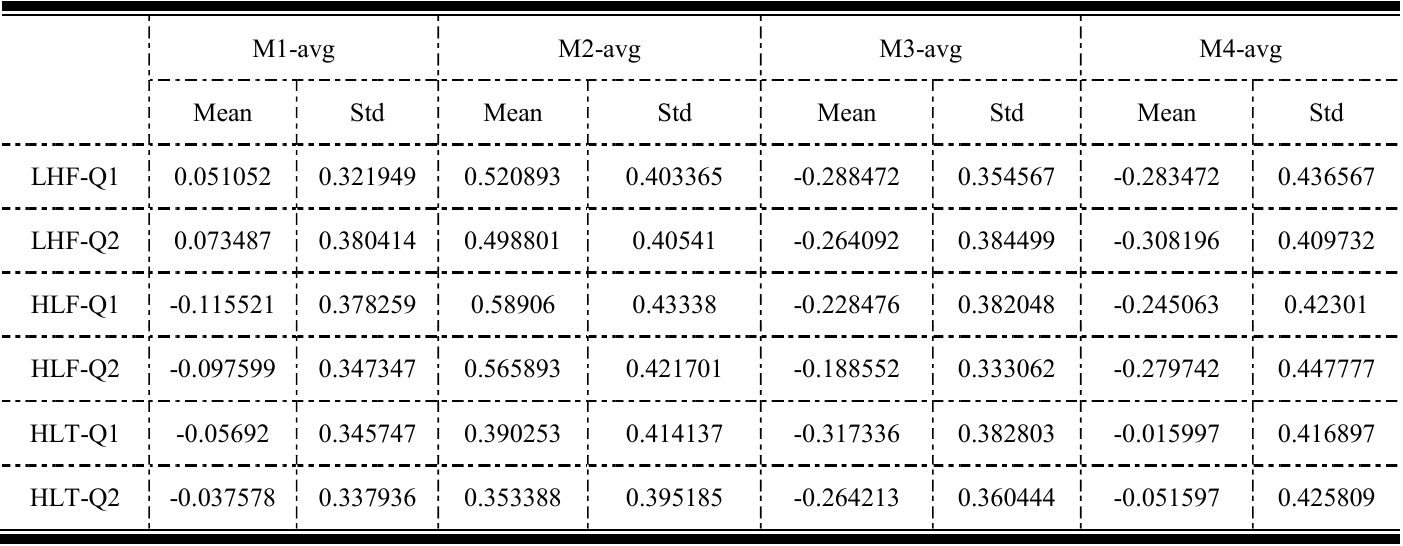}
	\caption{The statistics of each motivation in each questionnaire.}
	\label{table:moti_mean}
\end{table}

\begin{table}
	\centering
	\includegraphics[]{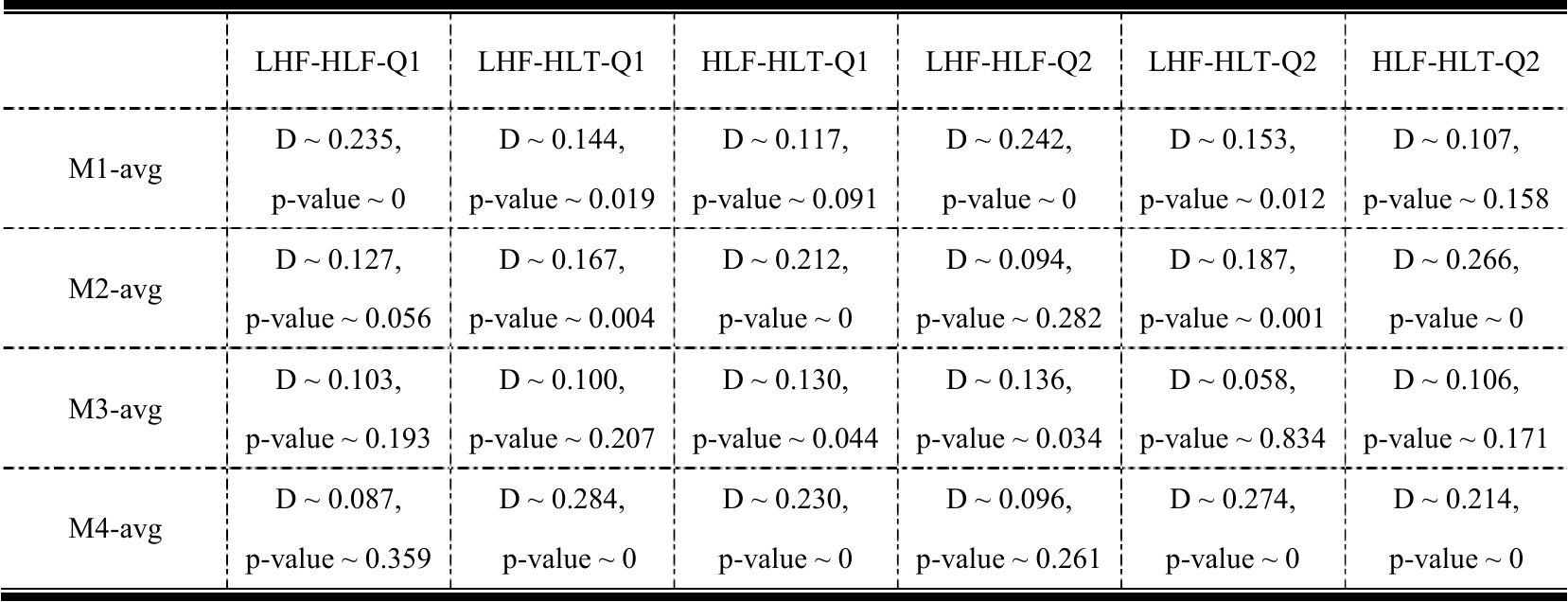}
	\caption{The results of K-S tests}
	\label{table:moti_ks}
\end{table}

\subsection*{S13.2 Differences in motivations between anger and joy}
Next, we divide the news in the questionnaires according to the emotions it carries with the largest occupation. News1 and News5 in LHF news are dominated by anger. Joy dominates News2 in LHF news and News1, 3, 4, 5 in HLT news. The rest of the news is dominated by other emotions. In the analysis in S13.1, we found that the marked keywords play a role in widening differences. Hence, we directly combine the responses without keywords and those with keywords according to their dominant emotions to further examine the emotions’ stimuli with respect to retweeting motivation. The results are analyzed in the main text, and the K-S test results are shown in Table S26. Furthermore, in terms of neglecting emotion dominance, all the data of the questionnaires are divided into two groups randomly to analyze the difference in motivations. Surprisingly, no significant differences were observed in the four motivations (Fig. S23E-H) (anxiety management: K-S test $\sim$ 0.040, P $\sim$ 0.673; information sharing: K-S test $\sim$ 0.062, P $\sim$ 0.168; relationship management: K-S test $\sim$ 0.053, P $\sim$ 0.317; self-enhancement: K-S test $\sim$ 0.059, P $\sim$ 0.200), suggesting the significance of the different incentives provoked by anger and joy. 

\begin{table}
	\centering
	\includegraphics[]{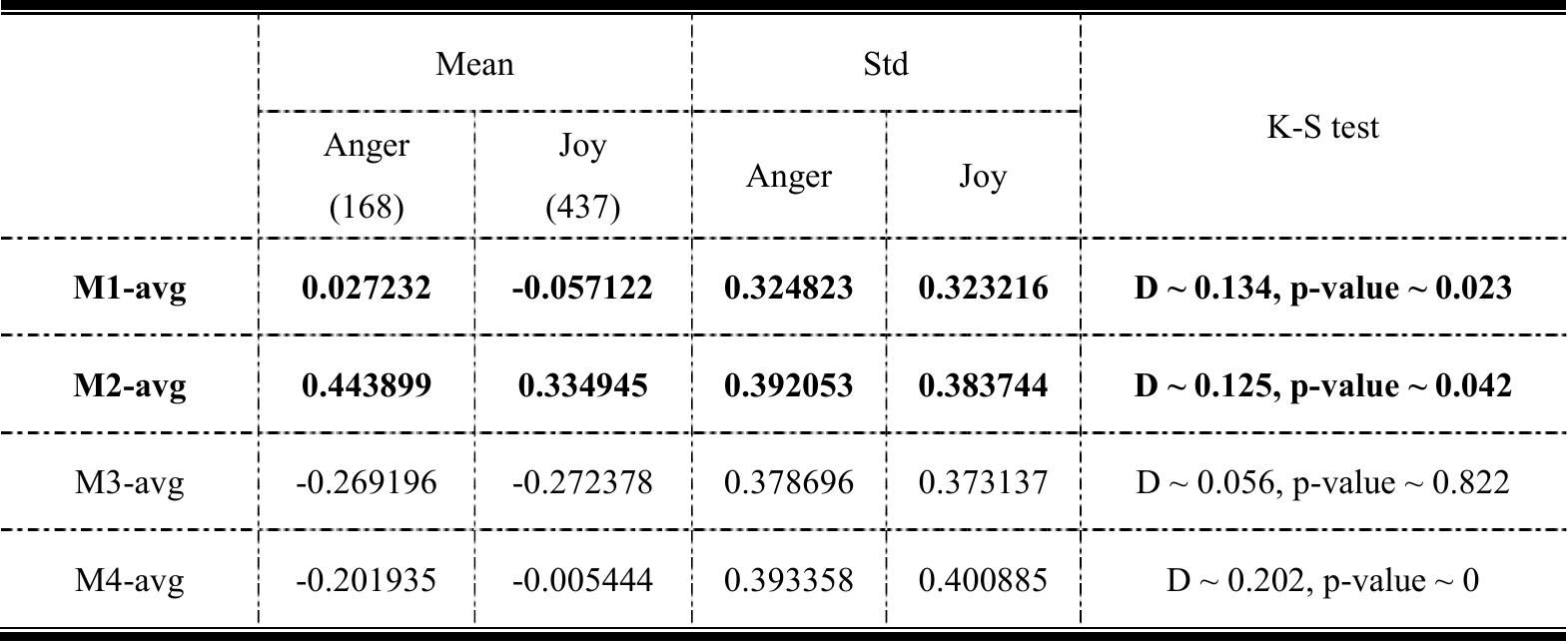}
	\caption{The statistics and K-S tests for anger and joy.}
	\label{table:anger_joy_ks}
\end{table}
\clearpage

\section*{S14 Preventing fake news by tagging and warning of anger}
Carrying more anger makes fake news more viral than real online news. According to this conclusion, instead of determining new features in fake news detection, developing new cues of tagging anger on social media is a promising approach to restrain the spread of fake news at the source. Because the intervention can be implemented immediately after posting, there will be no lag in the fight against fake news. More importantly, the principle of guaranteeing the freedom of speech will be respected, and an acceptable trade-off between free sharing and fake news prevention can be achieved. By alerting users of angry tweets, audiences can be persuaded to assess them more critically before emotionally retweeting, consequently leading to less emotional and more rational retweeters. Specifically, for tweets (news) that deliver too much anger, e.g., the occupation of anger surpasses a predetermined threshold ($\theta$), a retweeting warning could be provided on platforms such as Twitter, Facebook, and Weibo. According to a report from Facebook on battling misinformation related to COVID-19, warning labels can effectively prevent 95\% users from further accessing items (\url{https://about.fb.com/news/2020/04/covid-19-misinfo-update/}). In accordance with this, it is very optimistically assumed here that no angry tweets with warning tags from the platform will be retweeted. To determine the value of $\theta$, we focus on news with high volumes of retweets (HT news and HF news in our data) and define a measure to optimize $\theta$, i.e., preventing fake news that will be highly retweeted but not real news that will be popular. The measure is denoted as $\beta$ and is defined as

$$\beta=\frac{N_{HF\left(\geq\theta\right)}}{N_{HF}}-\frac{N_{HT\left(\geq\theta\right)}}{N_{HT}},$$

\noindent where 
\begin{itemize}
	\item $N_{HF}$ is the number of HF news items.
	\item $N_{HF\left(\geq\theta\right)}$ is the number of HF news items with an occupation of anger greater than $\theta$.
	\item $N_{HT}$ is the number of HT news items.
	\item $N_{HT\left(\geq\theta\right)}$ is the number of HT news items with an occupation of anger greater than $\theta$.
\end{itemize}

The values of $\beta$ for $\theta$ values increasing with a step size 0.1 and 0.05 are shown in Fig. S24 and Fig. S25, respectively, and the values peak when $\theta=0.2$. In our dataset from Weibo, warning about news in which anger occupies more than 20\% will efficiently and effectively prevent 46\% of highly retweeted fake news and only influence the circulation of 22\% of popular real news. In addition, according to $\theta=0.2$, we defined a variable H-Anger, which is 1 for a news item if the ratio of delivered anger is greater than 0.2, and 0 otherwise. A logit model for HF news and HT news items was then built, and the results showed that a H news item with anger greater than 0.2 (i.e., H-Anger equals 1) is 165\% more likely to be fake news than real news (Table S27). Besides, for all of the highly retweeted news items in our dataset (i.e., HF+HT), HF news items account for 89\% of those with an occupation of anger higher than 0.2, implying further that our treatment can predominantly target highly retweeted fake news items. Though the fraction of prevented fake news items that otherwise would be widely circulated is not as high as expected, considering the intrinsic characteristics of the intervention, i.e., very low cost and timely, the newly presented treatment should be weighted with high priority in the toolbox of mitigation strategies against fake news. Hence, it is worth employing on social media platforms such as Weibo, Twitter or Facebook to prevent the spread of fake news online at the source through this new approach.

\begin{figure}
	\centering
	\includegraphics[]{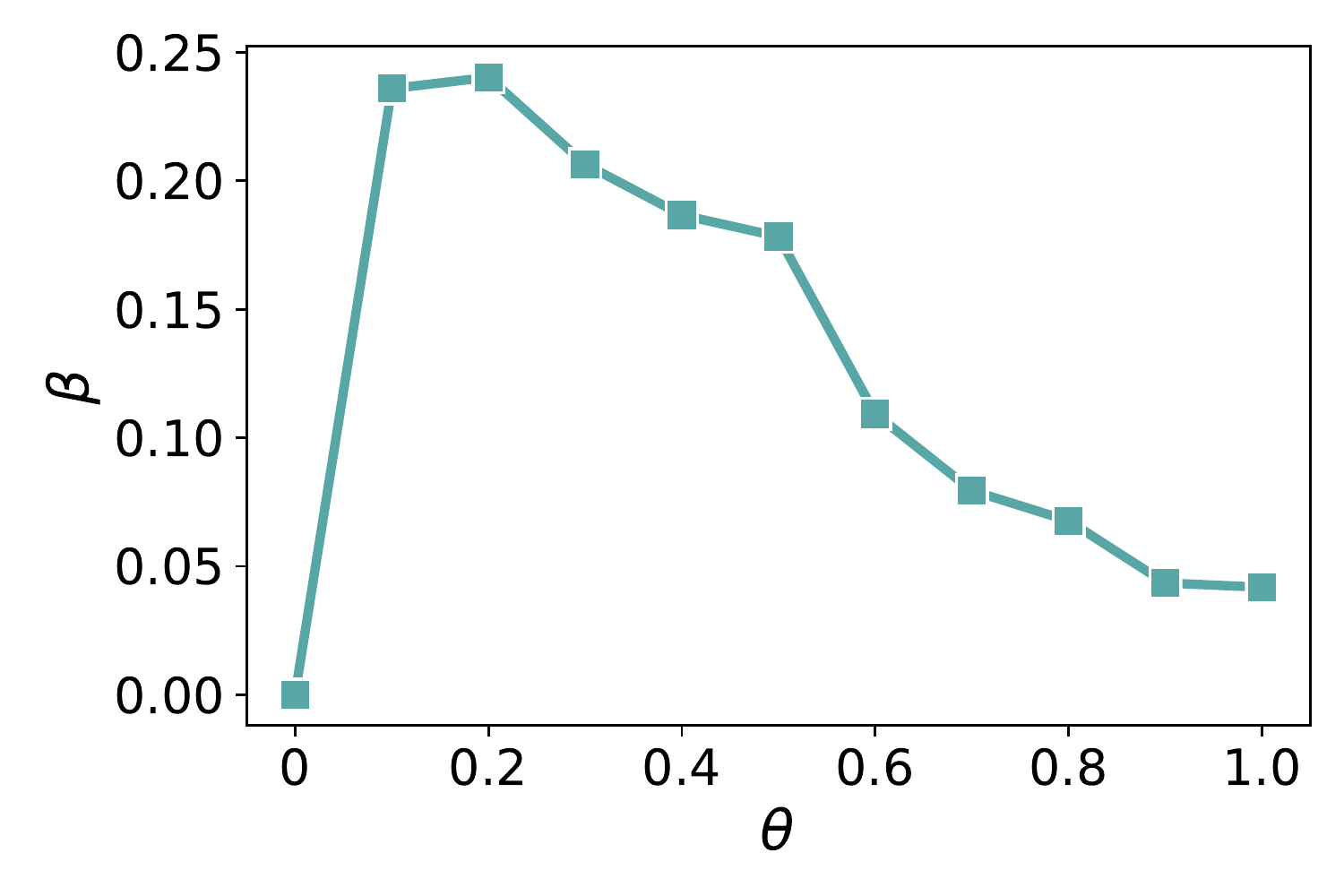}
	\caption{The value of $\beta$ with $\theta$ growing by 0.1.}
	\label{fig:theta_01}
\end{figure}

\begin{figure}
	\centering
	\includegraphics[]{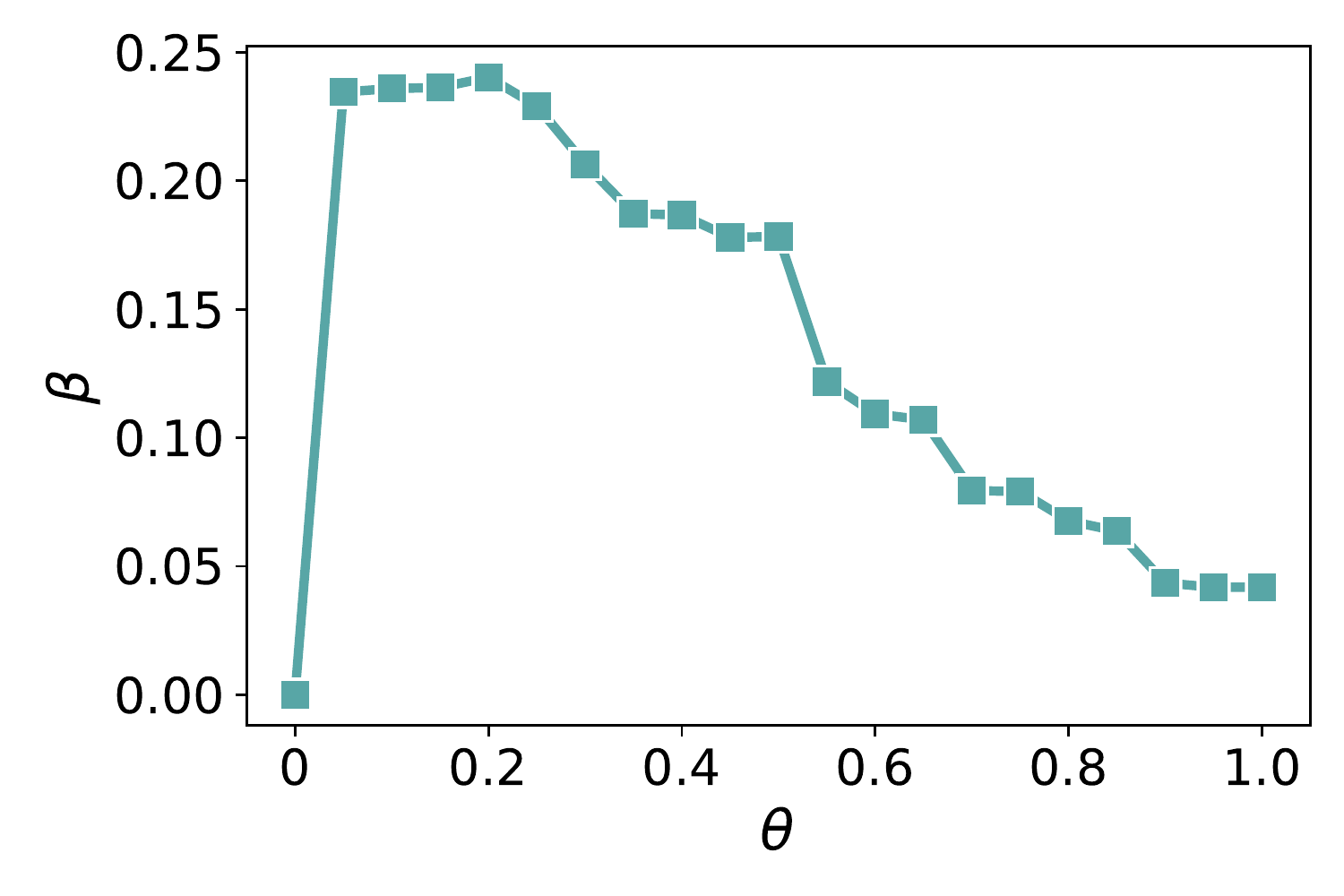}
	\caption{ The value of $\beta$ with $\theta$ growing by 0.05.}
	\label{fig:theta_005}
\end{figure}

\begin{table}	
	\centering
	\includegraphics[scale=0.8]{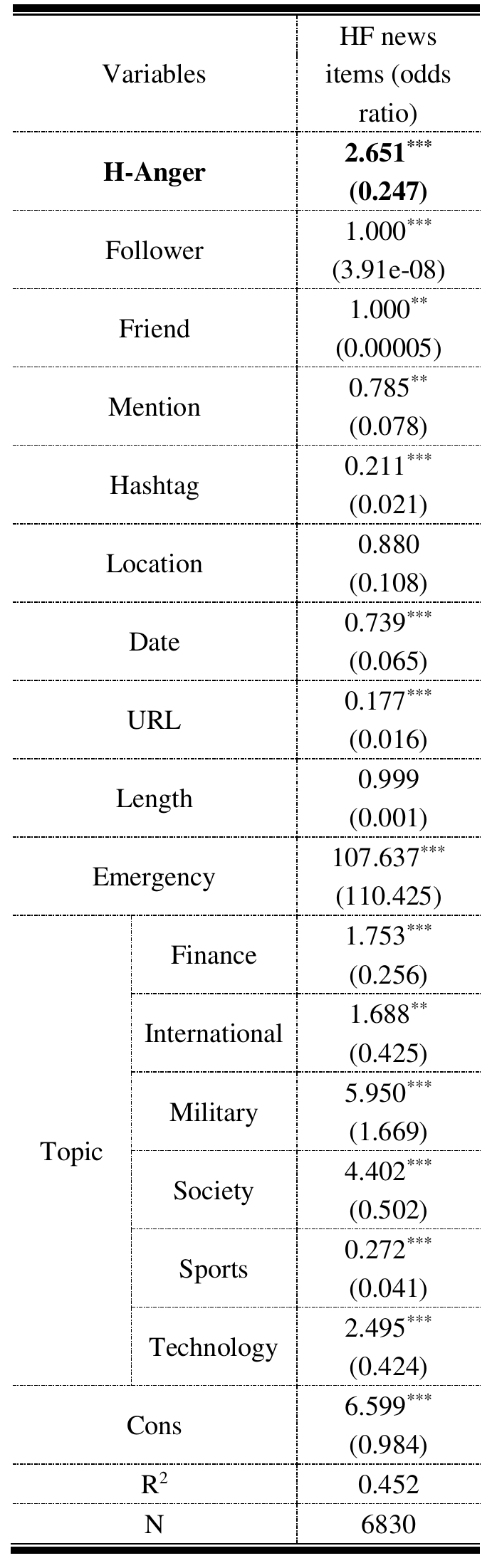}
	\caption{The logit model (odds ratio) for HF news and HT news items. $^\ast P<0.1, ^{\ast\ast} P<0.05, ^{\ast\ast\ast} P<0.01$.}
	\label{table:logit_hf_odds}
\end{table}

\end{document}